\documentclass[traditabstract]{aa}
\usepackage[labelfont=bf, labelsep=colon]{caption} 
\usepackage{url}
\usepackage{graphicx}
\usepackage{longtable,ltcaption}
\usepackage{rotating}
\usepackage{lscape}
\usepackage{sidecap}
\usepackage{amsmath} 
\usepackage{subcaption}
\usepackage{slantsc}
\usepackage{xcolor}
\usepackage[varg]{txfonts}
\usepackage{isotope}
\usepackage{siunitx}
\usepackage{ulem}
\usepackage{bm}
\usepackage{natbib}
\usepackage{url}
\usepackage{xspace}
\usepackage{placeins}
\usepackage{capt-of}
\usepackage{booktabs}
\usepackage{array}
\newcolumntype{V}{c} 
\bibpunct{(}{)}{;}{a}{}{,}

\usepackage[colorlinks=true,citecolor=blue,linkcolor=magenta,urlcolor=blue]{hyperref}
\usepackage[switch,columnwise]{lineno}

\newcommand{\msun}{$\text{M}_\odot$}
\newcommand{\teff}{$T_{\text{eff}}$}
\newcommand{\logg}{$\log g$}
\newcommand{\logy}{$\log{n(\mathrm{He})/n(\mathrm{H})}$}

\newcommand{\y}{${n(\mathrm{He}/\mathrm{H})}$}

\newcommand{\teffs}{effective temperatures }

\newcommand{\logL}{$\log L$}

\begin{document}


\title{Hot subdwarf stars from the Hamburg Quasar Survey}

\author{U.~Heber \inst{1}
   \and L. Kufleitner \inst{1}
   \and M.~Dorsch \inst{2}
\and M. Latour \inst{3}
\and H. Dawson\inst{2}
\and F. Mattig \inst{2}
\and S. Geier \inst{2}
 }

\offprints{U.\,Heber,\\ \email{ulrich.heber@fau.de}}

\institute{Dr.~Karl~Remeis-Observatory \& ECAP, Astronomical Institute, Friedrich-Alexander University Erlangen-Nuremberg, Sternwartstr.~7, 96049 Bamberg, Germany
\and 
Institut f\"ur Physik und Astronomie, Universit\"at Potsdam, Haus 28, Karl-Liebknecht-Str. 24/25, 14476 Potsdam-Golm, Germany
\and
Institut für Astrophysik und Geophysik, Georg-August-Universität Göttingen, Friedrich-Hund-Platz 1, 37077 Göttingen, Germany
}

\date{Received \ Accepted}

\abstract{Hot subluminous stars (sdO/B) are evolved low mass stars originating from red giants that lost their envelope almost entirely. The multitude of observed phenomena imply that several pathways may form hot subdwarfs, most involving close binary channels. The Hamburg Quasar Survey (HQS) led to the discovery of many faint blue stars including hot subdwarf stars. Many of the HQS-sdB stars have been studied in detail, but analyses of the helium-rich sdOB and sdO stars are lacking. The recent development of hybrid LTE/non-LTE model spectra (\textit{2$^{nd}$ generation Bamberg model grids}) enables us to improve the spectroscopic analyses of the sdB stars as well as of the previously unstudied sdO stars allowing precise atmospheric parameters to be derived, while consistently accounting for parameter correlations and systematic uncertainties.  The \textit{Gaia} mission provided astrometric data of unprecedented quality which allow fundamental stellar parameters to be derived from the atmospheric parameters via parallax measurements. 
We use spectral energy distributions to identify composite-colour sdB binaries and present the result of detailed spectroscopic analyses of {122} non-composite subdwarfs from the HQS to identify potential evolutionary pathways. Comparison to evolutionary tracks both in the Kiel (\teff-\logg) and the physical Hertzsprung-Russell (\teff-\logL) diagram finds the location of the sdB stars on the extreme horizontal branch. Their derived mass distribution and median mass of 0.45\,M$_\odot$ is consistent with the canonical EHB mass. We revisited the sample of known pulsating HQS-sdB stars and find no significant differences between their mass distributions and those of sdB stars not known to pulsate. The helium-rich sdOB and sdO stars, are found near the helium main-sequence (He-MS). The derived mass distribution of the extremely He-rich subdwarfs is broader (0.48 to 1.05\,M$_\odot$) and peaks at a median of 0.70\,M$_\odot$, significantly larger than those of the hydrogen-rich stars. Intermediate He-rich subdwarfs are also He-MS stars, but of lower mass (0.55\,M$_\odot$) than the extremely He-rich. This strongly supports the merger scenario for the origin of He-rich sdO stars, in which two helium white dwarfs merge following orbital decay driven by gravitational-wave emission, producing a He-rich sdO or sdOB star. From comparison to the results of similar studies we speculate that older populations produce more massive He-WD mergers.

}

\authorrunning{Heber et al.}
\titlerunning{Hot subdwarf stars from the Hamburg Quasar Survey}

\maketitle

\section{Introduction} \label{sec:intro}

Most hot subdwarf stars (sdB, sdO) are low mass stars powered by helium fusion in the core, or by shell burning for the more evolved stars. In the canonical evolutionary scenario, they form the progeny of red giant stars (RGB) whose hydrogen envelopes have been stripped off near the tip of the RGB, exposing the helium-burning core following helium ignition \citep[see][for reviews]{2009ARA&A..47..211H,2016PASP..128h2001H}. 
The sdB stars are located on the extreme horizontal branch (EHB), exhibiting temperatures ranging from 20,000 to 40,000 K, masses of about half a solar mass, radii from 0.1 to 0.25 solar, and luminosities of about 10 to 30 solar luminosities. The sdB phase lasts for on the order of a hundred Myr depending mostly on the mass of the sdB \citep{2003MNRAS.341..669H}.
Once helium is exhausted in the core, the stars evolve through the sdO phase, where temperatures exceed
40,000\,K and the luminosity increases. The mass distribution of these canonical hot subdwarfs is expected to peak at the core mass required for helium ignition under degenerate conditions of 0.45 to 0.5\,\msun\ depending on metallicity. 
The loss of the envelope is most easily achieved by mass transfer in close-binary systems via Roche lobe overflow (RLOF), or common envelope formation and ejection. Therefore, it has been suggested that hot subdwarfs should reside in close binaries \citep{1976ApJ...204..488M}, a prediction that was later confirmed by the detection of many hot subdwarfs in short period (hours to days) binaries with white dwarf or low mass main sequence companions \citep[e.g.][]{Maxted2001}. 
Apparently single hot subdwarfs can be formed by mergers of two helium white dwarfs driven by gravitational wave emission and magnetic braking  and should be located near the helium main sequence in the Hertzsprung-Russell diagram \citep[HRD,][]{1984ApJ...277..355W}. Their mass distribution is predicted to be much wider than that of the canonical subdwarfs \citep{2003MNRAS.341..669H} and may produce more massive and helium-rich hot subdwarfs up to about 0.9\msun. 

Disentangling the various evolutionary pathways requires the atmospheric parameters of homogeneous samples of stars via quantitative spectral analyses using sophisticated model atmospheres and synthetic spectra. 
{To} derive the fundamental stellar parameters, the distances of the stars need to be known. Here, parallax measurements by the 
\textit{Gaia} mission are crucial, because they allow the atmospheric parameters to be converted to stellar parameters (radius $R$, luminosity $L$, and mass $M$). 
Homogeneous samples of subdwarfs can be drawn from large surveys such as LAMOST 
\citep{Luo2021,2024ApJS..271...21L} and SDSS \citep{2024A&A...690A.368G}. A brief summary of these developments since the advent of the \textit{Gaia} data releases can be found in \citet{2024arXiv241011663H}. Early photographic surveys for faint blue stars at high Galactic latitudes such as the Palomar-Green \citep[PG, ][]{1986ApJS...61..305G}, the Byurakan \citep{1987IAUS..121...17L,1987IAUS..121...25M}, and the Hamburg objective prism surveys \citep{1995A&AS..111..195H,1996A&AS..115..227W}, and the associated spectroscopic follow-up campaigns have been important for the development of the field. It is timely to revisit these samples combining \textit{Gaia} data with state-of-the-art model atmospheres and analysis techniques. \citet{2025arXiv251102539L} did so for the PG subdwarf sample. Here we deal with the hot subdwarfs from the Hamburg Quasar Survey (HQS).  

\subsection{The Hamburg Quasar Survey}

The HQS was a photographic wide-angle objective prism survey to hunt for low redshift quasars using the 80 cm Schmidt telescope at the German-Spanish Astronomical Centre (DSAZ) on Calar Alto, Spain \citep{1995A&AS..111..195H, 1996A&AS..115..235R, 1999A&AS..134..483H}. 

The survey was carried out from 1980 to 1997 with
coverage of $\approx13\,600\,\mathrm{deg}^2$ of the Northern sky ($\delta>0\deg$)
at high Galactic latitudes ($|b|\gtrsim 20\deg$). The magnitude range of the
survey is $13 \la B \la 18.5$\,mag with spectral coverage from
3,400 to 5,400\,\AA\ and a resolution of $\sim45$\,\AA\ at $H_{\gamma}$.
The plates were scanned and are available through the APPLAUSE database \citep{2024A&A...687A.165E}. Blue objects were classified after visual inspection as candidates for quasars, hot stars, and narrow emission line objects.  

Quasars (quasi-stellar objects, QSO) are blue and faint objects.
Therefore, the QSO population is expected to be outnumbered by evolved stars such as hot white dwarfs, cataclysmic variables, and hot subdwarfs. QSOs can be identified from their strong emission lines, which can be seen in the objective prism spectra despite their low resolution \citep{1988ASPC....2..143E}. 
However, it may be difficult to distinguish objective prism spectra of QSO from those of metal-poor halo stars, if their redshift is incompatible with the wavelength coverage of the spectra \citep{1989A&A...223L...1G}. 
Classification of faint blue stars at such low resolution is difficult \citep[see][for details]{2001ASPC..226..397H}.
Hydrogen-rich white dwarfs of intermediate temperatures can be identified through their  characteristically  broad Balmer absorption lines. For other hot stars, the resolution is too low to resolve spectral features and, therefore, need follow-up spectroscopy to unravel their nature. 

Follow-up campaigns of visually selected hot star candidates were carried out at the Calar Alto observatory, mostly using the TWIN Spectrograph at the 3.5m telescope, resulting in a sample of 400 faint blue stars \citep{1991ASIC..336..109H}. About half of them turned out to be hot white dwarfs \citep[e.g.][]{1995A&A...303L..53D,1996A&A...311L..17H,1998A&A...338..563H,1998A&A...338..651R}. 
Others turned out to be cataclysmic variables \citep[more than 50 were found in the HQS survey; ][]{2002ASPC..261..190G,2005A&A...443..995A,2009ASPC..404..276A}.
Our follow-up spectroscopic observations also contributed significantly to the small class of 
PG~1159 stars \citep[e.g.][]{1994A&A...286..463D,1996A&A...309..820D}, 
and led to the discovery of a new subtype of hot helium-rich white dwarfs showing absorption lines of ultra-highly excited ionisation stages \citep[e.g. O\,{\sc viii}, Ne\,{\sc ix}, Ne\,{\sc x},][]{1995A&A...293L..75W}, pointing at the existence of wind-fed circumstellar magnetospheres \citep{2019MNRAS.482L..93R}. 

\subsection{Hot subdwarfs from the HQS survey}

\begin{figure}
\centering 
\includegraphics[width=\columnwidth]{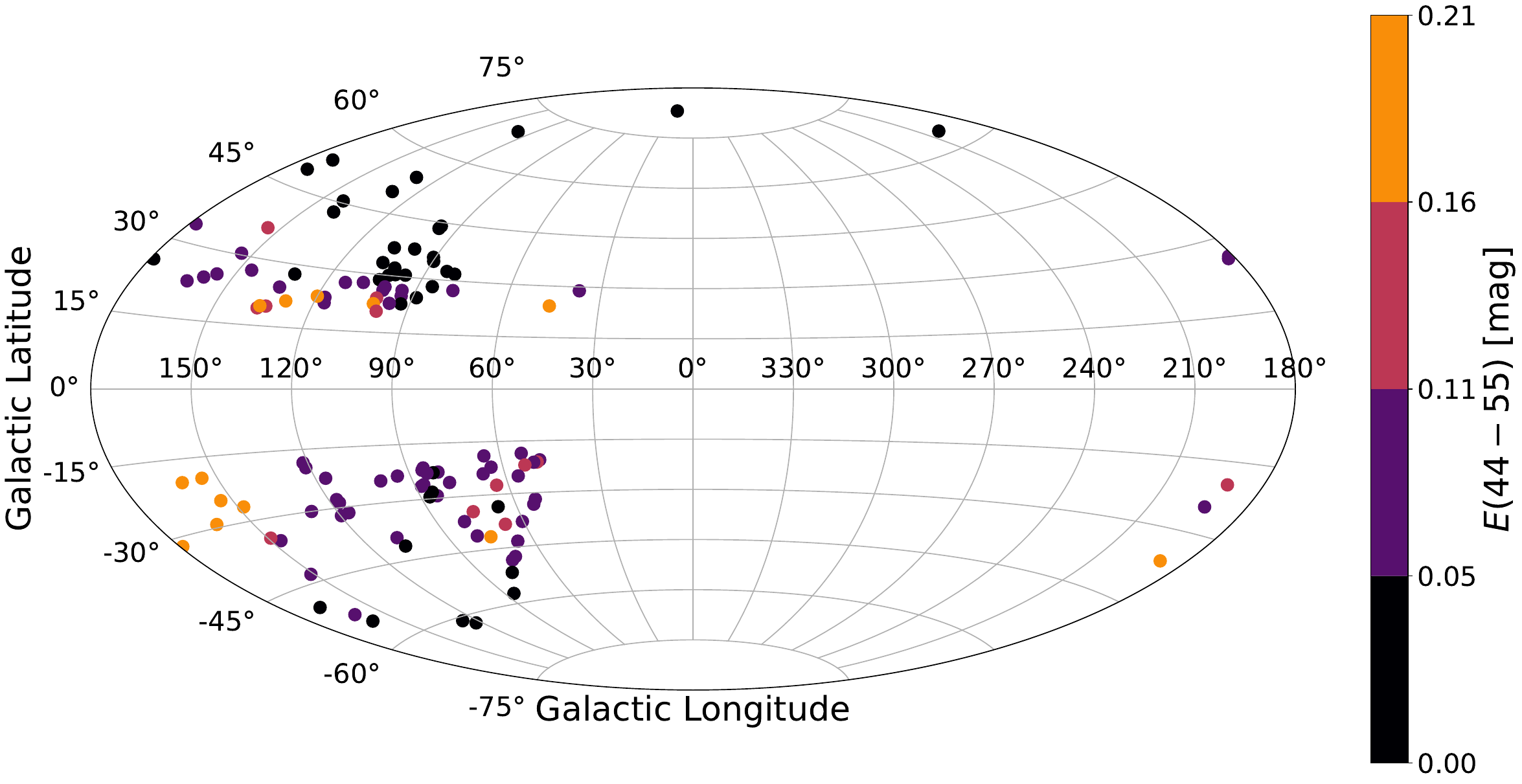}

\includegraphics[width=0.8\columnwidth]{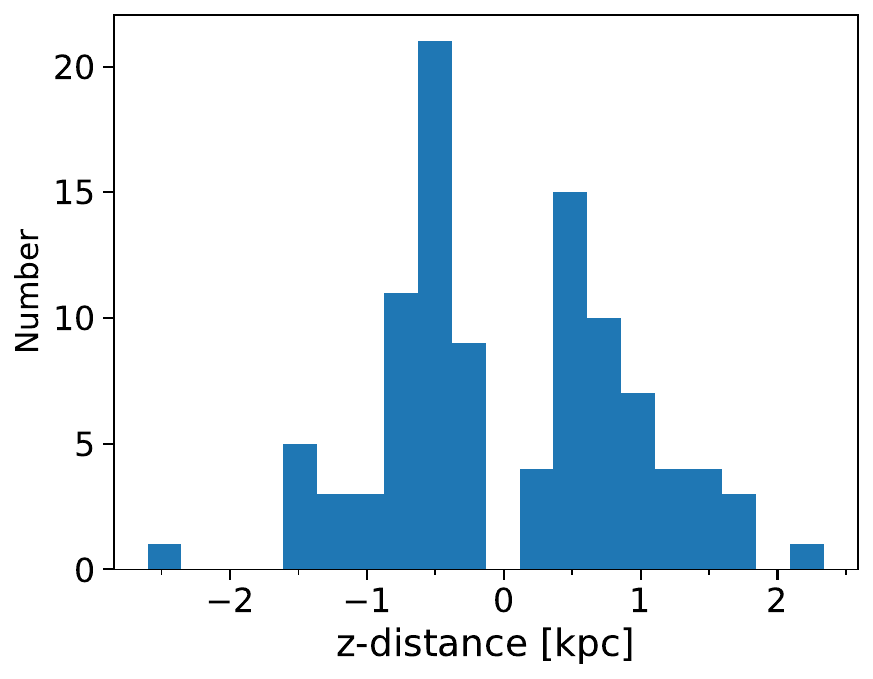} 
\caption{\textit{Top}: Distribution of the non-composite HQS subdwarfs in Galactic coordinates, colour coded with the monochromatic interstellar reddening parameter $E(44-55)$ (see Sect. \ref{sect:ism}); \textit{bottom}: a histogram of {distances from the Galactic plane $z$ for all stars with parallax errors better than 25\%} (bottom) calculated from their \textit{Gaia} parallaxes (see Sect. \ref{sect:astrometry} for details).
} 
\label{fig:sky1}
\end{figure}

The remainder of observed faint blue stars were classified as hot subdwarfs of B- and O-type. They are located in both the Southern and the Northern Galactic sky, mostly at intermediate latitudes ($|b|=20^\circ$ to 60$^\circ$) and longitudes between $l=45^\circ$ and 160$^\circ$ and distances from the Galactic plane up to $|z|\approx2.5$\,kpc (Fig.~\ref{fig:sky1}).

Optical spectra have been obtained in follow-up campaigns and 107 sdB and sdOB stars have been analysed by \citet{2003A&A...400..939E} 
using the same metal line-blanketed LTE atmospheres and non-LTE models as in \citet{2000A&A...363..198H}, while another 58 subdwarfs were classified as mostly helium-rich sdO types \citep{1997fbs..conf..375L}, but no quantitative spectral analyses have been carried out for all of them. 
High resolution spectra of another nine sdB stars were analysed in the same way by \citet{2005A&A...430..223L}. Composite hot subdwarf binaries with {F,G,K} type main-sequence companions have been found via infrared photometry 
\citep{2003AJ....126.1455S,2005A&A...430..223L}.

{In this work, we shall carry out a comprehensive spectroscopic and photometric 
analysis of a selected sample of non-composite hot subdwarf stars from the HQS project.}
We present the analyses of the stars, the derived results, and our interpretations as follows.
We first introduce the observational material (observed spectra and photometric light curves) and our final sample of {\bf
122} stars selected for the analyses in Sect.~\ref{sec:data}. 
The grids of hybrid LTE/non-LTE model atmospheres and synthetic spectra used in the global fitting procedure are described in Sect. \ref{sect:model_atmospheres}. We carefully study systematic uncertainties
(Sect. \ref{sect:systematics}), and compare the new results for sdB stars with that to a previous study in Sect. \ref{sect:sdb_compare}.
The atmospheric parameters derived and the distribution of the stars in the Kiel diagram are presented in Sect. \ref{sect:atmos_parameters}. We then discuss the spectral energy distributions (SEDs) and \textit{Gaia} parallaxes used to derive stellar parameters (Sect. \ref{sect:sed}). The distribution of the stars in the physical Hertzsprung-Russel Diagram (i.e. \teff--\logL) and the mass distributions of different subdwarf subtypes are compared in Sect. \ref{sect:stellar_parameters}. Conclusions from a comparison to similar studies are drawn in Sect. \ref{sect:comparison_studies} and 
 we conclude the paper and give an outlook in Sect. \ref{sect:conclusion}.

\section{Observational data and sample selection}\label{sec:data}

This study is based on optical spectra (Sect. \ref{sect:spec_class}), time series photometry (Sect. \ref{sect:lc}), photometric measurements from the ultraviolet to the infrared regime (Sect. \ref{sect:photometry}), and parallaxes from \textit{Gaia} DR3 (Sect. \ref{sect:astrometry}).

\begin{figure*}[h!]
    \centering
    \includegraphics[width=0.9\textwidth]{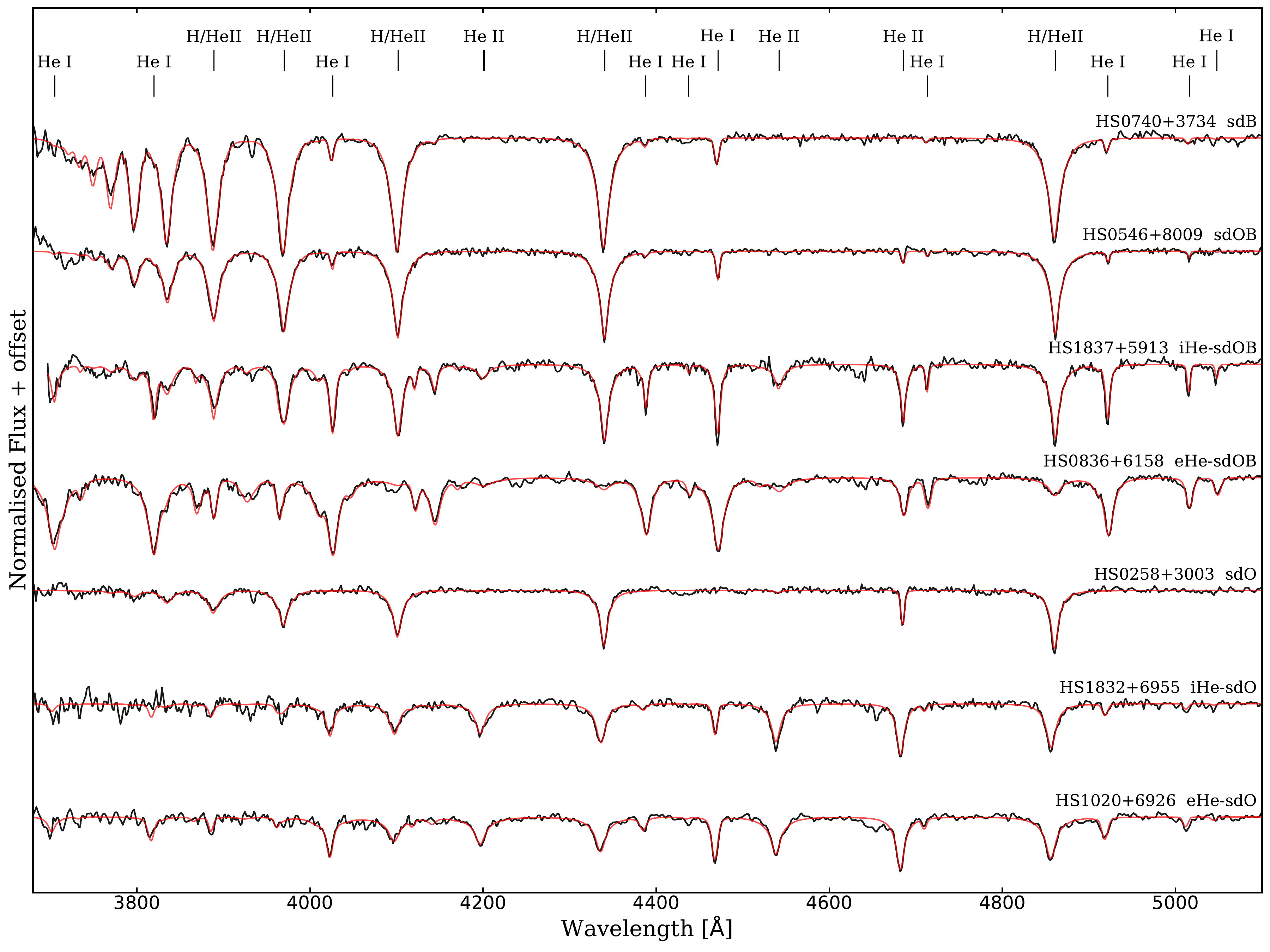}
    \caption{Example spectral fits (red) to DSAZ spectra (black) of the main types of hot subdwarfs in HQS.
    }
    \label{fig:example_spectra}
\end{figure*}

\subsection{Spectroscopic observations and classification}\label{sect:spec_class} 

Optical spectra have been obtained during 14 observing runs at the DSAZ, Calar Alto observatory in Spain between 1989 and 1998, using
the TWIN spectrograph mounted at the 3.5m telescope and CAFOS at the 2.2m, which provided spectral resolutions from 3.4\,\AA\ to 8.0\,\AA\ \citep[for details see Table 1  of][]{2003A&A...400..939E}. Their signal-to-noise ratio ranges from 17 to 160 at a median of 56. Additional medium-resolution DSAZ spectra were taken from \citet{2002A&A...389..180S} and \citet{2002A&A...386..249D}. The DSAZ spectra are complemented by data retrieved from the SDSS DR17 \citep{2023ApJS..267...44A} and LAMOST DR10 \citep{2022Innov...300224Y} databases, available for $\approx$30\% of the sample, only. 

Nine sdB stars, including six non-composite ones, have been {also} observed at high spectral resolution \citep[$R$=18,500 or better,][]{2005A&A...430..223L} with UVES at the ESO VLT at two epochs each, and three of them were found to be radial velocity variable \citep{2004Ap&SS.291..321N}. These spectra have been analysed in the same way as the low-resolution ones. 

Rather than attempting to classify the individual spectra in a classical procedure \citep{1990A&AS...86...53M,2013A&A...551A..31D,2024A&A...690A.368G} \footnote{
Spectral atlases at $\approx$1\AA\ resolution are provided by \citet{2025arXiv251102539L} and \citet{2021MNRAS.501..623J}  demonstrating their diversity.}, we use the derived \teff\ to formally distinguish sdB (\teff=20 to 30kK), sdOB (\teff=30 to 40kK) and sdO ($>$40kK) stars and the derived \logy\ to separate the sample into three helium categories: He-poor, intermediate He-rich (iHe) and extremely He-rich (eHe) subdwarfs.  Hence, we shall use nine categories.  It has become common practice to distinguish iHe from eHe subdwarf at an abundance of \y=$4$ \citep[see][]{2016PASP..128h2001H}. The solar helium abundance has traditionally been used to tell apart He-rich from He-poor subdwarfs. However, recent studies by \citet{2025arXiv251102539L} and \citet{Dawson2025} have shown that this limit should be somewhat lowered from the solar value (\logy=$-1.0$) to \logy=$-1.2$, as we shall do in this paper. Example spectra of HQS subdwarfs for all subtypes are shown in Fig. \ref{fig:example_spectra}.

\subsection{Photometric variability, pulsations, and binarity}\label{sect:lc}

{A significant fraction of hot subdwarfs show photometric variability, either due to pulsations or binarity, and, on occasion, from both effects at the same time. Many of the hot subdwarfs from the HQS have been followed-up with time series photometry to identify such variability effects. We reviewed the literature to identify the variable stars in our sample and we retrieved photometric data from the the Zwicky Transient Facility \citep[ZTF,][]{2014htu..conf...27B}
to search for additional large-amplitude variables (see Appendix \ref{sect:pulsation_reflect}). 

\paragraph{Pulsations:} Multi-periodic radial and non-radial pulsations have been observed in hot subdwarfs. Two types of pulsation can be distinguished: p-modes for which the restoring force is the pressure force and g-modes with buoyancy as restoring force
\citep[see][for a review]{2021RvMP...93a5001A}. The observed oscillation periods of p-modes are short (few minutes) and longer for g-modes ($\approx$45--250 min). P-mode pulsators are found at higher \teff\ ($\approx$28,000 to 36,000 K), while g-mode pulsators are somewhat cooler (22,000 to 30,000 K). Hybrid pulsators showing both p-mode and g-mode pulsations are found mostly at intermediate temperatures. Ten p-mode pulsators and nine g-mode pulsators have been found among the HQS subdwarfs. 
{More details are given in Appendix \ref{sect:pulsation_reflect}.}  

\paragraph{Binarity: reflection effect and eclipsing hot subdwarfs. }

Eight close binaries with late-type main-sequence companions are known among the HQS subdwarfs and exhibit a reflection effect; two are also eclipsing systems. The analysis of the ZTF light curves, presented in Appendix \ref{sect:pulsation_reflect}, confirms all eight previously known cases, but reveals no additional objects with detectable photometric variability. In four of the reflection-effect binaries, contamination of the optical spectra by the companion was found to be too severe; these systems were therefore excluded from the sample.

\subsection{Multi-wavelength photometry}\label{sect:photometry}

Photometric measurements are obtained by querying several public databases \citep[see][for a list of the photometric surveys used]{2024A&A...685A.134C,2025AJ....170...86S} and are used to construct spectral energy distributions (see Sect. \ref{sect:sed}).

Hot subdwarfs in binaries with F-, G-, K-type main sequence companions are easy to detect because their spectral energy distributions are composite. The subdwarf dominates the blue part while the cool star contributes in the red and infrared part (see Sect. \ref{sect:composite}). The analysis of the 29 composite spectrum binaries, known from the literature and identified in the course of our analyses (see Sect. \ref{sect:composite}), is beyond the scope of this paper. We also excluded those close binary systems {with strong reflection effect mentioned in the previous subsection}, because their strong light variations indicate that the optical spectra might be contaminated by light scattered back from the companion.

\subsection{Gaia astrometry}\label{sect:astrometry}

\begin{figure}
\centering 
\includegraphics[width=\columnwidth]{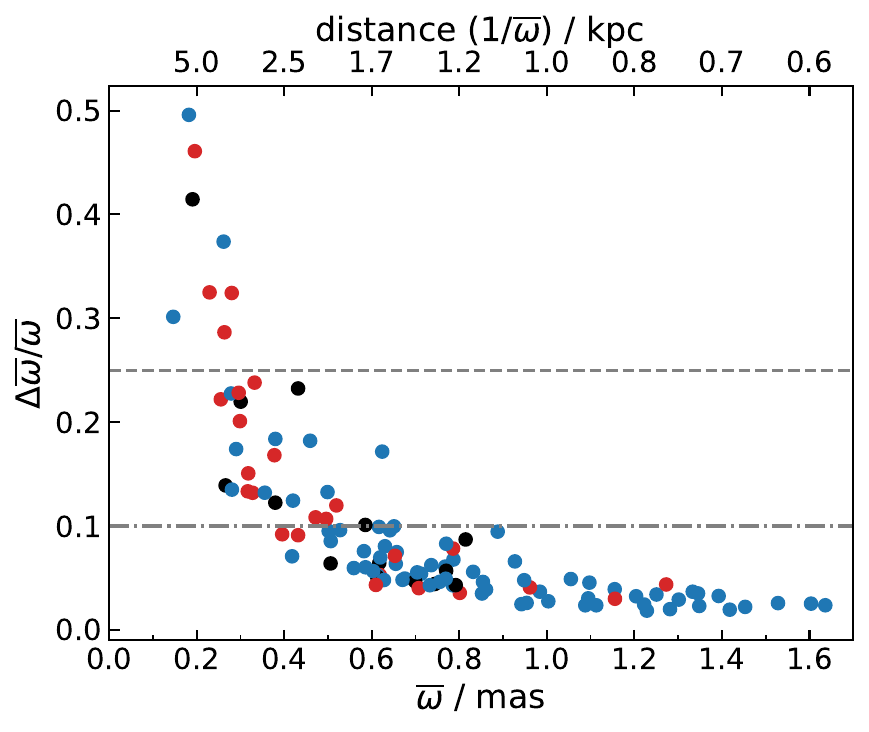}
\caption{
Distribution of parallax uncertainties as a function of parallax. The dashed dotted, and the dashed lines mark the 10\% and 25\% uncertainty levels, respectively. HS0941+4649 and HS1000+4704 are off the scale; their parallax uncertainties are larger than the parallax. He-poor stars are shown in blue, iHe rich ones in {black} and eHe-rich in red.
} 
\label{fig:sky}
\end{figure}

\textit{Gaia} parallaxes \citep{2016A&A...595A...1G,2023A&A...674A...1G} provide the stellar distances, which are crucial to derive stellar parameters. Because some stars are at large distance, their parallaxes are small and relative uncertainties are large, increasing strongly with decreasing parallax. From Fig. \ref{fig:sky} we infer that for about 70\% of the sample the uncertainties are smaller than 10\% and for 90\% lower than 25\%, respectively. The distance distribution peaks at 1.5 kpc with the nearest object at $\approx$ 600 pc (Fig. \ref{fig:sky}). The most distant stars lie beyond 5 kpc.

\subsection{The selected sample of hot subdwarfs from the HQS}\label{sect:sample}

{Our final sample of non-composite hot subdwarfs amounts to {122} stars. About two thirds of them }
are {hydrogen-rich 
and one third is helium-rich. }
While the sdBs have been studied before, similar quantitative spectral analyses of the spectra of the He-rich hot subdwarfs have not been published. Seven stars were excluded {from the sample} because their spectral type turned out to be unrelated to the EHB after performing the spectral analyses (see Sect. \ref{sect:individual}). To derive 
stellar parameters, accurate parallaxes are required. We consider the stars with relative parallax uncertainties larger than 25\% unreliable, which allows us to investigate the stellar parameters for {bf 103} hot subdwarf stars.  

\section{$2^{nd}$ generation Bamberg models: Hybrid LTE/non-LTE model atmospheres and synthetic spectra} \label{sect:model_atmospheres}

Previously, metal line blanketed LTE model atmospheres \citep{2000A&A...363..198H} and metal-free non-LTE models \citep{2007A&A...462..269S} or combinations thereof \citep{2003A&A...400..939E, 2005A&A...430..223L} were used in many studies by our team \citep[see][]{2016PASP..128h2001H}. Through the last decade, we developed large grids of model atmospheres and synthetic H/He line spectra that combine metal-line blanketing and non-LTE effects in an approximate way \citep{2011JPhCS.328a2015P} to cover the entire parameter space of the hot subdwarfs homogenously. These \textit{$2^{nd}$ generation Bamberg model grids}   
of hybrid LTE/non-LTE atmospheres cover a wide range of atmospheric parameters (see Fig. \ref{fig:grid}). 
Because the He abundances of hot subdwarfs vary by seven orders of magnitudes from star to star, grids were calculated to cover the entire range.

In short, our approach makes use of the LTE code ATLAS12, the non-LTE code DETAIL, and the spectrum synthesis code SURFACE. The metal composition of hot subdwarfs is far from solar but shows a characteristic pattern, where low mass elements are underabundant while heavier elements such as the iron group are overabundant \citep{2011PhDT.......261P,2013A&A...549A.110G}. Our model atmosphere calculations used the mean abundance pattern derived by \citet{2011PhDT.......261P} as reported by \citet{2013MNRAS.434.1920N}. 
The ATLAS\,12 code \citep{kurucz2013} was used to compute the atmospheric structure in LTE with the most recent Kurucz line lists to incorporate the line blanketing effect with the opacity sampling technique based on the mean chemical abundance patterns of hot subdwarfs. The hydrogen, He\,\textsc{i}, and He\,\textsc{ii} line spectra were calculated with the DETAIL \citep{Giddings1981,ButlerGiddings85} and SURFACE codes \citep{ButlerGiddings85} allowing for departure from local thermodynamic equilibrium. Recently, all three codes have been modified \citep[see][for details]{2018A&A...615L...5I,2021A&A...650A.102I} to incorporate the occupation probability formalism \citep{1988ApJ...331..794H} for level dissolution. Consistent line broadening tables for hydrogen \citep{2009ApJ...696.1755T} have been incorporated. Both  modifications are important to model the confluence of lines near the Balmer jump \citep{1939ApJ....90..439I}. Helium line broadening has also been improved by implementing the corrected tables of \cite{1997ApJS..108..559B}, \cite{2009A&A...503..293G}, and \cite{2012A&A...542A..75L}. A rigorous treatment of non-LTE effects on the atmospheric structure is not possible in the ADS approach. However, ATLAS\,12 offers the opportunity to include departure coefficients, which were calculated for small model atoms of hydrogen and helium with DETAIL and passed back to ATLAS\,12 refining the atmospheric structure iteratively.  This allowed us to incorporate non-LTE effects on the atmospheric structure in an approximate way.

To account for the diversity of the abundance patterns of individual subdwarfs the grids of H/He spectra are also available for the mean pattern scaled by a factor of 1/100, 1/10 and 10 (henceforth $z$ = $-2$, $-1$, 0, $+1$, on a logarithmic scale). The \textit{$2^{nd}$ generation Bamberg model grids} have been used recently, e.g. by \citet{2024A&A...690A.368G}, \citet{2025arXiv251102539L}, and 
\citet{Dawson2025}. 

\begin{figure}
\centering
\includegraphics[width=\columnwidth]{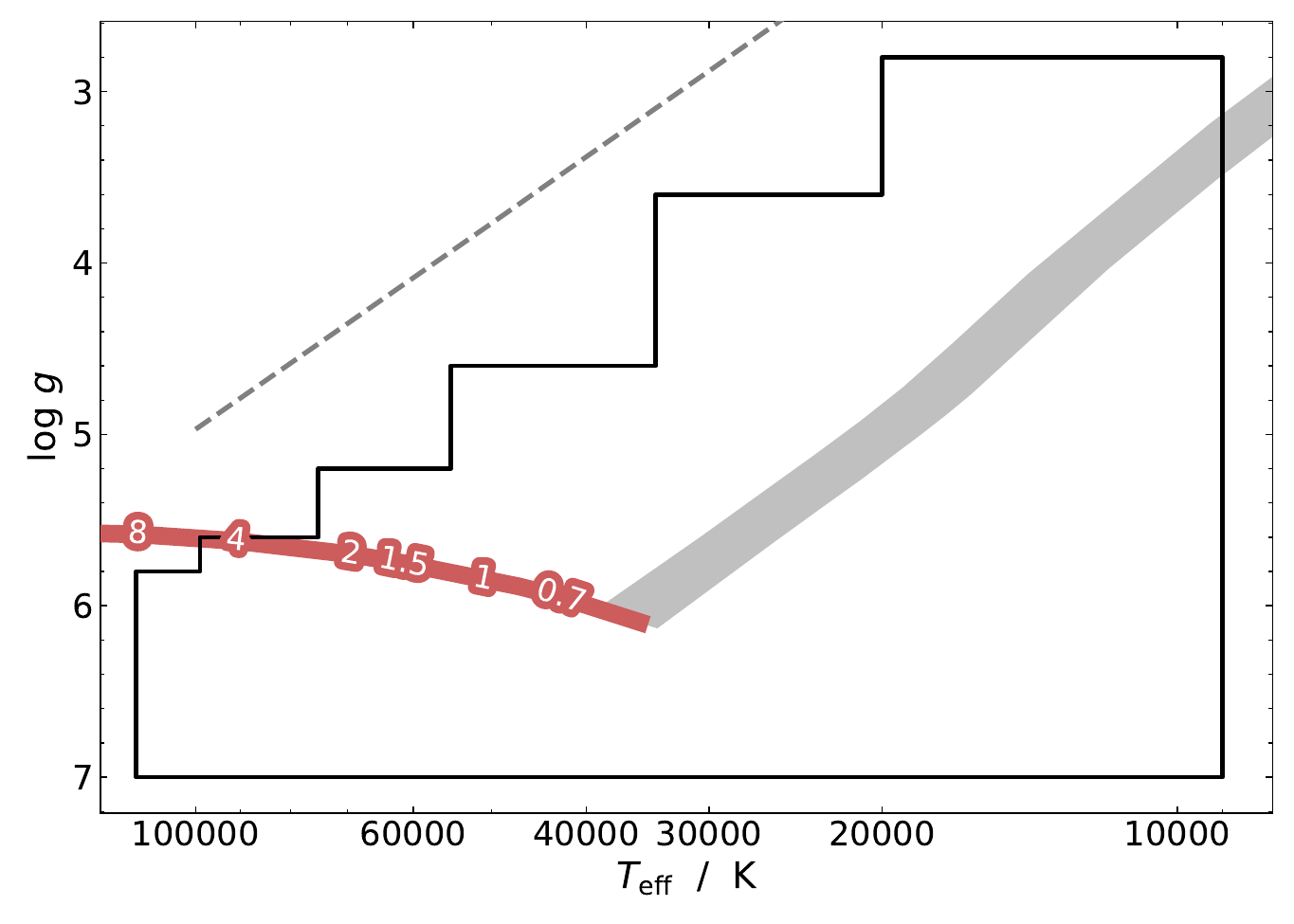} 
\caption{Dimension of the model atmosphere grid in the \teff-\logg\ plane (black line). The grey band denotes the EHB for solar composition \citep{Dorman1993} for comparison. The helium main sequence \citep{1971AcA....21....1P} is shown in red with masses labelled. The dashed line shows the Eddington limit for solar composition \citep{1988ApJ...324..279L}. Grids are constructed at metal composition of 1/100, 1/10, 1 and 10 times the standard sdB metallicity pattern and $n$(He)/$n$(H) of 1/10,000 to $\approx$ 100 times solar, depending on metal content.}
\label{fig:grid} 
\end{figure}

\subsection{Spectral analysis procedure}\label{sect:spec_analysis}

We used a global fitting procedure developed by \citet{2014A&A...565A..63I}, which replaced the previously used selective $\chi^{2}$ minimisation technique \citep{1999A&A...350..101N,phd_hirsch2009}. The entire useful spectral range, usually from 3600 or 3700\,\AA\ to 7300\,\AA, was used rather than pre-selecting spectral lines. Interstellar lines, such as the \ion{Ca}{ii}, H \& K and the \ion{Na}{i} D lines, as well as artefacts are automatically excluded. The continuum is modelled with an Akima spline \citep{1970JACM...17..589A} anchored at preselected intervals. The entire multi-parameter space (effective temperature, surface gravity, helium abundance, projected rotational velocity, and radial velocity) is fitted simultaneously with the continuum. Poorly fitting spectral regions are automatically excluded. 
Because of the low spectral resolution of the TWIN and CAFOS spectra, the projected rotation velocities were uncertain but consistent with zero. 
We note that the same procedure was applied by \citet{Dawson2025} and \citet{2025arXiv251102539L}  
for the sample of nearby hot subdwarfs (volume complete to 500\,pc) and the Arizona-Montr\'{e}al sample, respectively.
If multiple spectra are available, they are fitted simultaneously. To ensure homogeneity, we restrict ourselves to 
spectra that cover the Balmer jump to at least 3700\,\AA.

\begin{figure}
\centering
\includegraphics[width=\columnwidth]{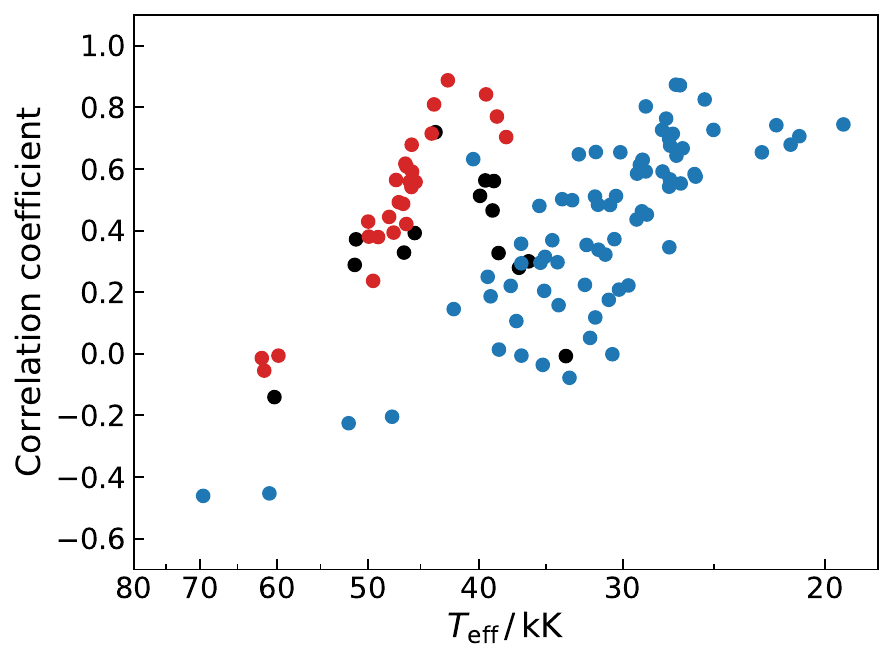} 
\caption{Correlation of \teff\ with \logg: correlation factors as a function of \teff\ for He-poor (blue), intermediate helium-rich ({black}), and extremely He-rich subdwarfs (red). 
}
\label{fig:teff_logg_corr}
\end{figure}

The atmospheric parameters are not independent, but correlated.
The most important correlation among them is between the effective temperature and the gravity. We calculate covariance matrices and confidence maps and derive the correlation coefficients using the Cholesky-Banachiewicz decomposition algorithm \citep[see][ and references therein]{1995A&A...300L..25P} for the HQS sample. 
This correlation is at the level of statistical uncertainties, only, which means that the uncertainties are small on an absolute scale, including the correlations. 
The correlation is positive for all but the very hottest stars (see Fig. \ref{fig:teff_logg_corr}). It decreases with \teff\ both for the eHe rich sdOB/Os and for the He-poor subdwarfs, with the largest correlation factors at the lowest \teffs\ (20kK and 40kK, respectively). 
Fig. \ref{fig:hqs_conf_maps} shows illustrative examples of confidence maps for different spectral types.

\begin{figure*}
\centering
\includegraphics[width=0.325\textwidth]{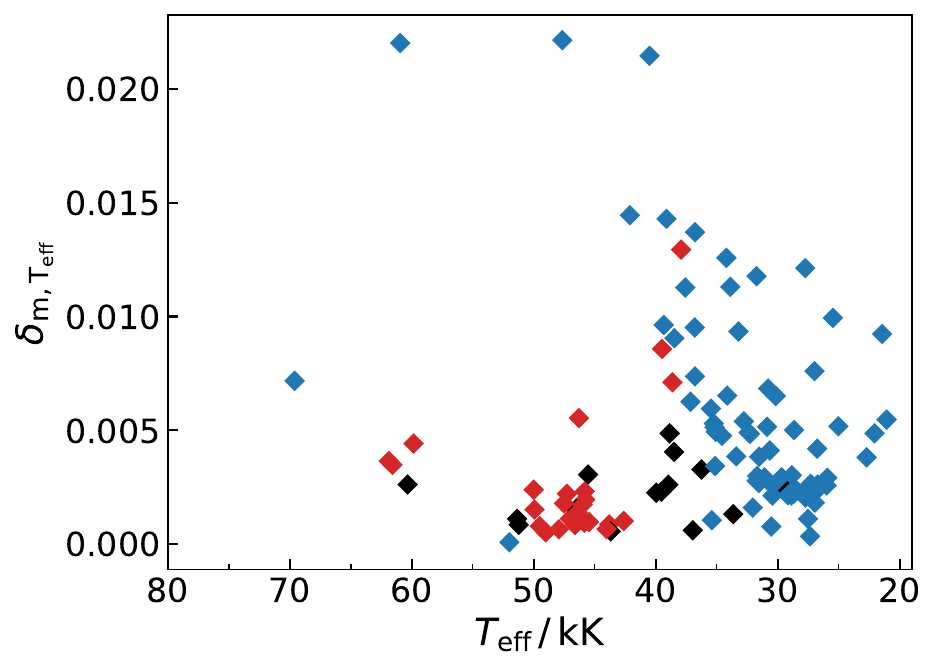} 
\includegraphics[width=0.325\textwidth]{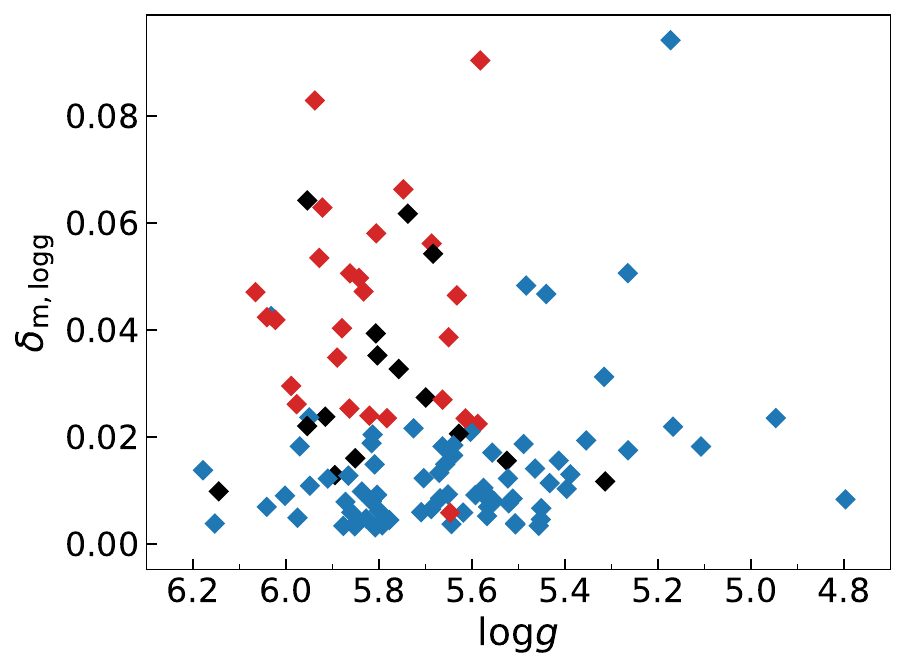} 
\includegraphics[width=0.325\textwidth]{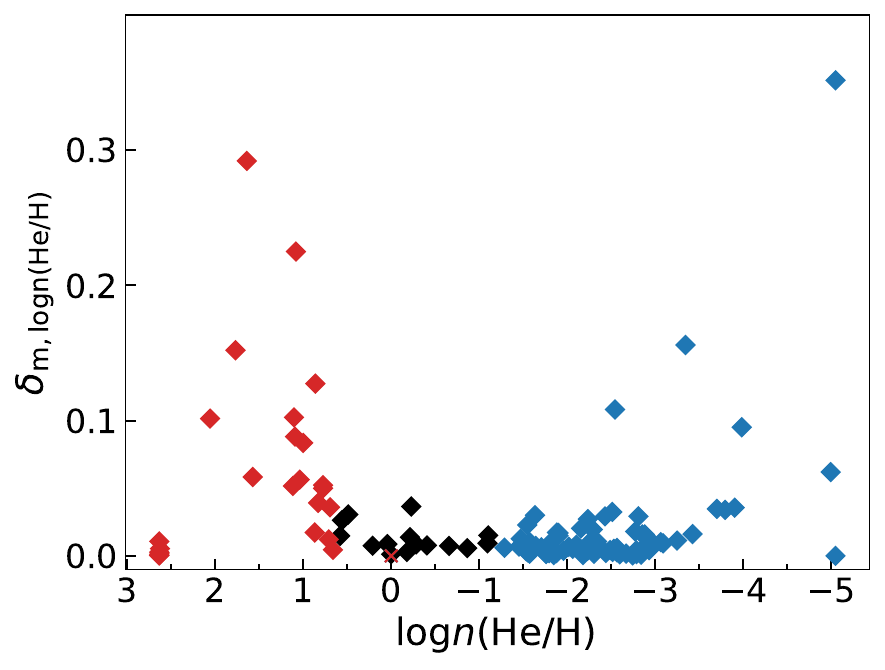} 
\caption{ 
The impact of metallicity for the determination of atmospheric parameters for \teff, \logg, and \logy, from left to right. Shown are half of the maximum differences  between parameters derived using models with a standard metal composition and those with $z=\pm 0.3$, for helium-poor (blue), helium-rich (red), and intermediate-helium ({black}) hot subdwarfs. For \teff\ the relative differences are shown.}
\label{fig:hqs_z_impact}
\end{figure*}

\begin{figure*}
\centering
\includegraphics[width=0.32\textwidth]{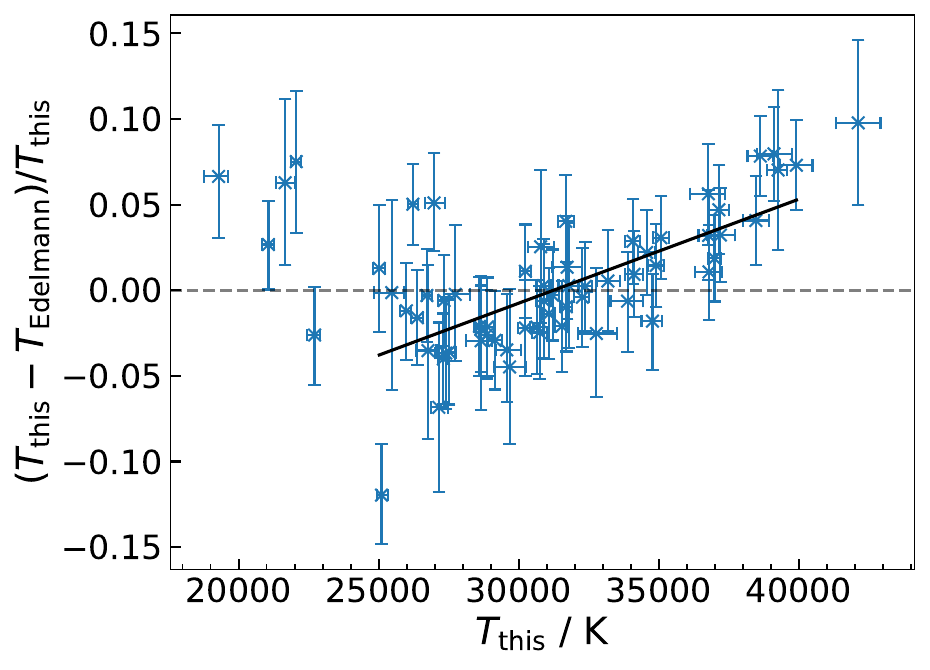} 
\includegraphics[width=0.32\textwidth]{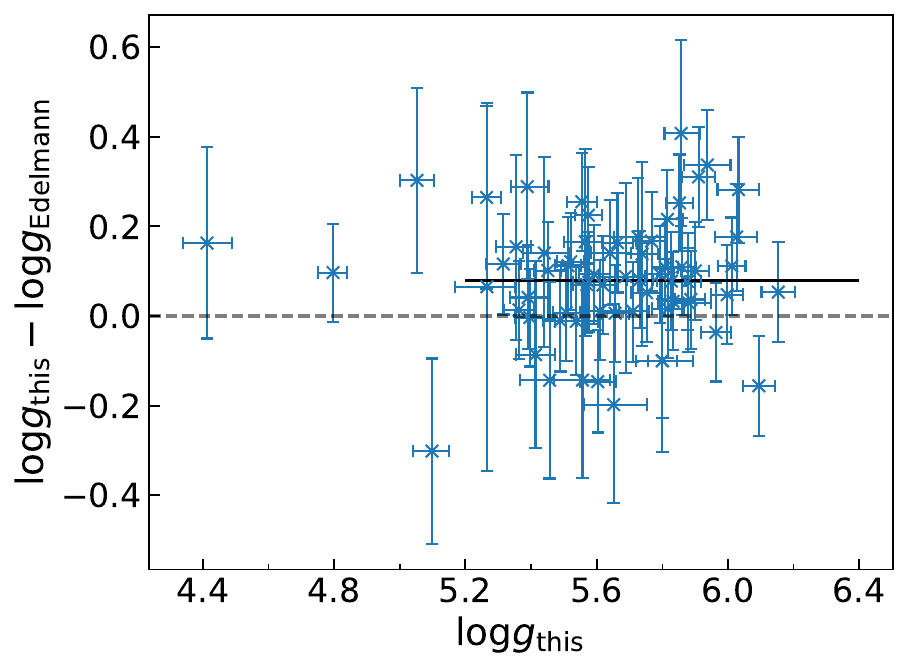} 
\includegraphics[width=0.32\textwidth]{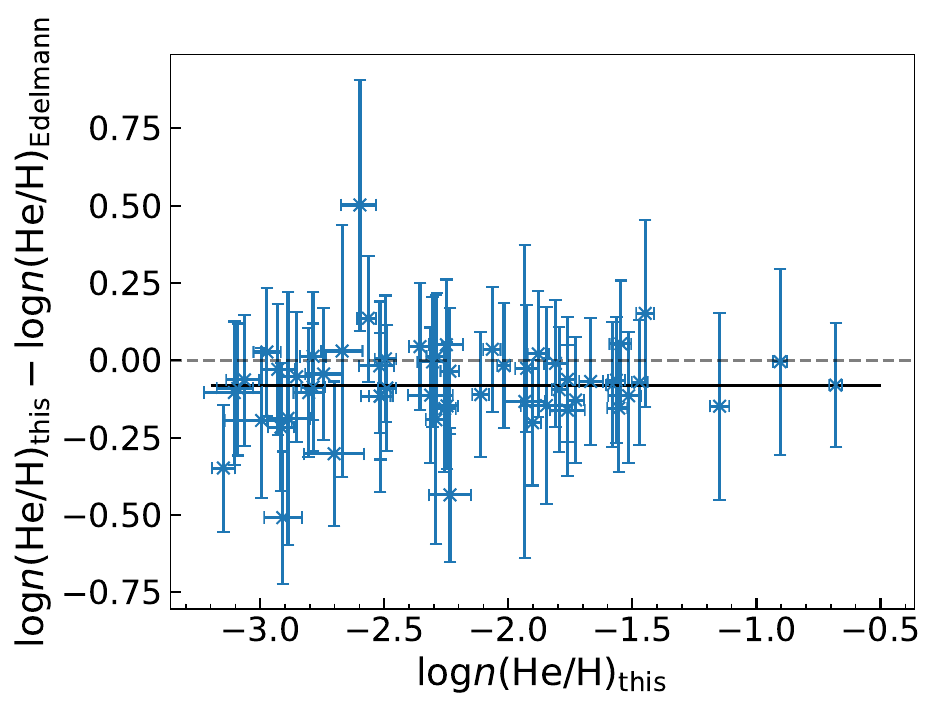} 

\caption{Comparison of the effective temperatures (left panel), surface gravities (middle panel), and helium abundance (right panel) with those derived by \citet{2003A&A...400..939E}. The median values are indicated by a black horizontal line.
}
\label{fig:comparison_sdb}
\end{figure*}

\subsection{Systematic uncertainties}\label{sect:systematics}

Besides statistical uncertainties from noise, systematic uncertainties also arise from different sources. 
\paragraph{Intrinsic scatter} may be caused by limitations in the instrumental set-up, the data reduction, and the normalisation procedure. Repeated observations of the same star often scatter more than expected from statistical noise. In their investigation of sdB stars from the SPY project \citet{2005A&A...430..223L} studied the distributions of differences in atmospheric parameters between two exposures of $\approx$50 non-composite sdBs and found $\Delta$\teff = 374 K, \logg = 0.049, and \logy = 0.044. Recently,  
\citet{Dawson2025} studied a larger sample based on low resolution spectra and found a polynominal representation of the intrinsic scatter with averages of 0.55\,\% in \teff, 0.039 in \logg, and 0.02 in \logy\ for the main parameter ranges. 
We adopted the latter as a contribution to the systematic uncertainties of the atmospheric parameters, because they are based on spectra of similar resolution to the DSAZ spectra.

\paragraph{Metallicity effect:} Another uncertainty arises from the unknown metallicity of the individual stars. 
The UV line blocking is dominated by the large number of iron and nickel lines, which are responsible for most of the metal line blanketing effect on the atmospheric temperature and density stratification. Since the strength of the lines depends on the abundance of the elements, their impact does as well. From high-resolution studies \citep[e.g.][]{2006A&A...452..579O,2008ApJ...678.1329B} we know that the average abundance of Fe is nearly solar, while Ni is about 10 times solar in hot subdwarf stars. However, the abundances vary significantly from star to star by about $\pm$0.3 dex \citep[$1\sigma$,][]{2013A&A...549A.110G}.
 To account for these uncertainties, we repeated the fitting procedure for enhanced ($z=+0.3$) 
 and depleted ($z=-0.3$) 
 metal contents (see Sect.\ \ref{sect:model_atmospheres}). 
This variation leads to systematic uncertainties on the atmospheric parameters, which we plot in Fig.\ \ref{fig:hqs_z_impact}
and Fig.\ \ref{fig:hqs_HRD_z_effect}\footnote{A similar study was carried out by \citet{Dawson2025} for their volume-complete subdwarf sample, cf. their Fig. A.1.}. These systematic uncertainties differ strongly among subtypes. While for helium poor subdwarfs the \teff\ uncertainties 
are larger than for He-rich ones, the opposite holds for \logg. The He-rich stars show absorption lines from both neutral and ionised helium, allowing the ionisation equilibrium to constrain \teff\ more precisely than for the He-poor stars, which lack this information. On the other hand, the determination of surface gravity is largely based on the Stark-broadened Balmer and helium lines as well as the level-dissolution of high-lying H levels. The effect is more pronounced for the hydrogen Balmer line than for the helium line profiles. For the helium-to-hydrogen ratio (\logy), the effect remains small over a large range of values, while it is large for both very low and very high He abundances. For low values, the He lines are very weak or absent, while at the high end the hydrogen Balmer line contribution to the blend with He\,{\sc ii} is very small.  


The additional contributions to the systematic uncertainties resulting from metallicity uncertainty $\delta_{\textrm{m},j}^2$ (for $j$ = \teff, \logg, \logy) were added along with the observational (intrinsic $\delta_{\textrm{i},j}$) ones in quadrature to the statistical ($\delta_{\textrm{s},j}$) ones, which defines the adopted total uncertainty of the atmospheric parameters:

\begin{center}
$\sqrt{\delta_{\textrm{s},j}^2 + \delta_{\textrm{m},j}^2 + \delta_{\textrm{i},j}^2}$, 
\end{center}
where $\delta_{\textrm{i},j}$ is the same for all stars. 

\subsection{The impact of the \textit{$2^{nd}$ generation Bamberg model} approach: Revisiting the sdB stars in the HQS}\label{sect:sdb_compare}
To quantify the impact of the new model grid and analysis procedure, we revisited the sample of sdB stars from the HQS survey.
\citet{2003A&A...400..939E} analysed the same spectra using a mix of metal line blanketed LTE atmospheres and non-LTE model atmospheres, applying the former to stars cooler than $T_{\rm eff}=27\,000$\,K, the latter for stars hotter than $T_{\rm eff}=35\,000$\,K, and the mean of both for intermediate temperatures. Selected spectral lines of He and H were used to derive the atmospheric parameters.  
We compare their results to our new results from the homogeneous hybrid LTE/non-LTE model grid and the global analysis strategy in Fig.\ \ref{fig:comparison_sdb}, based on the same observations. 

The new results are systematically different from the previously published ones as demonstrated in Fig. \ref{fig:comparison_sdb}. A trend for the \teff\ difference to increase with the revised ones is obvious from the left panel. A linear regression gives 

\(T_{\mathrm{eff}}^{new} = 1.26 \times T_{\mathrm{eff}}^\mathrm{publ} - 8\ [\mathrm{kK}]\), \\[1mm]
\noindent valid for temperatures between 25\,kK and 40\,kK. This can be traced back to the model grids used in the previous analysis. The new grid treats metal line blanketing and departures from LTE in a homogeneous way while the previous analysis made use of a mix of model grids as explained above. Some outliers are evident in the surface gravity values, but no clear trend is observed. On average, the revised values are higher than the previous ones by 0.08\,dex. The revised helium abundances are lower by 0.08\,dex.  
These results can be used to convert published results for sdB stars based on the old models \citep[e.g.][and many more]{Maxted2001,2005A&A...430..223L,2003MNRAS.338..752M,2011MNRAS.415.1381C,Geier2011} to the new scale.

\section{Results of the spectroscopic analyses}\label{sect:atmos_parameters}

The sdO stars from the HQS survey follow-up have not been studied before.
Because of their high helium abundance, both iHe and eHe subdwarfs stars show rich line spectra of neutral and ionised helium lines. 
Illustrative examples for spectral fits of the TWIN spectra of the iHe-sdOB HS\,1837+5913, the eHe-sdOB HS\,0836+6158, the iHe-sdO HS\,1832+6955, and the eHe-sdO HS\,1020+6926 are shown in Fig.\ \ref{fig:example_spectra}.


There are some spectroscopically unusual stars that should be mentioned before we turn to the results for the full sample. 

\subsection{Individual Objects}\label{sect:individual}

A few sdO stars turned out to be very hot and/or of unusually low surface gravity. Although these stars are of great interest, we report only preliminary results in Table \ref{tab:individual_sd}; a sufficiently detailed investigation would require improved observations and additional atmospheric modelling. 

\subsubsection{Very hot stars}

HS0216+0313 and HS1215+6247
(off-scale in Fig. \ref{fig:hqs_kiel_hrd}) have effective temperatures exceeding 100 kK and high surface gravities (\logg=$>$7.0 and 6.5, respectively). Because \ion{He}{ii} is detected at near solar abundance in both stars, they are likely DAO white dwarfs {and, therefore, excluded from the sample}.

\subsubsection{Four post-AGB star candidates}

The very hot He-rich sdO stars HS0237+0342, HS0657+5333, HS0736+3952, and HS1736+5521 have unusually low gravities (\logg\ $\approx$ 5.2 to 5.4) 
indicating that they either
belong to the rare {group of} He-rich hot post-AGB stars, similar to the trio of stars LSE153, LSE259 and LSE263 \citep{1989A&A...222..150H,2024A&A...683A..80K}, though at somewhat higher gravity, or to the class of O(He) stars \citep{2023MNRAS.519.2321J,2025A&A...693A.167W}, though at somewhat lower \teff. Strong C\,{\sc iv} lines are present in the spectra of the first three stars.
Further investigations are necessary to draw firm conclusions. {We therefore exclude them from the sample.}

\subsubsection{A possibly magnetic sdO star}

HS0312+2225 
is an iHe-sdO with a strong 4620\,\AA\ absorption feature. Because such a feature has been found in magnetic He-sdOs \citep{2024A&A...691A.165D} we speculate that the star might be magnetic. 
However, no Zeeman splitting or broadening can be seen. HS0312+2225 has atmospheric parameters (\teff=45.5 kK, \logg=5.95, \logy=0), {very similar} to the known magnetic sdOs. High resolution spectra are required to search for evidence of magnetism. {We keep the star in the sample.}

\subsubsection{HS2123+0048, the central star of PN G053.9-33.2}

This star has been suggested to be the central star of the planetary nebula  G053.9-33.2 and to be marginally light variable with a period of 2.546 d, and might be an eclipsing binary \citep{2024A&A...690A.190A}. Further observations and modelling are required to improve the results.
Although the object is very interesting we exclude it from the sample here.

\begin{table}
\centering
\caption{Estimated atmospheric parameters of unusual {HQS hot} stars {that are excluded from the final sample} and not shown in Fig. \ref{fig:hqs_kiel_hrd}.
}
\label{tab:individual_sd}
\renewcommand{\arraystretch}{1.15}
        \begin{tabular}{crrrc}
                \toprule\toprule
               Star  & \teff\ [kK] & \logg\ & \logy\ & Comment  \\
                \midrule 
HS0216+0313               & $\gtrapprox$100 & 6.5 & $-1.1$ &DAO\\
HS1215+6247               & $\gtrapprox$100 & $>7.0$ & $-1.3$ & DAO\\
HS0237+0342               & 64.9 & $\approx$5.2 & $\gtrapprox$0.7 &pAGB\\
HS0657+5333               & 65.5 & 5.4 & $\gtrapprox$ 0.7 & pAGB \\
HS0736+3952               & $>$75 & $\approx$5.3 & $\gtrapprox$0.7 & pAGB\\
HS1736+5521               & 63.5 & 5.3 & $\gtrapprox$0.8 & pAGB\\
HS2123+0048               & 69.4 & 5.6  & $-1.0$ & CSPN\\
\bottomrule
\end{tabular}
\end{table}

\subsection{Kiel diagram and helium abundance distribution}

The results of the spectral analysis are displayed in Fig.\ \ref{fig:hqs_kiel_hrd} as $T_{\rm eff}-\log{g}$ (Kiel) and \teff-\logy\ diagrams. The observed positions in the Kiel diagram are compared to predictions for the evolution of EHB stars \citep{Dorman1993}. Evidence is growing that the distribution of stars along the EHB band is not homogenous \citep{2012MNRAS.427.2180N,Geier2022,2025A&A...693A.121H}. \citet{Geier2022} defined three different structures named EHB1 (cool end, \teff\ $<$ 25\,kK), EHB2 (moderate \teff), and EHB3 (hot end, \teff\ $>$ 33\,kK), sdB and sdOB stars lying beyond or below the EHB band were named postEHB and bEHB, respectively.
Most of the He-poor sdB and sdOB stars in our HQS sample lie within the predicted core helium burning EHB band, but only four sdBs are found to be cooler than 25\,000\,K, belonging to the EHB1 group. 
The surface gravities for some He-poor sdBs and sdOBs are lower than predicted for the EHB band, which hints at a more evolved He-shell burning state. He-poor stars are also found among the sdO stars at temperatures well beyond the hot end of the EHB. 
 They must have evolved even further towards the white dwarf cooling sequence. Many eHe-sdOs are nearly hydrogen-free; at helium abundances exceeding \logy\ = +2, hydrogen becomes undetectable in our spectra. 
 The distribution of He-rich stars differs significantly from that of the He-poor ones. 
 Most cluster near the helium main sequence at masses significantly higher than the canonical mass for the core helium flash. Projecting their position onto the He-ZAMS of \cite{1971AcA....21....1P} would predict masses from 0.45 M$_\odot$ to 1.0 M$_\odot$. 
 He-rich subdwarf stars have been proposed to originate from mergers of two helium white dwarfs in close binaries caused by gravitational wave emission \citep{1984ApJ...277..355W,1984ApJS...54..335I}. 
\citet{2012MNRAS.419..452Z} and \citet{2021MNRAS.504.2670Y} modelled the evolution of such mergers. As an example, Fig.\ \ref{fig:hqs_kiel_hrd} shows a track for a merger {model} of two He white dwarfs of 0.35\,M$_\odot$, ending up on the helium main sequence in the region where the He-rich sdOs are found. Because the mass of a He white dwarf may be as high as 0.45\,M$_\odot$, the merger products may be considerably more massive than 
half a solar mass, which is typically assumed for EHB stars.

\begin{figure*}
\sidecaption
\centering
\begin{minipage}[b]{12cm}
\includegraphics[width=\linewidth]{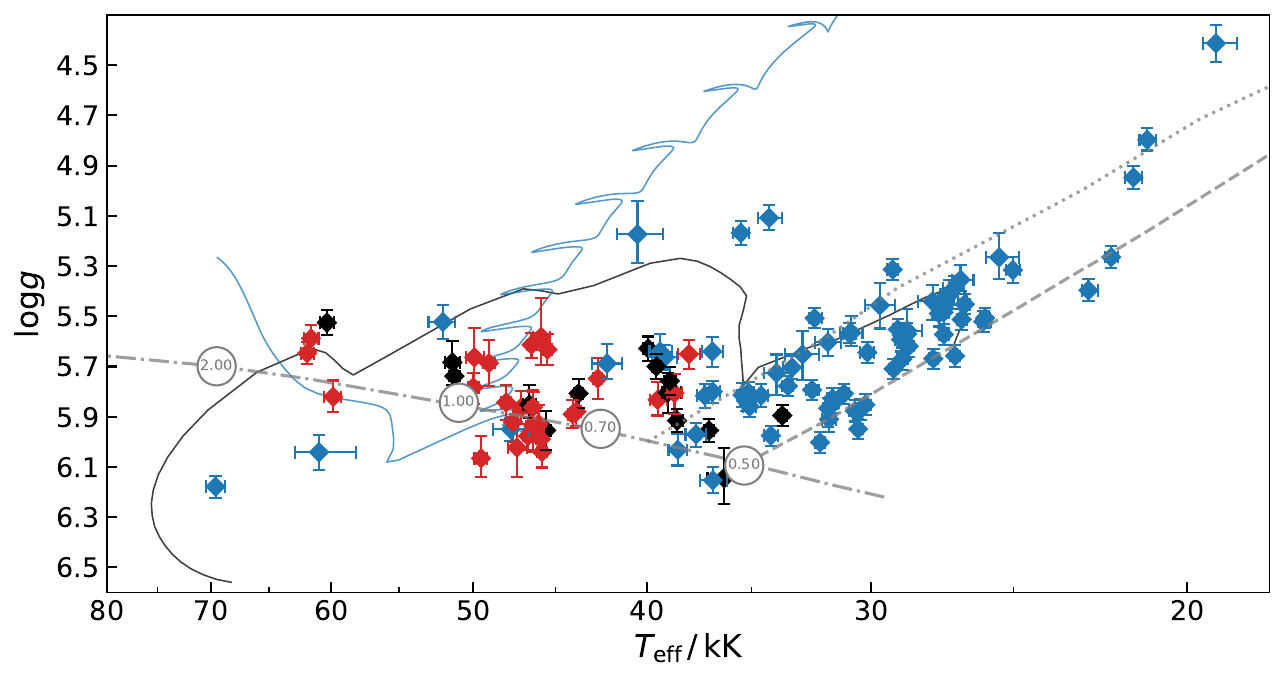}\\
\vspace{3mm}
\includegraphics[width=\linewidth]{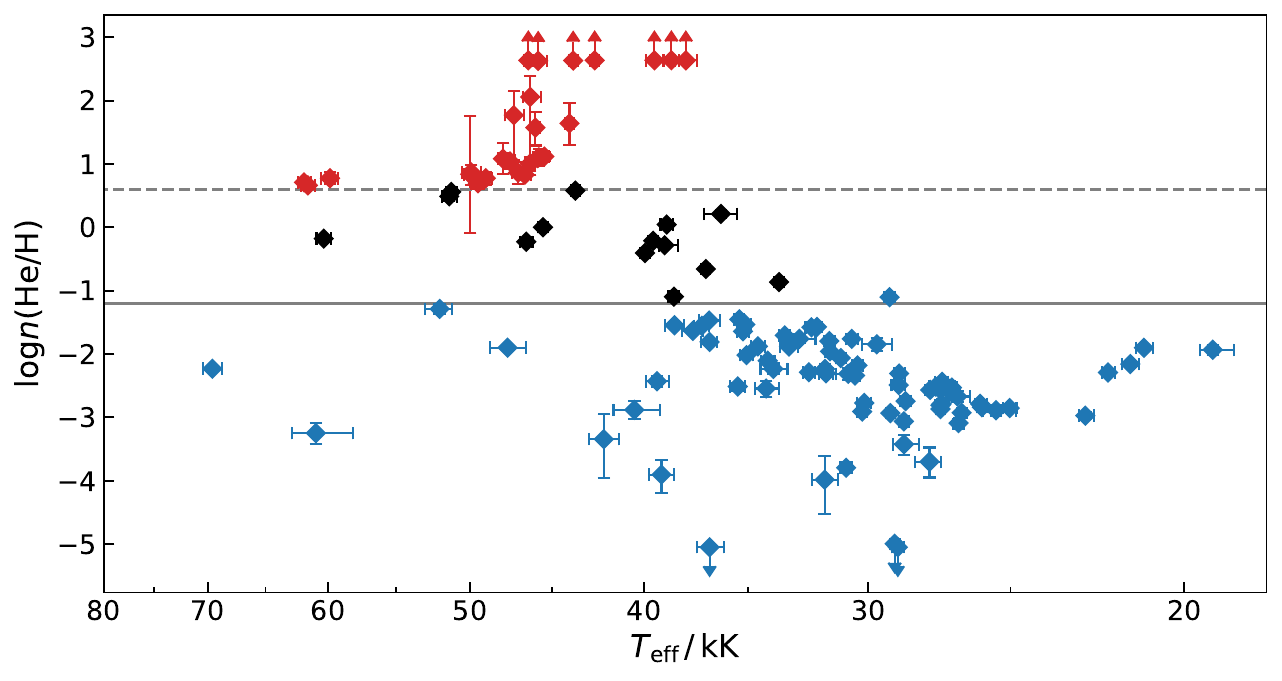}
\end{minipage}
\caption{\textit{Upper panel:} Kiel diagram: $T_{\rm eff}-\log{g}$ diagram. Helium poor stars are shown as blue circles, iHe subdwarfs as {black} and eHe subdwarfs as red ones.
The zero-age and terminal-age EHBs for a subsolar metallicity of $-1.48$ (solid grey lines) have been interpolated from evolutionary tracks by \citet{Dorman1993}. The helium main sequence (dashed line) is taken from \citet{1971AcA....21....1P}. The evolution of a merging He-WD system \citep{2021MNRAS.504.2670Y} is shown as a light blue line. \textit{Lower panel:} Helium abundance as a function of \teff. Lower/upper limits to \logy\ are marked by upward/downward arrows. The iHe-rich subdwarfs are distinguished from the He-poor ones by the solid line and from eHe-rich ones by the dashed line.
}
\label{fig:hqs_kiel_hrd}
\end{figure*}

\section{Spectral energy distribution }\label{sect:sed}

Model grids of spectral energy distributions (SEDs) allow us to analyse photometric measurements from the FUV through the optical to the near- and mid-infrared (see Sect. \ref{sect:photometry}). The synthetic flux distributions are calculated from the same model atmospheres used in the spectral analysis. 
The observed magnitudes were matched with synthetic SEDs keeping the atmospheric parameters fixed at the values determined from spectroscopy \citep[see][for details]{2018OAst...27...35H}. Hence, the angular diameter $\Theta$, and the interstellar colour excess $E(44-55)$ remained as free parameters. The interstellar extinction was accounted for by using the function given by \citet{2019ApJ...886..108F}, assuming a standard monochromatic extinction coefficient of $R(5500\AA)=3.02$. {The presentation of the flux distribution as $f_\lambda \times\lambda^3$ in Fig. \ref{fig:SED} eases the strong flux gradient towards the blue part of the SED.} The SEDs of hydrogen-rich subdwarfs are characterised by the presence of the Balmer jump. Because of the lack of hydrogen, the Balmer jump is replaced by an absorption edge of \ion{He}{i} in eHe-sdOBs (e.g. HS1843+6343), while in the hot sdOs ($\gtrsim$50,000K) both edges vanish (see Fig. \ref{fig:SED}). 

\subsection{Detection of cool companions to hot subdwarfs from spectral energy distributions}\label{sect:composite}

F-, G-, or K-type companions to hot subdwarfs can easily be detected from {the characteristic shape of} their SED, where the blue part is dominated by the subdwarf and the red and infrared part by the cool companion star. To detect late K- and early M-type companions, IR fluxes are crucial because their contributions in optical filters is very small. More than  80\% of the sample have IR coverage, mostly from the 2 micron all sky survey (2MASS) \citep{2003tmc..book.....C} and the UKIRT hemisphere survey \citep{2025AJ....170...86S}.

To model the flux distributions of cool companions, we interpolate in a grid of LTE Phoenix model spectra provided by \citet{2013A&A...553A...6H}. This adds additional parameters to the fitting routine. We fix the surface gravity at \logg\ = 4.5 and the metallicity at the solar value; these parameters cannot be constrained by the SED. This leaves us with the effective temperature of the cool star and the surface ratio as additional fit parameters. 

We confirm the composite spectrum nature of the 18 spectrum sdB binaries identified by \citet{2003A&A...400..939E} and the three found by \citet{2005A&A...430..223L} from spectroscopy. 
Another sdB, HS0016+0044, has been identified as a spectroscopic binary by \citet{2012MNRAS.425.1013G}, which is also confirmed by our SED analysis. In addition, we find IR excesses to five sdB stars. {Displaying the SED as $f\times\lambda^3$ in Fig. \ref{fig:SED} accentuates the flux contribution of F,G,K type stars, which are characterised by the bound-free absorption of H$^-$ ($\lambda<1.4\mu m$) in the H-band.}
HS0232+3155, HS1824+5745 (see Fig. \ref{fig:SED}), HS2151+0857, and HS2303+0152 host early K-type stars, whereas  
the IR excess of HS1813+7247 is very weak and the companion does not contribute at optical wavelengths.
The companion is an early M-type star, 
and we keep the star in our sample. 
{HS1813+7247 further shows light variations indicating that it is a g-mode pulsator \citep[TIC 229593795;][]{2024A&A...684A.118U}. Among the other HQS pulsating hot subdwarfs identified in the literature (see Sect~\ref{sect:lc}), we found that}
the p-mode pulsators HS1824+5745 (LS Dra), HS2303+0152, and HS2151+0857
are also composite spectrum objects (see Fig. \ref{fig:SED}) and, {thus, are not included in our final sample.}

Among the sdO stars we identified two composite objects, HS0735+4026 and HS2308+0942. The IR excess of the former is weak and we keep that star in the sample. 

It is worth noting that {15} of our targets lack infrared {H and K} magnitude measurements. Consequently, a slight IR excess (as for HS1813+7247) may not be detected \textbf{in} our analysis of the stars. 

\begin{figure*}
     \begin{subfigure}[b]{0.32\textwidth}
        \centering
       \includegraphics[width=\textwidth]{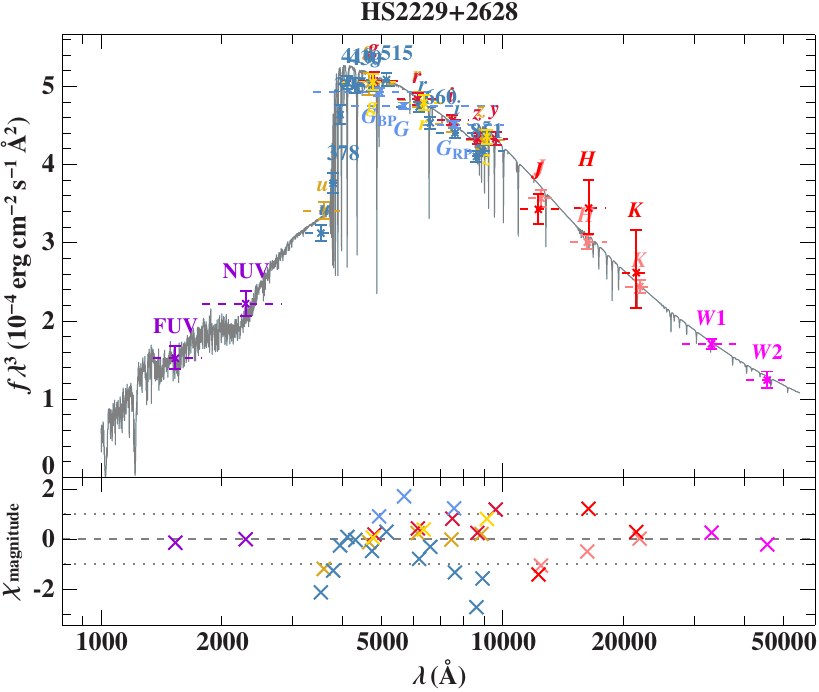}
    \end{subfigure}
    \begin{subfigure}[b]{0.32\textwidth}
         \centering
         \includegraphics[width=\textwidth]{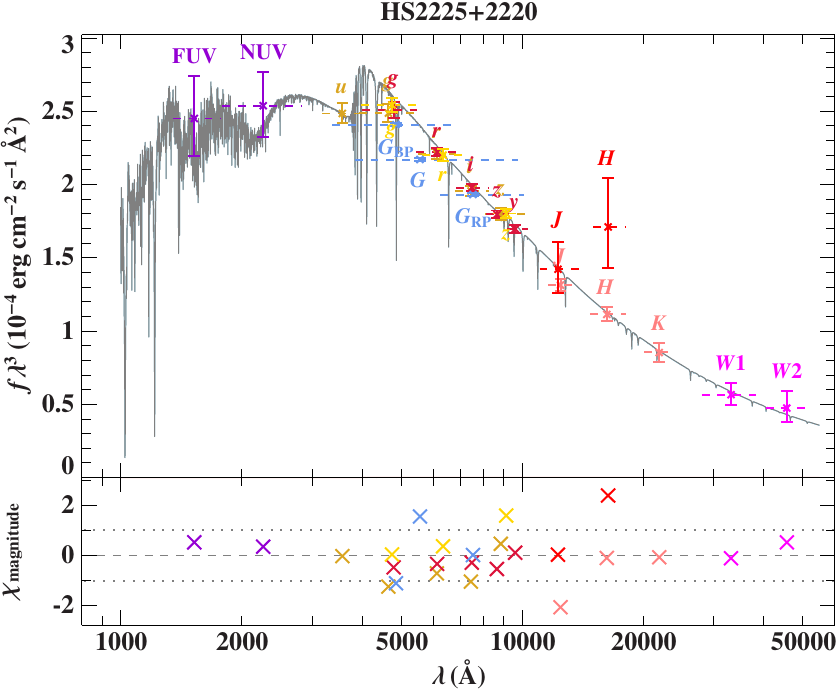}%
    \end{subfigure}
     \begin{subfigure}[b]{0.32\textwidth}
         \centering
        \includegraphics[width=\textwidth]{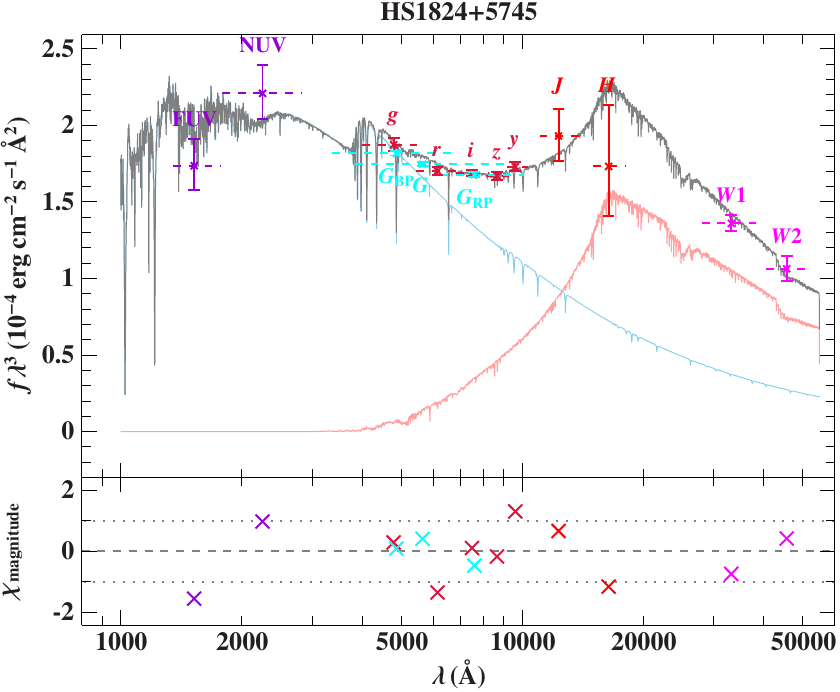} 
     \end{subfigure}
     \hfill
     \begin{subfigure}[b]{0.32\textwidth}
         \centering
        \includegraphics[width=\textwidth]{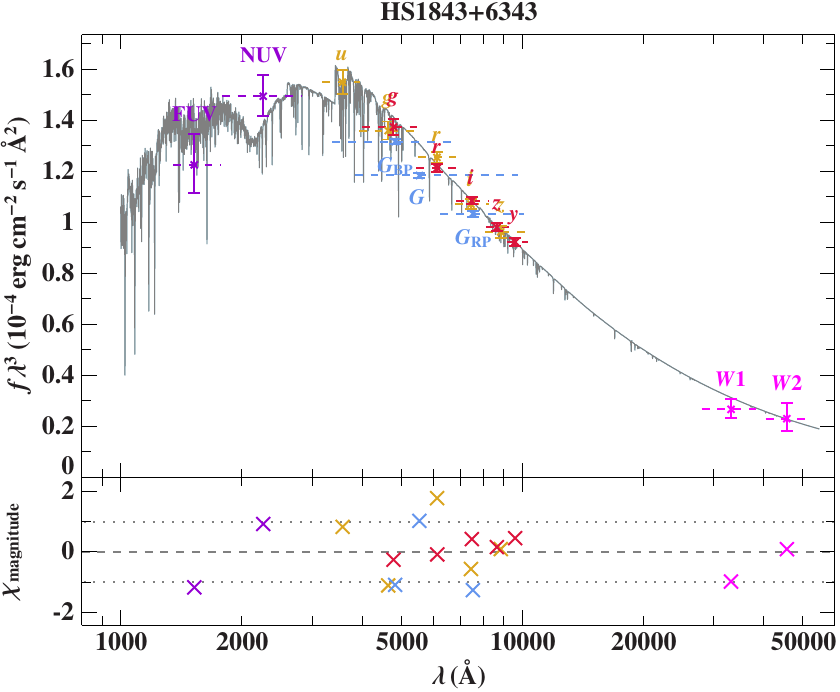} 
     \end{subfigure}
    \hfill
    \begin{subfigure}[b]{0.32\textwidth}
        \centering
         \includegraphics[width=\textwidth]{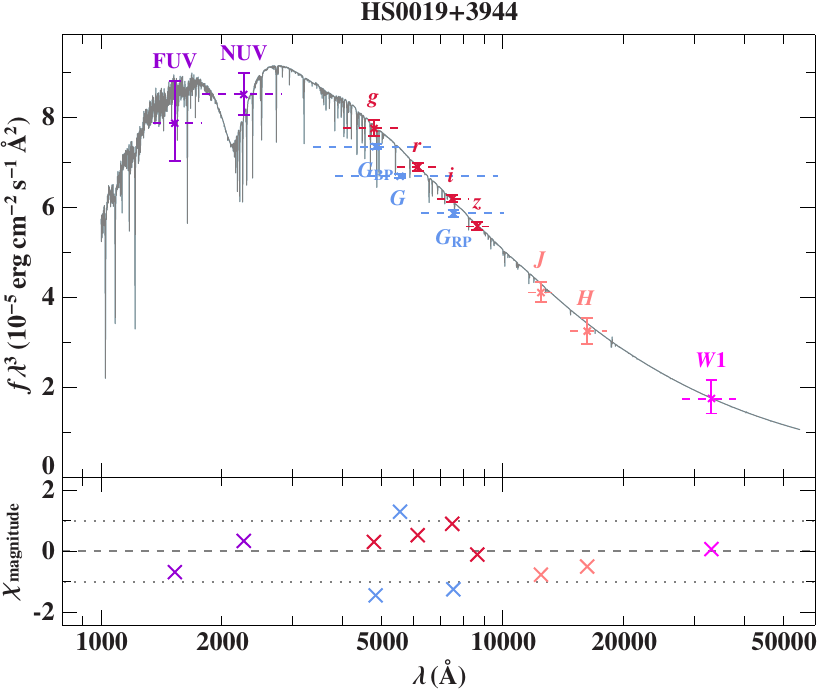}
     \end{subfigure}
     \hfill
     \begin{subfigure}[b]{0.32\textwidth}
         \centering
          \includegraphics[width=\textwidth]{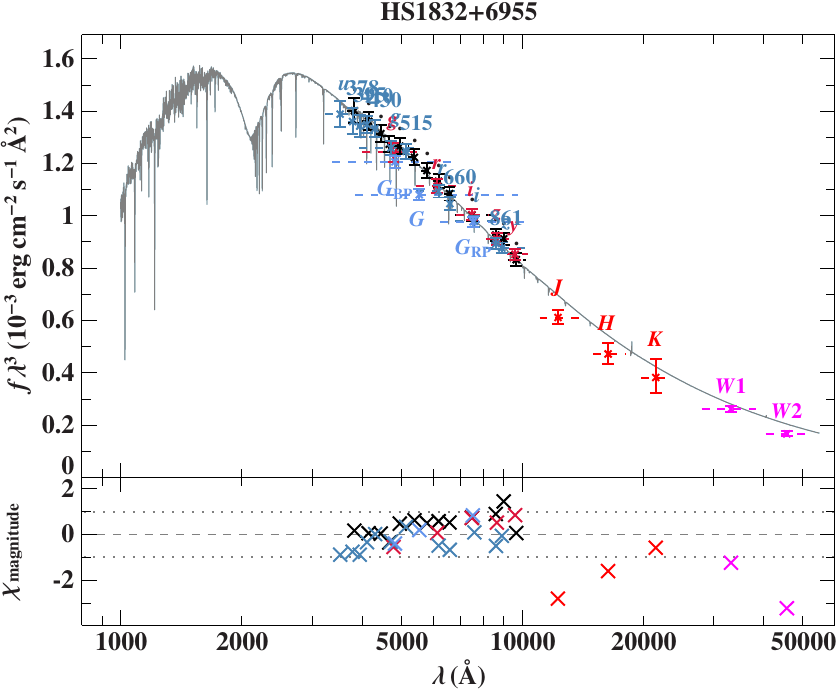} 
     \end{subfigure}
     \caption{Fits of the SEDs of selected programme stars. Each plot consists of two panels; the upper one compares the observed fluxes to the synthetic SED. To ease the slope of the distribution, the flux $f_\lambda$ is multiplied by the wavelength to the power of three. 
    Photometric fluxes are displayed as coloured data points 
    with their respective uncertainties and filter widths (dashed lines). The best-fit models are drawn as grey full drawn lines. The lower panels give uncertainty-weighted residuals $\chi$ to demonstrate the quality of the fit. The 2200\,\AA\ bump in the interstellar extinction curve is obvious in significantly reddened objects. 
    \textit{Top row:} He-poor hot subdwarfs (from left to right): The sdB HS2229+2628, the sdOB HS2225+2220, and the composite sdB pulsator HS1824+5745; 
    \textit{Bottom row}: Single He-sdOB/O stars (from left to right): the eHe sdOB HS1843+6343, the eHe sdO  HS0019+3944, and the iHe-sdO HS1832+6955.
    }\label{fig:SED}
\end{figure*}

\subsection{Interstellar reddening}\label{sect:ism}

Interstellar reddening as derived from the SED fit is found to be significant for most stars; its distribution on the sky is shown in Fig. \ref{fig:sky1}.  
The high reddening between the Galactic longitudes $l=150^\circ$ and $190^\circ$ at southern Galactic latitudes is caused by the Taurus-Perseus-California complex of the Gould belt \citep{2020Natur.578..237A}.
We compare the results to the line-of-sight reddening maps of \citet{1998ApJ...500..525S} and \citet{2011ApJ...737..103S} and find our results to be consistent with the predictions of the maps. 

\section{Stellar parameters from spectral energy distributions and parallaxes}\label{sect:stellar_parameters}

Radii, luminosities, and masses are derived by combining the atmospheric parameters determined in Section \ref{sect:atmos_parameters} with angular diameters and \textit{Gaia} DR3 parallaxes \citep{2023A&A...674A...1G}. Parallax zero-point offsets are corrected following \citet{2021A&A...649A...2L}. The relative parallax uncertainties ($\delta\varpi$) increase rapidly with decreasing parallax, as illustrated in Fig.\ \ref{fig:sky}. Stars with $\delta\varpi \geq 25$ per cent are excluded from the analysis, as such large uncertainties lead to unreliable stellar parameters. 
The re-normalised unit weight error (RUWE) is a quality control parameter that should be below 1.4 for good astrometric solutions \citep{2021MNRAS.506.2269E}.  
Thresholds on RUWE depend on the position of the object on the sky because of \textit{Gaia}'s scanning law \citep{2024A&A...688A...1C}.
All stars in our sample have RUWE $< 1.2$ and therefore pass this astrometric quality criterion, with the exception of HS2233+1418 (RUWE=1.52), which is excluded from the analysis.
This leaves a final sample of {103} stars for which stellar parameters are derived.


Absolute stellar radii, luminosities, and masses were computed through the basic relations:
\begin{equation}
    R = \frac{\Theta}{2 {\varpi}}, \quad
    L = 4\pi R^2 \sigma_{\mathrm{SB}} T_{\mathrm{eff}}^{4}, \quad
    M = \frac{g R^2}{G}
    \label{sed_params}
\end{equation}
\noindent
where $\Theta$ is the angular diameter, $\varpi$ the parallax,  $G$ is the gravitational constant, and  $\sigma_{\mathrm{SB}}$ is the the Stefan-Boltzmann constant.

We employed a Monte Carlo (MC) method to assess the uncertainties of the error propagation 
from the input parameters \teff, \logg, $\varpi$ and $\log \Theta$, which were represented by Gaussian distributions. The correlation of \teff\ and \logg\ was applied to the systematic uncertainties as well, {but turns out to have very little impact on the resulting stellar masses ($\Delta$M$<$0.005 M$_{\odot}$)}  (see Sect. \ref{sect:spec_analysis}).  
For each star, $10^6$ samples were generated, with each array carried through the full calculation to the final derived parameter. 
The results listed in Table \ref{tab:good_params} allow us to construct the arguably most important diagrams to test stellar evolution, that is the physical HRD (\teff, \logL) and the mass distribution.

\subsection{The Hertzsprung-Russell (\teff, \logL) diagram}

Fig. \ref{fig:hrd_mass_distribution} shows the HRD. The helium-poor stars mostly lie on the EHB band up to 
\teff\ = 40\,kK. At the hot end, some are found on the helium main sequence for half a solar mass. The He-rich stars mostly lie close to the HeMS at somewhat higher masses, irrespective of whether they belong to the iHe or eHe classes. 

The structure of the EHB, already discussed in the context of the Kiel diagram, is also evident in the HRD. In particular, there is a pronounced scarcity of sdB stars at the cool end (EHB1).

{In \citet{Dawson2025}, the region below the canonical EHB is defined using the $0.45~\mathrm{M_\odot}$ evolutionary tracks of \citet{Han2002}, which correspond to a luminosity of $\log L/L_{\odot} = 1.05$. Using a similar definition, \citet{Geier2022} and \citet{2025A&A...693A.121H} identified a distinct population of sdB stars referred to as bEHB or EHBb stars that are located below the canonical EHB band. This population is most prominent in \citet{Dawson2025}, but also present in the sample of \citet{2025arXiv251102539L}. By contrast, no stars in our sample fall below the EHB luminosity limit. Two sdOB stars are located on the $\log L/L_{\odot} = 1.05$ boundary, still consistent with degenerate core ignition. The absence of bEHB stars is also evident in the Kiel diagram. }

The bEHB stars are believed to have evolved from intermediate mass progenitors of 2.0 to 2.5 M$_\odot$ \citep[e.g.][]{Han2002} {which ignite helium burning in non-degenerate conditions}, that is, from relatively young stars. The lack of bEHB stars here may thus indicate that the HQS sample consists of an older stellar population than the other samples.   

\begin{figure*}
\centering
\includegraphics[width=0.80\textwidth]{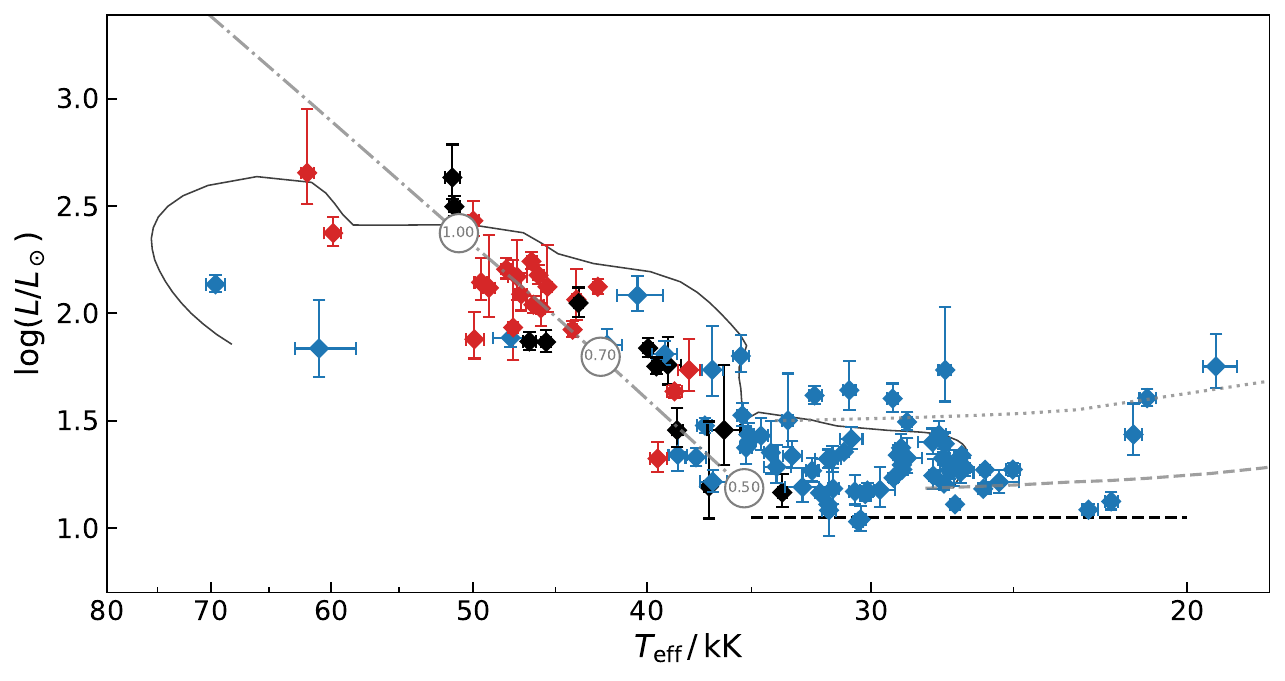} 
\includegraphics[width=0.49\textwidth]{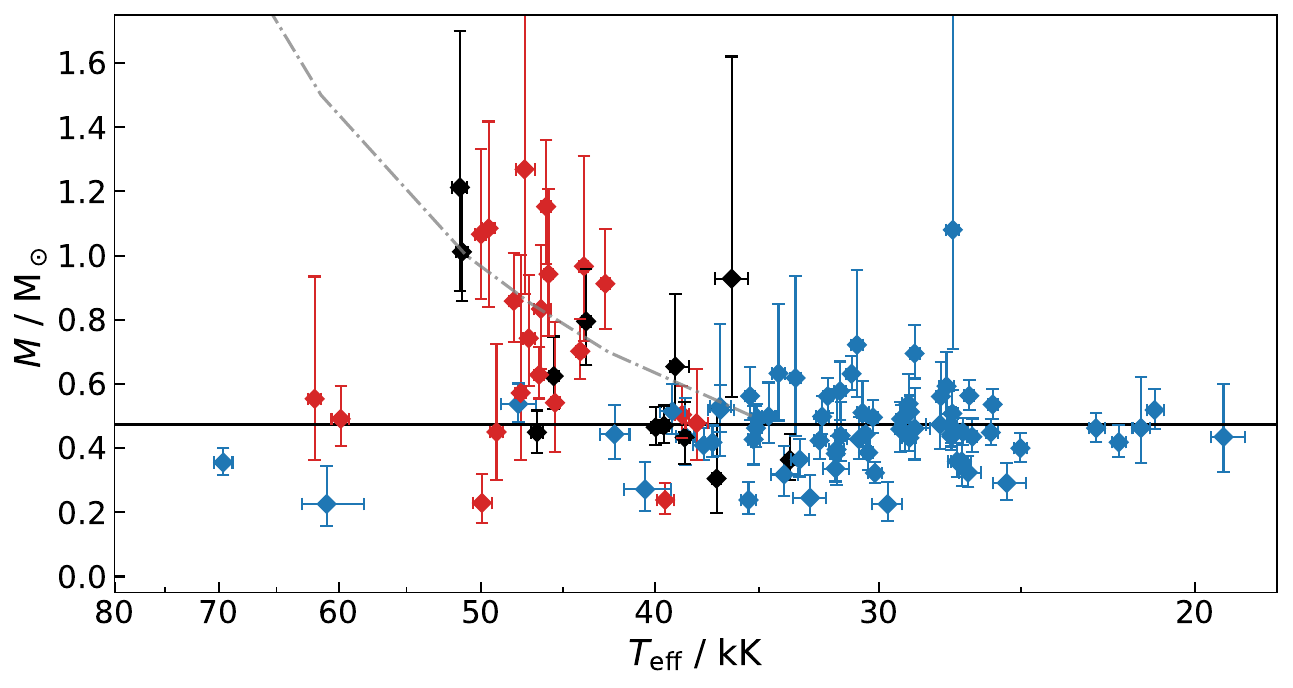}
\includegraphics[width=0.49\textwidth]{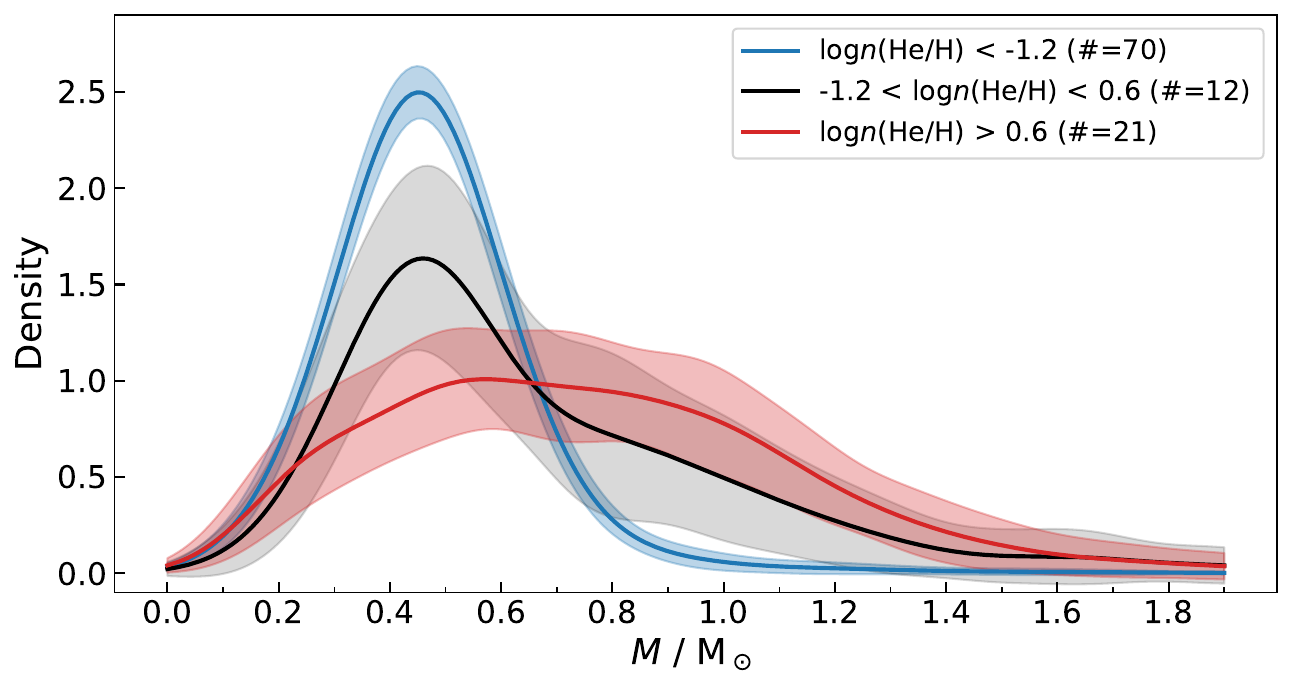}
\caption{HRD and mass distributions; \textit{Upper panel}: HRD ( $T_{\rm eff}-\log{L}$):
{The gray dashed and dotted lines mark the standard EHB band for a metallicity of $-1.48$ \citep{Dorman1993}.
The dashed black} horizontal line {at \logL = 1.05} marks the luminosity limit separating bEHB stars from EHB stars \citep[see][]{Dawson2025}. {While the lowest luminosity stars HS2029+0301 and HS2240+0136 lie on this line, none of the other stars lie below it.}  
\textit{Lower left}: Mass vs.\ \teff. 
Helium-poor stars are marked in blue, extremely helium-rich sdOB and sdO stars in red and iHe stars in {black}. 
\textit{Lower right}: Monte Carlo mass distribution as normalised Kernel Density Estimates for the three He subclasses. The shaded bands denote the uncertainty ranges.}
\label{fig:hrd_mass_distribution}
\end{figure*}

\subsection{Mass distribution}\label{sect:mass_distribution}

The stellar mass, which may be considered the most fundamental stellar parameter, {is computed} from \teff, \logg, and the parallax. {Since} the beginning of the  \textit{Gaia} era the parallax $\varpi$ has turned from an unknown to a quite precise quantity, at least for sufficiently nearby stars \citep[see][]{2024A&A...686A..25D}. 
To derive the mass, the gravity needs to be well constrained, but the accuracy of the spectroscopic results is limited by systematic uncertainties of the order of 0.07 dex, which are difficult to overcome, even for high quality data (see Sect. \ref{sect:systematics}).
In Fig.\ \ref{fig:hrd_mass_distribution} we plot the derived masses as functions of \teff, 
by separating He-poor (\logy\ $<$ $-1.2$), iHe  ($-1.2$ $<$ \logy\ $<$ +0.6) and eHe subdwarfs (\logy\ $>$ +0.6). 
Because the sample is flux-limited, more luminous He-sdO stars are seen to larger distances, and, therefore, have larger parallax uncertainties. They will benefit from improved parallaxes to be expected from \textit{Gaia} DR4. While the average masses of the helium-poor stars do not change with \teff, those of the He-rich subdwarfs increase with increasing \teff, consistent with their position in the HRD. 
In the lower right panel of Fig.\ \ref{fig:hrd_mass_distribution} we show the mass distribution from Monte Carlo simulations  as normalised Kernel Density Estimates (KDEs) with a bandwidth of 0.1\,\msun\ for a clear comparison between subclasses. 


Table~\ref{tab:mass_distributions} lists the average mass ($\overline{M}$) as a weighted mean and the median mass ($\widetilde{M}$) of the distributions for each {individual} spectral type, similar to the approach of \citet[][]{2025arXiv251102539L}. We {also provide the statistics by helium abundance, so for}  He-poor, intermediate helium (iHe-), and extreme helium (eHe-) hot subdwarfs{, representing the properties of the distributions shown in Fig.~\ref{fig:hrd_mass_distribution}.} 

\begin{table}
\footnotesize
\caption{Mass properties by spectral type in our sample. }\label{tab:mass_distributions}      
\centering                    
\begin{tabular}{l c c c c c c}        
\toprule\toprule
Type & $N$ & $\overline{M}$ & $\widetilde{M}$ &  $M_{\rm err}$ & $\sigma$ & $Q_{16}$-$Q_{84}$\\ 
 &  & [\msun] & [\msun] & [\%] & [\%] & [\msun] \\
 &  & (1) & (2) & (3) & (4) & (5)\\ 
\midrule
sdB & 32 & 0.48 & 0.46 & 13 & 30 & 0.40--0.54 \\
sdOB & {33} & 0.47 & 0.44 & 15 & 25 & 0.37--0.56 \\
pulsators & 15 & 0.49 & 0.47 & 12 & 23 & 0.40--0.61 \\
sdO & 5 & 0.43 & 0.36 & 19 & 32 & 0.26--0.48 \\
iHe-sdOB & 7 & 0.47 & 0.47 & 22 & 42 & 0.36--0.67 \\
eHe-sdOB & 3 & 0.41 & 0.48 & 20 & 25 & 0.32--0.50 \\
iHe-sdO & 5 & 0.73 & 0.80 & 18 & 34 & 0.56--1.08 \\
eHe-sdO & 18 & 0.80 & 0.79 & 24 & 34 & 0.53--1.07\\
\midrule
He-poor & {70} & 0.47 & 0.45& 15 & 28 & 0.36--0.56 \\
iHe-rich & 12 & 0.58 & 0.55 & 19 & 50 & 0.42--0.95 \\
eHe-rich & 21 & 0.74 & 0.70 & 23 & 41 & 0.48--1.05 \\
all & {103} & 0.51 & 0.47 & 17 & 47 & 0.36--0.70 \\
\hline
\end{tabular}
\tablefoot{
 (1) Weighted average. (2) Median. (3) Median of the individual mass uncertainties. (4) Standard deviation. (5) Interquartile range, 16$^{th}$-84$^{th}$ percentile. Both $M_{\rm err}$ and $\sigma$ are expressed as a percentage to remove the correlation between the masses and their absolute uncertainties.
}

\end{table}

Helium-poor sdB and sdOB stars are the most numerous members of the sample. Both show very similar median values (0.46 and 0.45\,\msun, respectively) consistent with predictions of the canonical evolution theory \citep{Dorman1993}. Given that the width of their distributions {($\sigma$)} is similar to the median of the individual mass uncertainties {($M_{\rm err}$)}, it is consistent with a narrow intrinsic distribution predicted by binary population synthesis models \citep[BPS,][]{2003MNRAS.341..669H}.     
\footnote{We note that HS2033+0507 has a mass of 1.1~\msun\ but with large uncertainties (see Fig.~\ref{fig:hrd_mass_distribution}). It is among the most distant sdBs in our sample. While this one star has little influence on the weighted average or the median mass of the sdBs, it strongly affects the standard deviation ($\sigma$). When removing the star, the $\sigma$ of the sdB distribution is equal to 22\%, thus fully consistent with that of the sdOB distribution. }
On the other hand, the mass distribution of the eHe-rich subdwarfs is very different from that of the He-poor hot subdwarfs. It is much broader, with 68\% of the stellar masses falling between 0.48 and 1.05~\msun\ with a median value of 0.70\,\msun. Furthermore, it does not show a distinct peak (see Fig.~\ref{fig:hrd_mass_distribution}). The mass distribution of the iHe subdwarfs is similar to that of the extreme ones (0.42\,--\,0.95\,\msun), at a lower median mass of 0.55\,\msun, which is still larger than that of the He-poor subdwarfs.  The standard deviation $\sigma$ is significantly larger than the typical uncertainty $M_\mathrm{err}$ for iHe- and eHe stars. This finding supports the He-WD merger scenario for He-sdO stars, for which BPS models predict a broad
mass distribution with masses up to $\approx$ 0.9\,\msun.


\paragraph{Mass distribution of pulsating subdwarf B stars}


\begin{figure}
\centering
\includegraphics[width=0.49\textwidth]{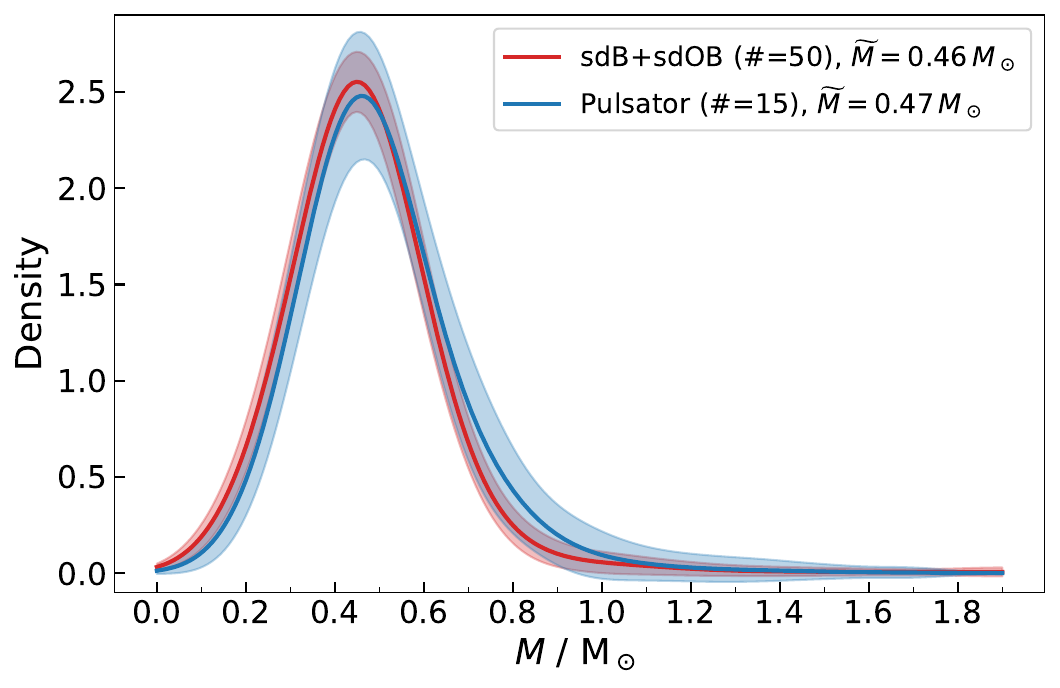}
\caption{Smooth mass distribution (normalised Kernel Density Estimates) for the non-composite pulsating {sdB/sdOB} stars (blue) from Monte Carlo calculations. For comparison
the mass distribution of the hot helium-poor subdwarf stars not known to pulsate is shown in red. The shaded bands denote the uncertainty ranges.}
\label{fig:mass_pulsators}
\end{figure}

The resulting mass distribution for the 15 non-composite pulsators is shown in Fig. \ref{fig:mass_pulsators} and compared to the distribution of the helium-poor sdBs and sdOBs, not known to pulsate. No difference in their mass distribution can be seen, and the median masses agree.








                 







\section{Conclusions from comparison to similar studies}\label{sect:comparison_studies} 

\begin{table*}
\centering
\caption{Median masses and quantiles for mass distributions from the HQS sample compared to those from the Arizona-Montr\'{e}al \citep[AM, ][]{2025arXiv251102539L}, and the volume limited 500pc sample \citep{Dawson2025} grouped by He abundance. $n$ is the number of stars per subclass.}
\label{tab:mass_distribution_comparision}
\renewcommand{\arraystretch}{1.15} 
        \begin{tabular}{cccc}
                \toprule\toprule
                 Type & HQS & AM & 500~pc   \\
                \midrule 
                He-poor  & 0.45 (0.37--0.56) ($n$=70) & 0.46 (0.40--0.54) ($n$=208) & 0.48 (0.38-0.60) ($n$=214) \\
                 iHe-rich  &  0.55 (0.42-0.95) ($n$=12) & 0.42 (0.34--0.51) ($n$=14)  &  0.48  (0.38--0.61) ($n$=17)  \\ 
                 eHe-rich  &  0.70 (0.48-1.05) ($n$=21) & 0.76 (0.48--1.18) ($n$=40)  & 0.52 (0.39--0.85) ($n$=15)  \\
%
        \bottomrule
        \end{tabular}

\end{table*}

The conclusions drawn from the HQS sample may be compared to those of two studies using the same analysis strategy. The Arizona-Montr\'{e}al sample \citep[AM;][]{2025arXiv251102539L} is a magnitude-limited sample drawn mostly from the Palomar Green survey, and thus comparable to the HQS sample, although it covers a larger area of sky at a shallower depth. The second study is a detailed analysis of the volume-complete sample of hot subdwarf stars within 500\,pc presented by \citet{Dawson2025}.} 
In Fig. \ref{fig:M_hqs_vs_am_500pc} we compare the mass distributions from the HQS to those from the AM and the 500pc samples. The only volume-limited sample available is that of the hot subdwarfs within 500\,pc. Because of the differences in brightness limits, there is no overlap between the HQS sample and the 500\,pc sample and little {(19 hot subdwarfs)} with the Arizona-Montr\'{e}al sample. 

The results for the H-rich subdwarfs from all three samples are similar (see Table \ref{tab:mass_distribution_comparision}). There is no correlation of mass with \teff, with median massses of 0.45 (HQS), 0.465 (AM), and 0.48 \msun\ (500pc) close to the prediction of canonical evolution models. There is a significant difference concerning the presence of low mass subdwarfs evolving from intermediate mass progenitors (2--3 M$_\odot$), which are present in both the 500\,pc sample (10\%) and to smaller fraction (6\%) in the Arizona-Montr\'{e}al sample, but are lacking in the HQS sample. 

{The eHe-sdOs show the trend of the masses to increase with increasing \teff\ in all samples as predicted for He-MS stars. Significant differences should be noted, though. In the 500\,pc sample, the eHe-sdOs found close to \teff=40\,kK; only one star is hotter than 45\,kK, and their median mass is smaller ($\widetilde {M}$=0.52\,\msun) than in the AM ($\widetilde {M}$=0.76\,\msun) and the HQS sample 
($\widetilde {M}$=0.70\,\msun). Hence, there is a lack of hot massive eHe-sdOs in the 500\,pc sample compared to those in the AM and HQS samples. It is also worthwhile to inspect the distribution along the He-MS. In the AM sample, the sequence appears to  be evenly populated, while there seems to be a gap separating the lower near-canonical masses (0.5\,\msun) He-sdOs from the higher mass ones (0.7 to 1.0\,\msun). This gap is even more obvious in the HRD (Fig.\ \ref{fig:hrd_mass_distribution}). 

{\citet{2024A&A...686A..25D} studied the space distribution of the 500\,pc sample and derived a scale height of 281$\pm$62\,pc. A kinematical study by \citet{Dawson2025} showed that 90\% of the sample belong to the thin disk population. 
Because the average distances of the stars above the plane in both the AM and HQS samples are larger than that scale height (see 
Fig. \ref{fig:sky1}), we conclude that, on average, the AM and HQS hold populations of older stars than the volume-complete 500pc sample. We speculate, that, if eHe-rich subdwarfs are indeed formed from He-WD mergers, the more massive ones are produced in older populations than the less massive ones.
Since the most massive He-WDs are formed in large red giants close to the tip of the RGB, the orbital periods of the post-CE double-He-WDs might be longer than the ones for the less massive He-WDs stripped in the subgiant stage. Since the merging time due to the emission of gravitational wave is a strong function of the orbital period, it might have taken the massive He-WD binaries longer to merge and form the He-sdOs we see now. The population of the 500\,pc sample might therefore just be too young to form them. 
}

\begin{figure}
\centering 
\includegraphics[width=\columnwidth]{figures/Teff_M_median_kombi.pdf}
\includegraphics[width=\columnwidth]{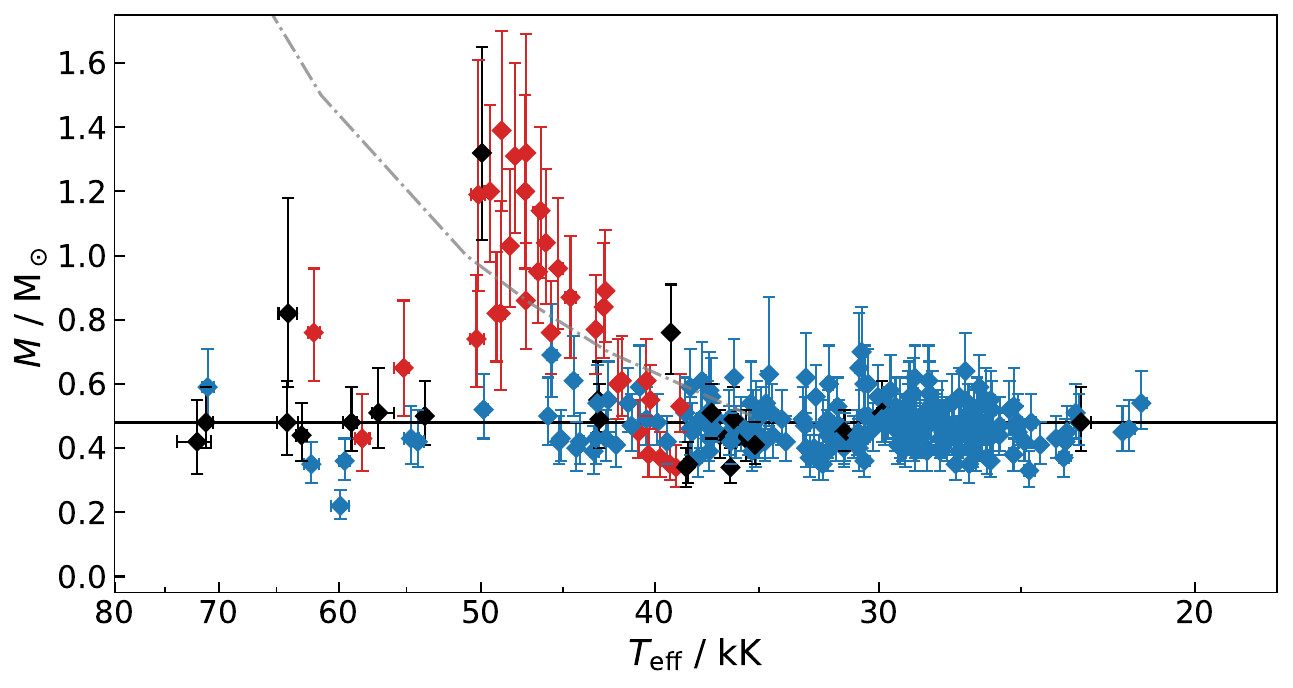} 
\includegraphics[width=\columnwidth]{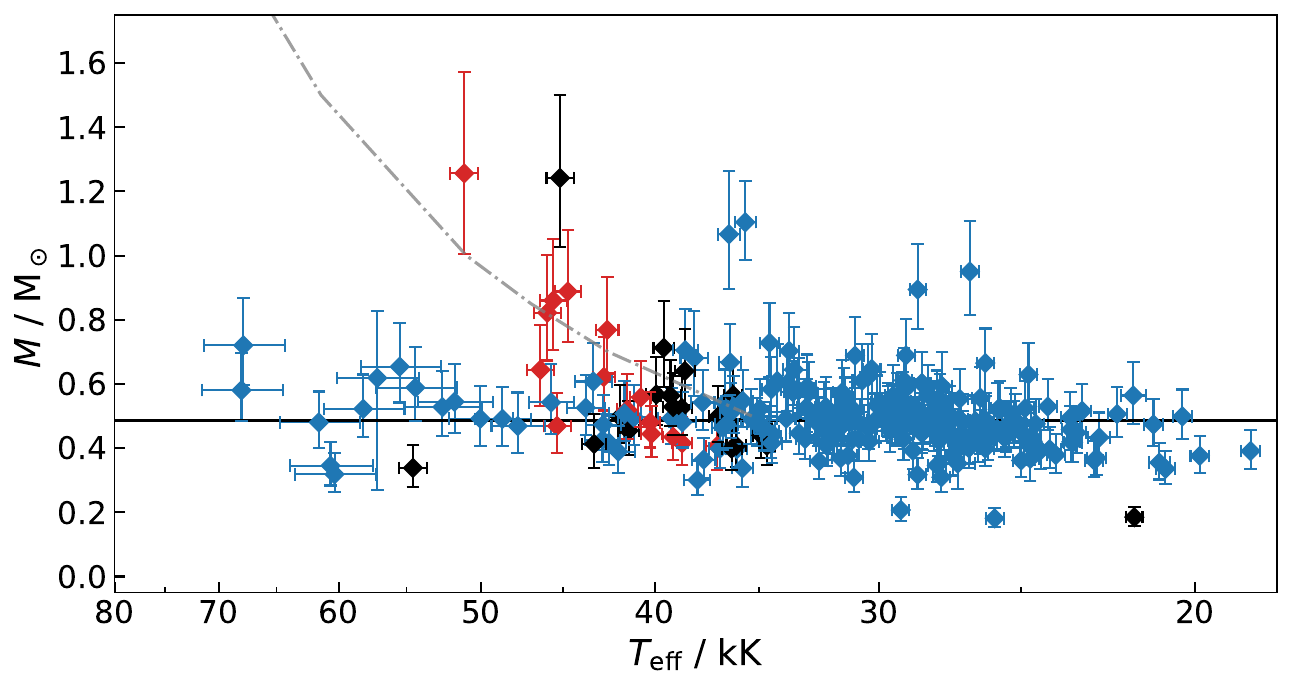}
\caption{Comparison of the mass distribution of the HQS sample (top) to the AM sample \citep[middle,][]{2025arXiv251102539L} and that of the 500pc sample \citep[bottom,][]{Dawson2025}. 
Helium poor stars are marked as blue circles, extremely helium-rich sdOB and sdO stars as red and iHe stars as {black} ones. 
} 
\label{fig:M_hqs_vs_am_500pc}
\end{figure}

\section{Summary and outlook}\label{sect:conclusion}

We studied 152 hot subdwarfs discovered in selected fields of the Hamburg Quasar Survey (HQS). The data include optical spectra, spectral energy distributions and \textit{Gaia} DR3 astrometry. The stars are located in the Northern and Southern Galactic sky, mostly at intermediate Galactic latitudes. Because of the brightness limitations, the survey covers stellar distances from 600\,pc up to a few kpc, which probably includes both young and old stellar populations. 

We detected low mass main-sequence companions of spectral type F, G, or K from their IR excess in 27 sdB/sdOB stars, of which 22 were already known; five sdBs are newly identified as composite binaries. We also discovered two new composite systems among the sdO stars, with HS0735+4026 being the only helium-rich subdwarf showing an IR excess in our survey. 
Light curves from the ZTF database revealed that four sdB binaries show periodic light variations due to reflected light from dM close companions. Because the amplitudes of their light variations were considered small, they were included in the sample for spectral analysis. 

Low resolution spectra of the non-composite systems taken at the Calar Alto observatory were analysed using extensive grids of hybrid LTE/non-LTE model atmospheres and synthetic spectra (the \textit{$2^{nd}$ generation Bamberg model grids}) by making use of a global fitting strategy. Spectra were included only when they cover the blue part towards the Balmer jump to ensure that the spectral data set is homogenous in terms of wavelength coverage. 

This allowed us to derive atmospheric parameters for 76 non-composite He-poor sdB/OB/O stars, 14 He-sdOB, and 26 He-sdO stars and compare them to predictions from evolutionary models in the Kiel diagram. In  addition, we estimated atmospheric parameters for two very hot ($>$100~kK) DAO white dwarfs, four He-rich post-AGB stars, and one CSPN.

The sdB stars in the sample had been analysed before with a mixed set of LTE and non-LTE synthetic spectra \citep{2003A&A...400..939E}. The comparison revealed a trend of the differences between the new and the old \teff\ results, but constant offsets for the gravity and helium abundance. This information may be used to correct previously published results to the new \teff\ scale.

The sdB stars are found along the EHB band in all three previously discovered EHB substructures, with a low population of the coolest regime (EHB1). Most He-rich sdOB and sdO stars line up near the helium main sequence for masses between 0.5 and 1.0 M$_\odot$. 

The spectral energy distributions allowed us to determine stellar angular diameters and the interstellar extinction. In combination with \textit{Gaia} parallaxes, the stellar parameters radius, luminosity, and mass are derived from  the atmospheric parameters. The error budget for the latter is often dominated by systematic uncertainties. For the interpretation of the stellar parameters we restricted the sample further to include stars with parallax uncertainties better than 25\%, only. The resulting (\teff, \logL) distribution (physical HRD) and the mass distribution consistently provide strong evidence that the He-poor hot subdwarfs have masses (0.45 M$_\odot$) consistent with the canonical mass for the core helium flash of low mass RGB stars. We also provided atmospheric and stellar parameters for 15 known pulsating sdB stars and find no significant difference of their distribution of masses and location in the physical HRD.
In previous studies \citep{Geier2022,2025A&A...693A.121H} sdBs were found below the EHB band (bEHB), which harbour subdwarfs with masses less than canonical {($<$0.45 \msun)}.
We found no bEHB star in the HQS sample, though. 

The proximity of the He-rich sdOB and sdO stars to the helium main sequence is as obvious as in the Kiel diagram.
The derived masses of He-rich sdOB stars (median 0.47\,M$_\odot$) are similar to those of the sdB stars, while the He-rich sdO stars are more massive (median 0.81\,M$_\odot$). Both results are consistent with the double helium white dwarf merger scenario. The He-rich sdO stars form the more massive and luminous part, and the He-rich sdOB stars the lower mass and luminosity part of the sequence.

A comparision to similar studies \citep[AM sample][]{2025arXiv251102539L} and the \citep[500pc volume complete sample][]{Dawson2025} revealed mostly consistent results. Significant differences, however, are the absence of low mass subdwarfs evolving from intermediate mass progenitors (2--3 M$_\odot$), in the HQS smaple, which are present in both the 500\,pc sample (10\%) in the AM sample (6\%). For the He-rich subdwarfs significant difference occur for their mass distribution as well as location in the HRD. Most interesting is the lack of hot massive eHe-sdOs in the 500\,pc sample compared to those in the AM and HQS samples. A gap in the HQS-eHe subdwarfs' mass distribution separating the lower near-canonical masses (0.5\,\msun) He-sdOs from the higher mass ones (0.7 to 1.0\,\msun) is not found in the AM sample. We speculate, that the more massive eHe-rich subdwarfs formed from He-WD mergers are produced in older stellar populations than the less massive ones.

We excluded the composite spectrum binaries as well as some reflection effect systems from the study presented here. Those systems are important to understand the stable RLOF binary evolution channel for the formation of hot subdwarfs. We shall apply the models and analysis strategy outlined here, to those objects as well, which requires a larger parameter set to fit and additional observations. 

\section*{Data availability}\label{sect:data_end}

Tables \ref{tab:good_params} and \ref{tab:bad_params} shall be available online at the CDS via anonymous ftp to \url{cdsarc.u-strasbg.fr (XXXX)} or via \url{http://cdsarc.u-strasbg.fr/viz-bin/cat/J/A+A/XXX/zzz}.
All the TWIN and CAFOS spectra analysed here shall be available as fits files at the CDS.

\begin{acknowledgements}

{We thank Andreas Irrgang for developing analysis tools.}
HD was supported by the DFG through grants GE2506/12-1,  GE2506/17-1 and GE2506/9-2, FM through grant GE2506/18-1 and ML through grant LA4383/4-1.
Based on observations collected at the Centro Astron\'omico Hispano Alem\'an (CAHA) at Calar Alto, operated jointly by the Max-Planck Institut f\"ur Astronomie and the Instituto de Astrof\'isica de Andaluc\'ia (CSIC). 
This work has made use of data from the European Space Agency (ESA) mission
{\it Gaia} (\url{https://www.cosmos.esa.int/gaia}), processed by the {\it Gaia}
Data Processing and Analysis Consortium (DPAC,
\url{https://www.cosmos.esa.int/web/gaia/dpac/consortium}). Funding for the DPAC
has been provided by national institutions, in particular the institutions
participating in the {\it Gaia} Multilateral Agreement.
Funding for the SDSS and SDSS-II has been provided by the Alfred P. Sloan Foundation, the Participating Institutions, the National Science 
Foundation, the U.S. Department of Energy, the National Aeronautics and Space Administration, the Japanese Monbukagakusho, the Max Planck Society, 
and the Higher Education Funding Council for England. The SDSS Web Site is http://www.sdss.org/. The SDSS is managed by the Astrophysical Research Consortium for the Participating Institutions. The Participating Institutions are the American Museum of Natural History, Astrophysical Institute Potsdam, University of Basel, University of Cambridge, Case Western Reserve University, University of Chicago, Drexel University, Fermilab, the Institute for Advanced Study, the Japan Participation Group, Johns Hopkins University, the Joint Institute for Nuclear Astrophysics, the Kavli Institute for Particle Astrophysics and Cosmology, the Korean Scientist Group, the Chinese Academy of Sciences (LAMOST), Los Alamos National Laboratory, the Max-Planck-Institute for Astronomy (MPIA), the Max-Planck-Institute for Astrophysics (MPA), New Mexico State University, Ohio State University, University of Pittsburgh, University of Portsmouth, Princeton University, the United States Naval Observatory, and the University of Washington. Funding for the Sloan Digital Sky Survey IV has been provided by the Alfred P. Sloan Foundation, the U.S. Department of Energy Office of Science, and the Participating Institutions. SDSS-IV acknowledges support and resources from the Center for High Performance Computing  at the University of Utah. The SDSS website is www.sdss4.org. SDSS-IV is managed by the Astrophysical Research Consortium for the Participating Institutions of the SDSS Collaboration including the Brazilian Participation Group, the Carnegie Institution for Science, Carnegie Mellon University, Center for Astrophysics, Harvard \& Smithsonian, the Chilean Participation Group, the French Participation Group, Instituto de Astrof\'isica de Canarias, The Johns Hopkins University, Kavli Institute for the Physics and Mathematics of the Universe (IPMU) / University of Tokyo, the Korean Participation Group, Lawrence Berkeley National Laboratory, Leibniz Institut f\"ur Astrophysik Potsdam (AIP), 
Max-Planck-Institut f\"ur Astronomie (MPIA Heidelberg), Max-Planck-Institut f\"ur Astrophysik (MPA Garching), Max-Planck-Institut f\"ur Extraterrestrische Physik (MPE), National Astronomical Observatories of China, New Mexico State University, New York University, University of Notre Dame, Observat\'ario Nacional / MCTI, The Ohio State University, Pennsylvania State University, Shanghai Astronomical Observatory, United Kingdom Participation Group, Universidad Nacional Aut\'onoma de M\'exico, University of Arizona, 
University of Colorado Boulder, University of Oxford, University of Portsmouth, University of Utah, University of Virginia, University of Washington, University of 
Wisconsin, Vanderbilt University, and Yale University.
Guoshoujing Telescope (the Large Sky Area Multi-Object Fiber Spectroscopic Telescope LAMOST) is a National Major Scientific Project built by the Chinese Academy of Sciences. Funding for the project has been provided by the National Development and Reform Commission. LAMOST is operated and managed by the National Astronomical Observatories, Chinese Academy of Sciences.
The research has made use of TOPCAT, an interactive graphical viewer and editor for tabular data
\citep{TOPCAT_2005ASPC..347...29T}, of the SIMBAD database, and the VizieR catalogue access tool, both operated at CDS, Strasbourg, France.
                                                  
\end{acknowledgements}

\bibliographystyle{aa}
\bibliography{hqs_subdwarfs}

\clearpage

\begin{appendix}
\onecolumn
\section*{Supplementary material}

In Appendix \ref{sect:pulsation_reflect} we present details about the pulsating stars in the HQS, results of the search for light variation, and the light curve analyses. Appendix \ref{sect:correlation} presents confidence maps of some subdwarfs representing different spectral subtypes used to derive the correlation coefficient $r$ (see Sect. \ref{sect:spec_analysis}). The systematic uncertainties related to the uncertainty of the metal content has been discussed in Sect. \ref{sect:systematics}. In Appendix \ref{sect:systematics_z} we demonstrate its impact on the Kiel diagram.
Two tables of analysis results can be found in Appendix \ref{stellar_parameters}.

\section{Pulsating subdwarfs and reflection effect binaries}\label{sect:pulsation_reflect}

\begin{figure*}[b]
\centering
\includegraphics[width=0.49\textwidth]{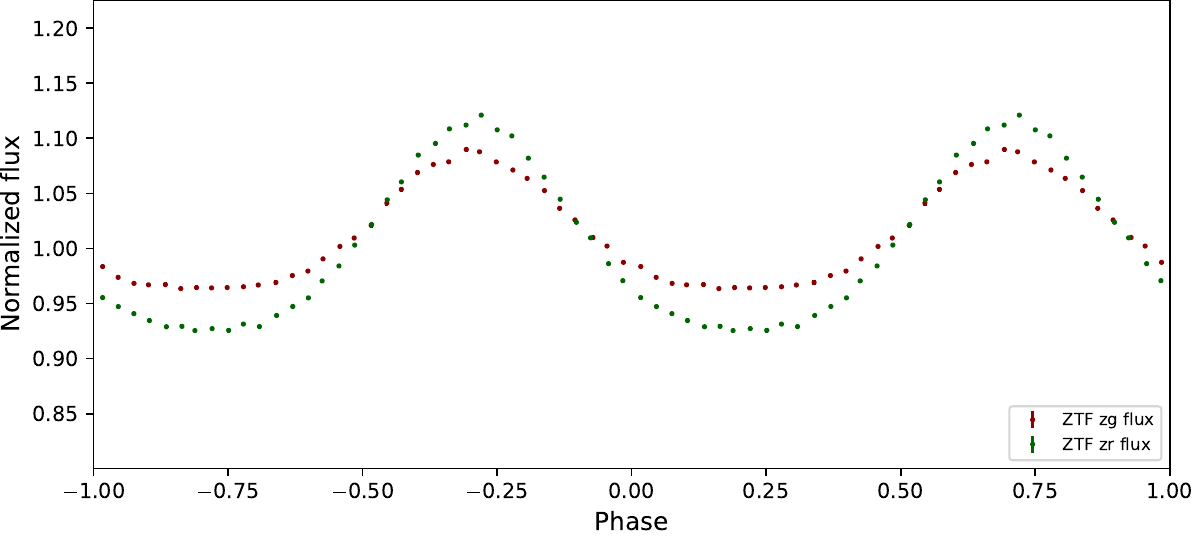} 
\includegraphics[width=0.49\textwidth]{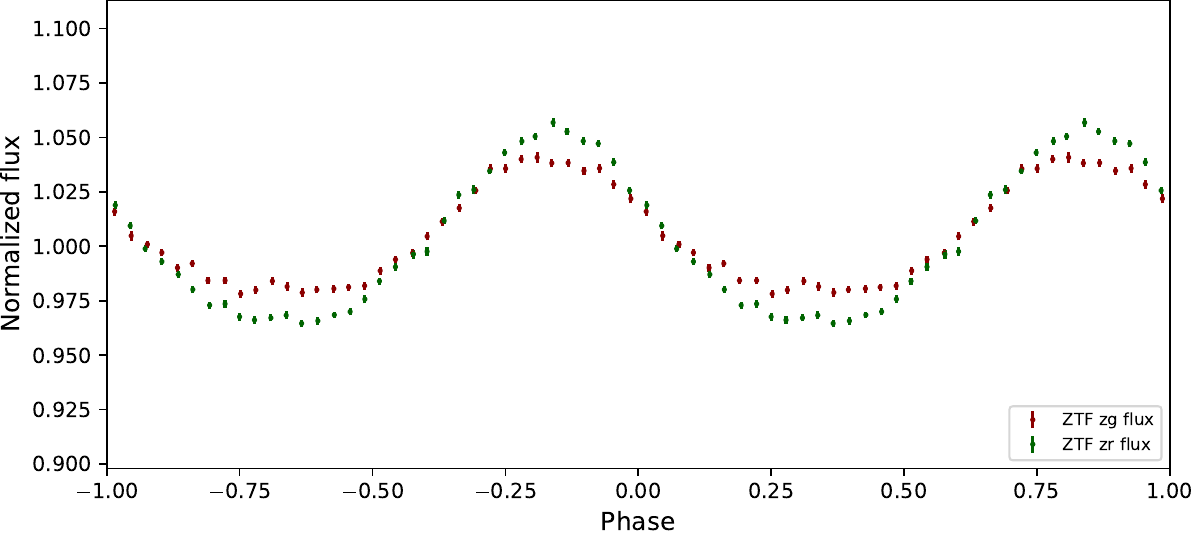} 
\includegraphics[width=0.49\textwidth]{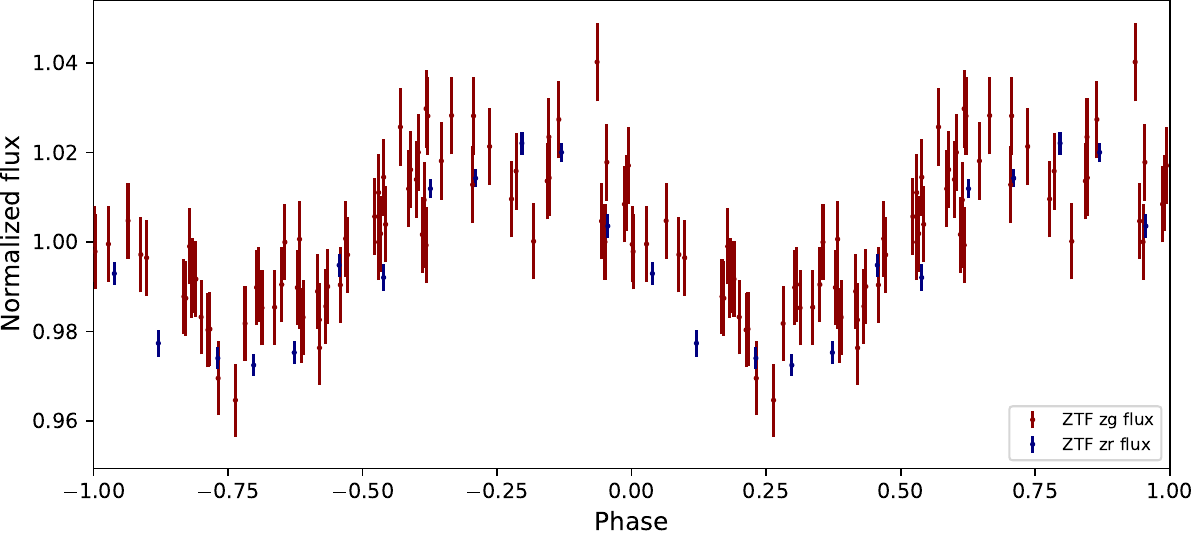} 
\includegraphics[width=0.49\textwidth]{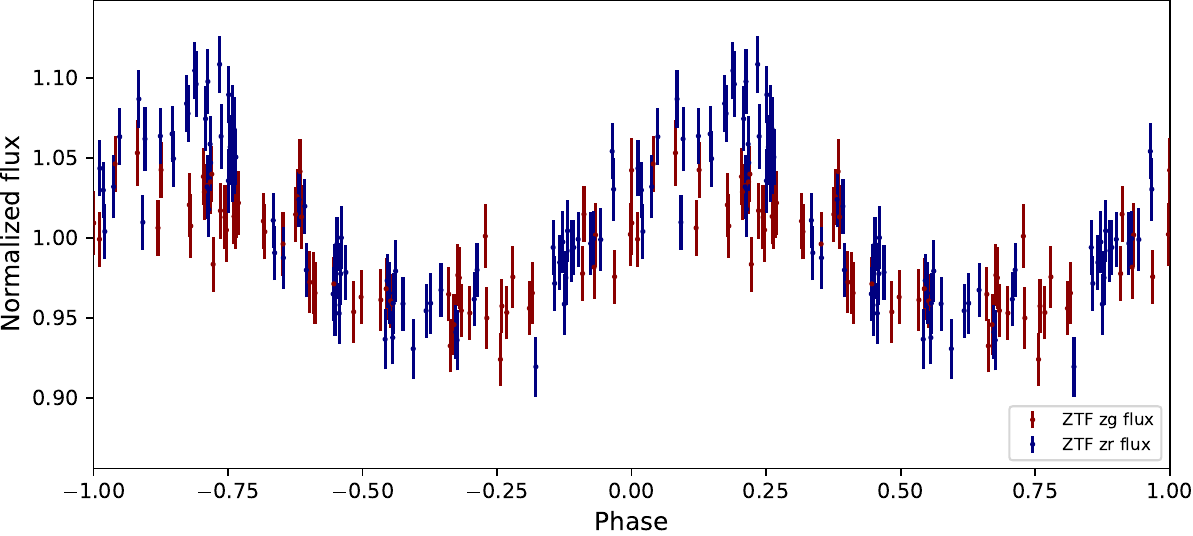}
\caption{ZTF light curves in the $g$ (blue) and $r$ (red) bands of reflection effect systems phased to the corresponding orbital periods: upper left: HS1843+6953 (P=0.3367d); upper right:  HS1909+7004 ($P$=0.3537d); lower left: HS2035+0418 ($P$=0.1310d); lower right: HS1320+2622 ($P$=0.1674d). No eclipses can be found for the other system. Note the different scales. 
}
\label{fig:lc}
\end{figure*}

A photometric follow-up campaign provided light curves for 80 sdB stars from the HQS \citep{2004Ap&SS.291..419S,2010A&A...513A...6O}. The cadence of 20 seconds and observing time of 30 minutes were tailored to detect $p$-mode pulsators, which show multimode oscillations with periods between 1 and 6 minutes. 
Pulsations were found in 10 out of the 80 surveyed stars \citep{2010A&A...513A...6O}.
Two of them,
HS0702+6034 (DW Lyn) and HS2201+2610 (V391 Peg), show additional $g$-modes and are thus hybrid pulsators \citep{
2006A&A...445L..31S,2009A&A...496..469L}.
Several of the HQS pulsators have been followed up to search for light travel time variations \citep[e.g.][]{2020A&A...638A.108M}.

Observations by the Transiting Exoplanet Survey Satellite \textit{TESS} \citep{2015JATIS...1a4003R} have also been successfully analysed to identify new pulsators. The longer observing baselines provided by \textit{TESS} is optimal for the detection of $g$-mode pulsations, which have longer periods and lower amplitudes than the $p$-mode pulsators. 
From the TESS data, four new HQS stars have been found to have $p$-modes \citep{2024A&A...686A..65B} and seven $g$-mode pulsators have been identified \citep{2021A&A...651A.121U,2022A&A...663A..45K,2024A&A...684A.118U,2024PASA...41...41S}.
Furthermore,  HS0212+1446 (TIC 58873368) and HS2233+2332 (TIC 242347840) are likely g-mode pulsators, but a binary nature could not be excluded \citep{2024A&A...684A.118U} because only a single period could be detected for each stars.

In addition to the pulsators, four short-period sdB binaries were discovered during ground-based campaigns, for which the light curves show a reflection effect caused by a close very low mass companion (dM or brown dwarf). Two of them are eclipsing systems of HW Vir type (HS0705+6700=V470 Cam, \cite{2001A&A...379..893D}, 
and HS2231+2441,  \cite{2008ASPC..392..221O}). For the former, eclipse time variations were detected indicating that it might have two circum-binary companions of low mass \citep[][and references therein]{2018haex.bookE..96M,2020MNRAS.499.3071S,2022MNRAS.514.5725P}, as for the latter, it does not show any evidence of period variations from more than 18 years of eclipse timing measurements \citep{2025AdSpR..76.1204E}. 
No eclipses have been found for the two other systems HS2333+3927 \citep{2004A&A...420..251H} and HS2043+0615 \citep{2014A&A...562A..95G}.
These four binaries were excluded from the spectral analysis because the reflection effect is quite large and their optical spectra may be contaminated by reflected light from the heated hemispheres of their companions \cite[see e.g.][]{2021MNRAS.501.3847S}, which is phase dependent.

\citet{2025A&A...700A..71K} analysed  \textit{TESS} observations  and 
found that HS2035+0418 (TIC 466750264) might be a reflection effect binary. Its period is short (0.1310\,d) but the amplitude is small \citep[see Fig. A.2 of][]{2025A&A...700A..71K}. 
HS1741+2133 is a known close binary with a WD companion in a 0.2\,d orbit \citep{2014ASPC..481..293K}. The \textit{TESS} light curve showed ellipsoidal variations with a period of 4.78\,h and an amplitude of $\approx$1 mmag \citep{2023A&A...673A..90S}.
HS1843+6953 and HS1909+7004 were identified as additional candidate reflection effect systems by \citet{2022A&A...666A.182S} with periods of
8.08\,h and 8.48\,h, respectively, but small amplitudes. HS1320+2622 was found to be variable by \citet{2022ApJ...928...20B}, which we confirm from our ZTF light-curve analysis and we derived an orbital period of 0.1674 d. 
Because of their small light variations (see Fig. \ref{fig:lc}) we keep those five stars in the sample. 

We searched for other large-amplitude light variables in our sample by constructing light curves from the ZTF database.
The ZTF is a ground-based photometric survey of
the Northern Hemisphere \citep{2014htu..conf...27B}. ZTF periodically scans the entire Northern Sky with a tree-day cadence, providing
photometric data in the \textit{g} and \textit{r} bands with an approximately two-day cadence. 

We combined all available photometric measurements from the ZTF survey, available for all but three stars of the sample, and used the lightkurve package 
for analysis. We recovered the known reflection effect systems (see Figs. \ref{fig:lc}) but did not find indications for light variations in additional stars at amplitudes exceeding $\approx$0.01 mag and periods from 0.1 to 30 days.



\section{Correlation of temperature and surface gravity in the spectral analysis}\label{sect:correlation}

The shape of the hydrogen and helium lines depends on temperature and density through atomic level excitation, ionisation and collisional line broadening, which leads to a correlation between the effective temperature and gravity.
To explore this correlation we generated 2D $\chi^2$ confidence maps for all stars. We present some examples in Fig. \ref{fig:hqs_conf_maps}, demonstrating that the shape of the maps changes with helium content. The resulting correlation coefficients are plotted in Fig. \ref{fig:teff_logg_corr}. They were included in the systematic uncertainties as well and used in the error propagation to compute the uncertainties on the stellar parameters.

\begin{figure*}[h]
    \begin{subfigure}[b]{0.32\textwidth}
         \centering      
        \includegraphics[width=\textwidth]{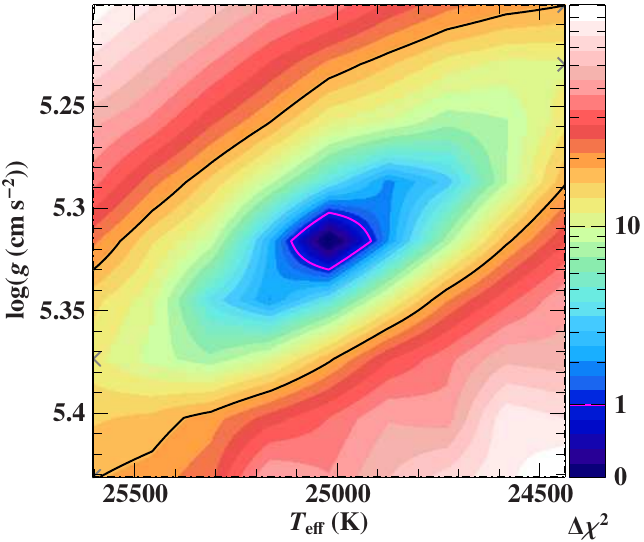}
     \end{subfigure}
     \hfill
    \begin{subfigure}[b]{0.32\textwidth}
        \centering
        \includegraphics[width=\textwidth]{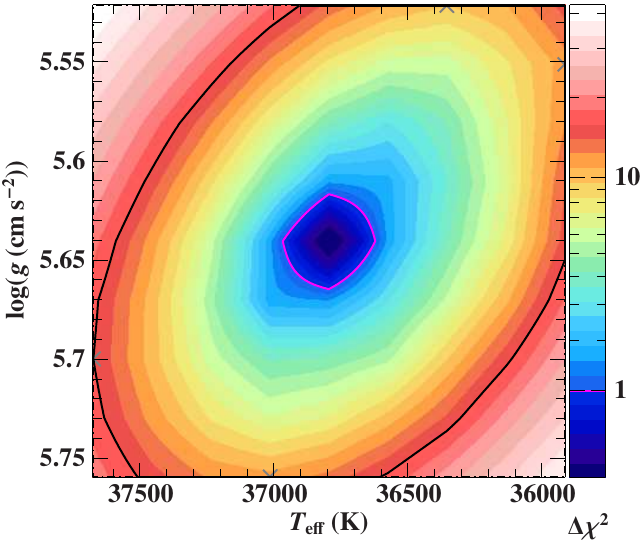}
     \end{subfigure}
     \hfill
     \begin{subfigure}[b]{0.32\textwidth}
         \centering
         \includegraphics[width=\textwidth]{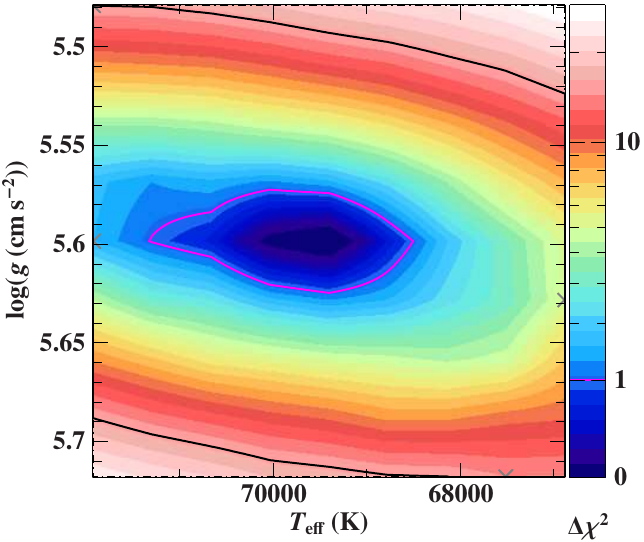}
         \end{subfigure}
     \begin{subfigure}[b]{0.32\textwidth}
         \centering
             \includegraphics[width=\textwidth]{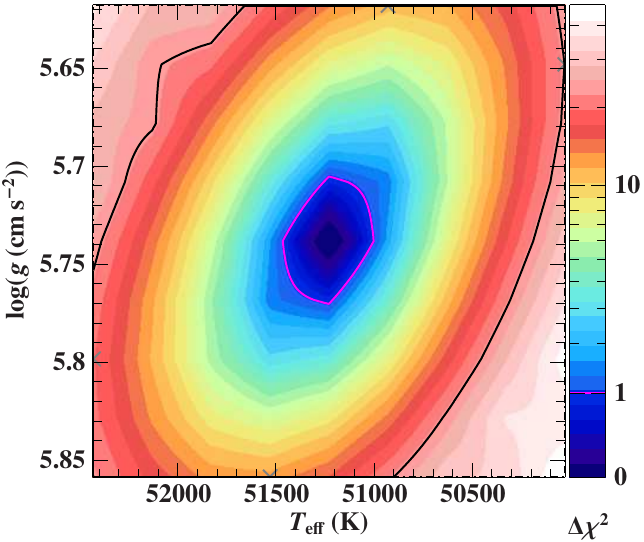}       
     \end{subfigure}
     \hfill
    \begin{subfigure}[b]{0.32\textwidth}
        \centering
     \includegraphics[width=\textwidth]{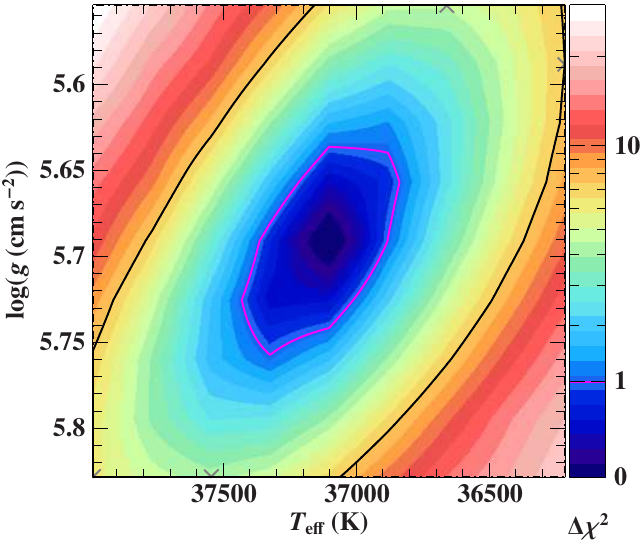}    

    \end{subfigure}
    \hfill
     \begin{subfigure}[b]{0.32\textwidth}
         \centering
         \includegraphics[width=\textwidth]{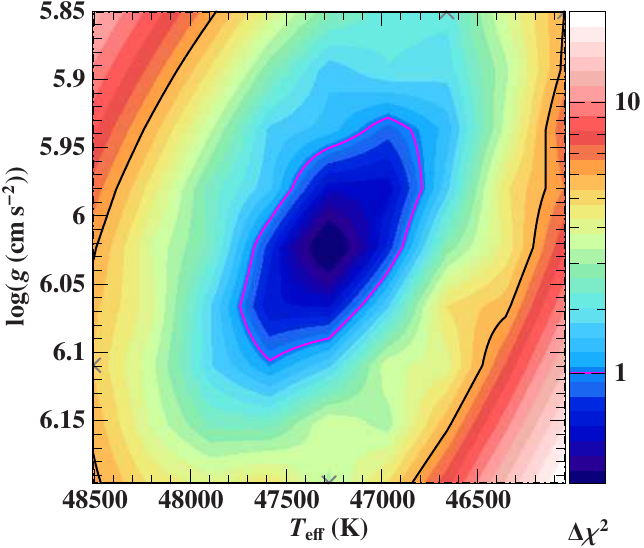}
         \end{subfigure}
    \caption{Confidence maps for atmospheric parameters derived from spectroscopy.
     Examples are selected to cover different spectral types.
The contour lines refer to single p0arameter uncertainties of 68\% (red contour) and 99.7\% (black contour), respectively. 
    \textit{Top row}: The He-poor sdB/O stars: The cool sdB, HS2218+0201 (correlation coefficient r=0.73), the sdOB star HS1806+5024 (r=0.36), and the hot sdO HS2123+0045(r=-0.52). \textit{Bottom row}: He-rich sdOs:
    the iHe sdO HS1832+6955 (r=0.37), the eHe sdOB HS1843+6343 (r=0.70), and the eHe sdO HS1638+6733 (r=0.56).} \label{fig:hqs_conf_maps}
\end{figure*}

\clearpage

\section{Systematic uncertainties due to metallicity differences}\label{sect:systematics_z}

The effect of different metal content has been discussed in Sect. \ref{sect:systematics}.
Fig. \ref{fig:hqs_HRD_z_effect} demonstrates the effect on stars' position in the Kiel-diagram and in the 
\logy\ versus \teff\ plane.

\begin{figure*}[h]
\centering
\includegraphics[width=0.45\textwidth]{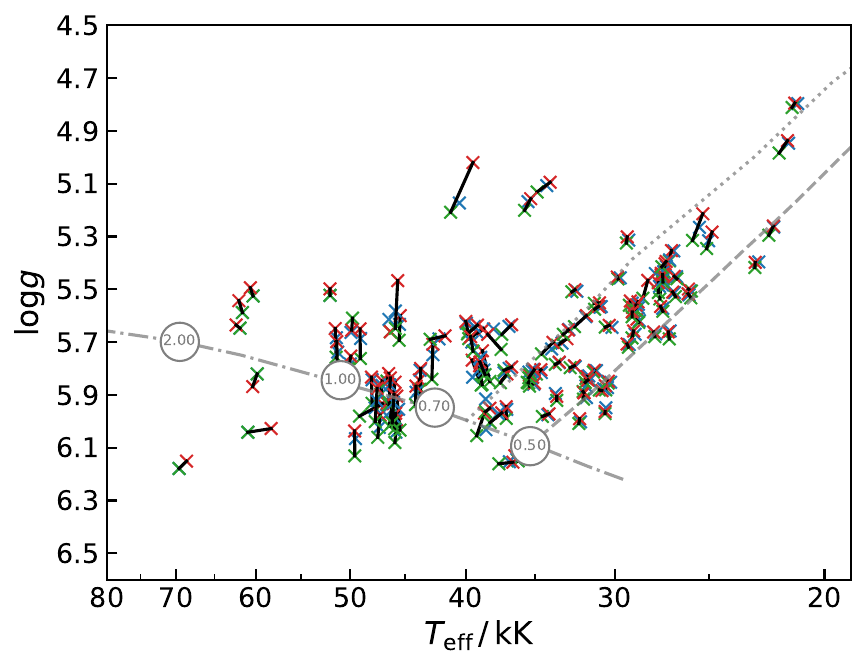} 
\includegraphics[width=0.45\textwidth]{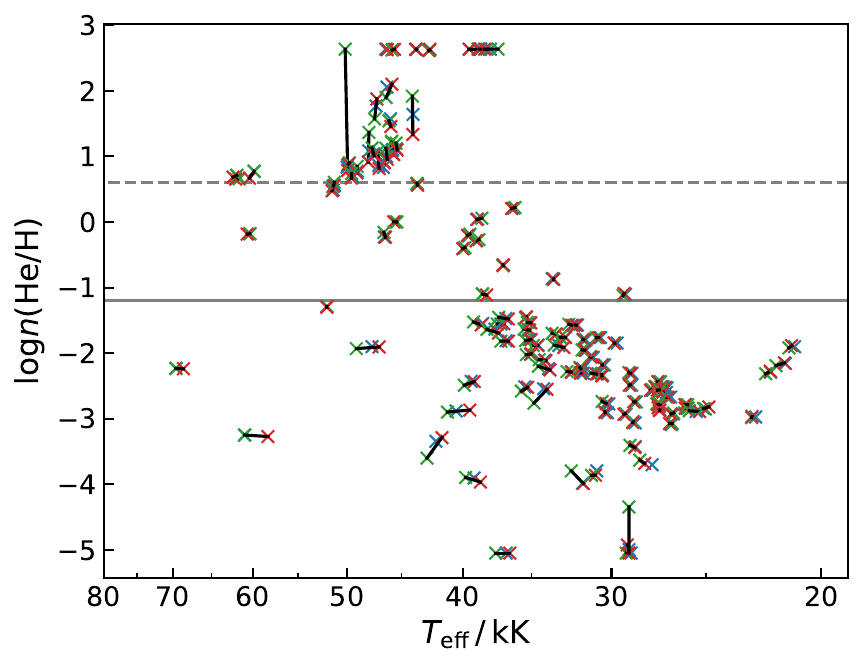} 
\caption{ 
Impact of metallicity for the determination of atmospheric parameters in the 
\teff\ versus \logg\ (left) and \logy\ versus \teff\ (right) plane. Results derived for standard metal composition (z = 0) are marked by blue crosses.  
Red and green symbols indicate the parameters
derived for these same stars when fixing the metal content to half (z=-0.3) and twice (z=+0.3) the standard composition, respectively, and are connected by black lines
to highlight the shifts. The position of the EHB band (dashed and dotted lines) and the zero age helium main-sequence (dashed-dotted line, marked with the stellar mass) is also shown in the \teff\ versus \logg\ plane.}
\label{fig:hqs_HRD_z_effect}
\end{figure*}

\section{Atmospheric and stellar parameters}\label{stellar_parameters}

Table \ref{tab:good_params} contains the atmospheric parameters obtained from spectroscopy and stellar parameters obtained from parallax and SED for the HQS subdwarfs, which have parallax uncertainties better than 25\%. Table \ref{tab:bad_params} gives the atmospheric parameters from spectroscopy and the angular diameter and interstellar reddening parameter from the SED analysis for the rest of the HQS subdwarf sample.


{\small
\begin{landscape}
\begin{longtable}{lllllllllllll}
\renewcommand{\arraystretch}{1.2}
\setlength{\tabcolsep}{0.25em}\\
\caption{Results for stars with suitable parallaxes ($\delta \varpi/\varpi<0.25$). 
Effective temperature (\teff), surface gravity (\logg), helium to hydrogen abundance
ratio ($\log{n(\mathrm{He}/\mathrm{H})}$) are from spectroscopy. Interstellar reddening ($E(44-55)$) and angular diameters ($\Theta$) are from SED fitting and the stellar parameters (radius $R$, luminosity $L$, and mass $M$) are from \textit{Gaia} DR3 parallax.
}\label{tab:good_params}\\ 
\toprule\toprule
Name & \textit{Gaia} DR3 ID & Class & $T_\mathrm{eff}$ & $\log{g}$ & $\log{n(\mathrm{He}/\mathrm{H})}$ & $E(44-55)$ & $\log \Theta$ & $R$ & $L$ & $M$ & comment \\
  &   &   & [K] & [cgs] &  & [mag] & [rad] & [R$_\odot]$& [L$_\odot$] & [M$_\odot$] & \\
\midrule
\endfirsthead 
\caption{Continued}\\
\toprule\toprule
Name & \textit{Gaia} DR3 ID & Class & $T_\mathrm{eff}$  & $\log{g}$ & $\log{n(\mathrm{He}/\mathrm{H})}$ & $E$(44$-$55) & $\log \Theta$ & $R$ & $L$ & $M$ & comment  \\
  &   &   & [K] & [cgs] &  & [mag] & [rad] & [R$_\odot$]& [L$_\odot$]& [M$_\odot$] & \\
\midrule
\endhead
\endlastfoot
\endlastfoot
HS0023+3049 & 2862194144817359872 & sdB & $29000^{+ 50}_{- 50}$ & +$5.55^{+ 0.01}_{- 0.01}$ & $<-4.5$ & +$0.102^{+ 0.003}_{- 0.020}$ & $-11.142^{+ 0.003}_{- 0.019}$ & +$0.185^{+ 0.010}_{- 0.010}$ & $21.8^{+ 2.5}_{- 2.3}$ & $0.45^{+ 0.07}_{- 0.06}$ & L,C2 \\
HS0025+3423 & 365939429892433920 & iHe-sdOB & $33640^{+ 80}_{- 100}$ & +$5.89^{+ 0.02}_{- 0.01}$ & $-0.86^{+ 0.01}_{- 0.01}$ & +$0.102^{+ 0.003}_{- 0.007}$ & $-11.382^{+ 0.003}_{- 0.007}$ & +$0.113^{+ 0.011}_{- 0.010}$ & $14.6^{+ 3}_{- 2.3}$ & $0.36^{+ 0.09}_{- 0.07}$ & T,L \\
HS0035+3034 & 2858607022491460096 & sdOB & $38470^{+ 210}_{- 210}$ & +$6.03^{+ 0.03}_{- 0.03}$ & $-1.55^{+ 0.03}_{- 0.04}$ & +$0.085^{+ 0.007}_{- 0.007}$ & $-11.509^{+ 0.005}_{- 0.007}$ & +$0.105^{+ 0.012}_{- 0.010}$ & $22^{+ 6}_{- 4}$ & $0.44^{+ 0.12}_{- 0.09}$ & T \\
HS0039+4302 & 387533868326227456 & sdOB & $32380^{+ 80}_{- 120}$ & +$5.79^{+ 0.02}_{- 0.02}$ & $-2.29^{+ 0.04}_{- 0.04}$ & +$0.070^{+ 0.004}_{- 0.008}$ & $-11.324^{+ 0.004}_{- 0.007}$ & +$0.137^{+ 0.010}_{- 0.009}$ & $18.5^{+ 2.6}_{- 2.2}$ & $0.42^{+ 0.07}_{- 0.06}$ & C,P \\
HS0040+4417 & 387992437690797696 & sdB & $29700^{+ 600}_{- 600}$ & +$5.46^{+ 0.09}_{- 0.09}$ & $-1.85^{+ 0.11}_{- 0.11}$ & +$0.078^{+ 0.007}_{- 0.016}$ & $-11.386^{+ 0.007}_{- 0.016}$ & +$0.147^{+ 0.017}_{- 0.014}$ & $15^{+ 4}_{- 2.8}$ & $0.23^{+ 0.07}_{- 0.06}$ & C \\
HS0048+0026 & 2537182630815412736 & iHe-sdOB & $38500^{+ 60}_{- 60}$ & +$5.92^{+ 0.01}_{- 0.01}$ & $-1.10^{+ 0.01}_{- 0.01}$ & +$0.028^{+ 0.003}_{- 0.003}$ & $-11.499^{+ 0.003}_{- 0.004}$ & +$0.120^{+ 0.014}_{- 0.012}$ & $29^{+ 7}_{- 6}$ & $0.43^{+ 0.12}_{- 0.09}$ & T,L,S \\
HS0055+0138 & 2537690089791382656 & sdOB & $33210^{+ 140}_{- 180}$ & +$5.70^{+ 0.02}_{- 0.03}$ & $-1.88^{+ 0.03}_{- 0.03}$ & +$0.020^{+ 0.003}_{- 0.004}$ & $-11.302^{+ 0.005}_{- 0.005}$ & +$0.141^{+ 0.011}_{- 0.010}$ & $22^{+ 4}_{- 2.8}$ & $0.36^{+ 0.07}_{- 0.06}$ & T,L \\
HS0209+0141 & 2514278566658235776 & sdOB & $30160^{+ 90}_{- 90}$ & +$5.64^{+ 0.02}_{- 0.02}$ & $-2.77^{+ 0.05}_{- 0.05}$ & +$0.034^{+ 0.002}_{- 0.003}$ & $-11.066^{+ 0.004}_{- 0.004}$ & +$0.142^{+ 0.006}_{- 0.005}$ & $14.9^{+ 1.3}_{- 1.1}$ & $0.32^{+ 0.04}_{- 0.04}$ & T,L \\
HS0212+1446 & 78846635958426496 & sdB & $27350^{+ 70}_{- 90}$ & +$5.48^{+ 0.02}_{- 0.01}$ & $-2.81^{+ 0.05}_{- 0.05}$ & +$0.099^{+ 0.005}_{- 0.013}$ & $-11.038^{+ 0.006}_{- 0.013}$ & +$0.205^{+ 0.009}_{- 0.009}$ & $21.2^{+ 2}_{- 1.8}$ & $0.47^{+ 0.06}_{- 0.06}$ & L,P \\
HS0213+2329 & 101206407500333184 & sdOB & $32270^{+ 70}_{- 170}$ & +$5.51^{+ 0.02}_{- 0.02}$ & $-1.58^{+ 0.02}_{- 0.02}$ & +$0.107^{+ 0.010}_{- 0.010}$ & $-10.992^{+ 0.006}_{- 0.006}$ & +$0.206^{+ 0.011}_{- 0.010}$ & $41^{+ 5}_{- 4}$ & $0.5^{+ 0.07}_{- 0.06}$ & T \\
HS0222+2334 & 101826669497182720 & sdO & $47660^{+ 70}_{- 110}$ & +$5.95^{+ 0.01}_{- 0.01}$ & $-1.90^{+ 0.02}_{- 0.01}$ & +$0.123^{+ 0.008}_{- 0.008}$ & $-11.093^{+ 0.007}_{- 0.007}$ & +$0.129^{+ 0.005}_{- 0.005}$ & $77^{+ 9}_{- 8}$ & $0.54^{+ 0.07}_{- 0.06}$ & T,L \\
HS0233+3037 & 132944115577925504 & sdB & $26950^{+ 70}_{- 60}$ & +$5.66^{+ 0.02}_{- 0.01}$ & $-2.53^{+ 0.02}_{- 0.02}$ & +$0.171^{+ 0.003}_{- 0.003}$ & $-10.915^{+ 0.003}_{- 0.003}$ & +$0.165^{+ 0.005}_{- 0.004}$ & $12.9^{+ 0.7}_{- 0.7}$ & $0.45^{+ 0.05}_{- 0.04}$ & T,L \\
HS0258+3003 & 123108816565129344 & sdO & $69600^{+ 600}_{- 600}$ & +$6.18^{+ 0.02}_{- 0.02}$ & $-2.23^{+ 0.03}_{- 0.03}$ & +$0.180^{+ 0.010}_{- 0.010}$ & $-11.378^{+ 0.007}_{- 0.006}$ & +$0.080^{+ 0.004}_{- 0.004}$ & $136^{+ 14}_{- 12}$ & $0.36^{+ 0.05}_{- 0.04}$ & T,L,N \\
HS0312+2225 & 62425258001804160 & iHe-sdO & $45540^{+ 70}_{- 90}$ & +$5.95^{+ 0.03}_{- 0.02}$ & +$0.00^{+ 0.01}_{- 0.01}$ & +$0.203^{+ 0.004}_{- 0.007}$ & $-11.320^{+ 0.003}_{- 0.006}$ & +$0.138^{+ 0.009}_{- 0.008}$ & $73^{+ 10}_{- 8}$ & $0.62^{+ 0.13}_{- 0.11}$ & T \\
HS0315+3355 & 125893776439505664 & eHe-sdOB & $39470^{+ 50}_{- 210}$ & +$5.83^{+ 0.02}_{- 0.04}$ & $>2.6$ & +$0.198^{+ 0.006}_{- 0.006}$ & $-11.459^{+ 0.006}_{- 0.006}$ & +$0.098^{+ 0.009}_{- 0.008}$ & $21^{+ 4}_{- 4}$ & $0.24^{+ 0.06}_{- 0.05}$ & T,L \\
HS0338+2946 & 120129277493243008 & sdOB & $30240^{+ 100}_{- 100}$ & +$5.85^{+ 0.02}_{- 0.02}$ & $-2.91^{+ 0.08}_{- 0.08}$ & +$0.243^{+ 0.007}_{- 0.007}$ & $-11.109^{+ 0.005}_{- 0.005}$ & +$0.138^{+ 0.006}_{- 0.005}$ & $14.4^{+ 1.2}_{- 1.1}$ & $0.5^{+ 0.06}_{- 0.05}$ & C \\
HS0339+8153 & 569988137807051136 & eHe-sdO & $42620^{+ 100}_{- 70}$ & +$5.75^{+ 0.03}_{- 0.03}$ & $>2.3$ & +$0.224^{+ 0.009}_{- 0.009}$ & $-11.117^{+ 0.007}_{- 0.007}$ & +$0.211^{+ 0.009}_{- 0.008}$ & $133^{+ 12}_{- 10}$ & $0.91^{+ 0.18}_{- 0.15}$ & T \\
HS0352+1019 & 3302955130527310464 & sdB & $25930^{+ 40}_{- 60}$ & +$5.51^{+ 0.01}_{- 0.01}$ & $-2.84^{+ 0.04}_{- 0.04}$ & +$0.207^{+ 0.003}_{- 0.003}$ & $-10.886^{+ 0.003}_{- 0.003}$ & +$0.214^{+ 0.006}_{- 0.005}$ & $18.6^{+ 1}_{- 1}$ & $0.54^{+ 0.05}_{- 0.05}$ & C,L,P \\
HS0357+0133 & 3259569020169965312 & sdB & $28840^{+ 90}_{- 100}$ & +$5.67^{+ 0.02}_{- 0.02}$ & $-2.31^{+ 0.03}_{- 0.03}$ & +$0.317^{+ 0.005}_{- 0.005}$ & $-11.020^{+ 0.004}_{- 0.004}$ & +$0.173^{+ 0.005}_{- 0.005}$ & $18.7^{+ 1.1}_{- 1}$ & $0.51^{+ 0.06}_{- 0.05}$ & T,L \\
HS0430+7712 & 552891694067107328 & sdB & $26720^{+ 50}_{- 40}$ & +$5.51^{+ 0.01}_{- 0.01}$ & $-3.09^{+ 0.06}_{- 0.09}$ & +$0.138^{+ 0.004}_{- 0.004}$ & $-10.856^{+ 0.003}_{- 0.003}$ & +$0.218^{+ 0.005}_{- 0.005}$ & $21.8^{+ 1.1}_{- 1}$ & $0.56^{+ 0.05}_{- 0.05}$ & C,P \\
HS0444+0458 & 3281870929912521984 & sdOB & $34560^{+ 180}_{- 170}$ & +$5.82^{+ 0.03}_{- 0.03}$ & $-1.88^{+ 0.04}_{- 0.04}$ & +$0.073^{+ 0.004}_{- 0.024}$ & $-11.382^{+ 0.004}_{- 0.017}$ & +$0.145^{+ 0.014}_{- 0.012}$ & $27^{+ 6}_{- 5}$ & $0.5^{+ 0.11}_{- 0.09}$ & AM,P \\
HS0445+7503 & 503632542352021376 & sdOB & $31070^{+ 70}_{- 80}$ & +$5.81^{+ 0.02}_{- 0.01}$ & $-2.06^{+ 0.03}_{- 0.03}$ & +$0.161^{+ 0.003}_{- 0.003}$ & $-11.084^{+ 0.003}_{- 0.004}$ & +$0.164^{+ 0.005}_{- 0.004}$ & $22.6^{+ 1.3}_{- 1.2}$ & $0.63^{+ 0.06}_{- 0.06}$ & C \\
HS0447+7545 & 551708207239316352 & sdOB & $30800^{+ 400}_{- 400}$ & +$5.57^{+ 0.04}_{- 0.06}$ & $-2.31^{+ 0.09}_{- 0.10}$ & +$0.241^{+ 0.011}_{- 0.011}$ & $-11.264^{+ 0.008}_{- 0.009}$ & +$0.179^{+ 0.010}_{- 0.010}$ & $26^{+ 4}_{- 2.8}$ & $0.43^{+ 0.07}_{- 0.07}$ & C,N \\
HS0457+0907 & 3291164582009541376 & sdOB & $37140^{+ 190}_{- 150}$ & +$5.82^{+ 0.03}_{- 0.02}$ & $-1.55^{+ 0.03}_{- 0.05}$ & +$0.149^{+ 0.003}_{- 0.003}$ & $-11.100^{+ 0.004}_{- 0.004}$ & +$0.132^{+ 0.006}_{- 0.005}$ & $30^{+ 2.6}_{- 2.3}$ & $0.42^{+ 0.06}_{- 0.05}$ & C \\
HS0546+8009 & 553641320479547520 & sdOB & $35080^{+ 130}_{- 120}$ & +$5.86^{+ 0.02}_{- 0.02}$ & $-2.02^{+ 0.02}_{- 0.02}$ & +$0.092^{+ 0.003}_{- 0.004}$ & $-11.102^{+ 0.003}_{- 0.005}$ & +$0.137^{+ 0.003}_{- 0.003}$ & $25.5^{+ 1.3}_{- 1.3}$ & $0.49^{+ 0.05}_{- 0.05}$ & T \\
HS0659+5734 & 1000554898377796864 & eHe-sdO & $45840^{+ 50}_{- 50}$ & +$5.99^{+ 0.02}_{- 0.01}$ & +$1.09^{+ 0.04}_{- 0.04}$ & +$0.066^{+ 0.003}_{- 0.016}$ & $-11.456^{+ 0.003}_{- 0.016}$ & +$0.163^{+ 0.021}_{- 0.017}$ & $105^{+ 28}_{- 21}$ & $0.94^{+ 0.27}_{- 0.2}$ & T \\
HS0702+6043 & 1002864074658491648 & sdB & $27310^{+ 70}_{- 100}$ & +$5.45^{+ 0.02}_{- 0.02}$ & $-2.79^{+ 0.05}_{- 0.05}$ & +$0.077^{+ 0.004}_{- 0.004}$ & $-11.147^{+ 0.004}_{- 0.004}$ & +$0.222^{+ 0.013}_{- 0.012}$ & $24.7^{+ 3}_{- 2.6}$ & $0.51^{+ 0.08}_{- 0.07}$ & C,P \\
HS0707+8225 & 1142767759538804608 & sdB & $27700^{+ 400}_{- 150}$ & +$5.44^{+ 0.05}_{- 0.03}$ & $-3.70^{+ 0.23}_{- 0.25}$ & +$0.042^{+ 0.003}_{- 0.006}$ & $-11.242^{+ 0.006}_{- 0.008}$ & +$0.216^{+ 0.015}_{- 0.013}$ & $25^{+ 4}_{- 4}$ & $0.47^{+ 0.1}_{- 0.08}$ & T \\
HS0708+6938 & 1109216024779190016 & iHe-sdOB & $38960^{+ 140}_{- 700}$ & +$5.80^{+ 0.07}_{- 0.06}$ & $-0.28^{+ 0.03}_{- 0.03}$ & +$0.070^{+ 0.007}_{- 0.007}$ & $-11.544^{+ 0.008}_{- 0.005}$ & +$0.167^{+ 0.024}_{- 0.019}$ & $57^{+ 18}_{- 13}$ & $0.65^{+ 0.23}_{- 0.16}$ & T,S,N \\
HS0740+3734 & 895907607894131840 & sdB & $22060^{+ 90}_{- 120}$ & +$5.26^{+ 0.02}_{- 0.02}$ & $-2.29^{+ 0.03}_{- 0.04}$ & +$0.059^{+ 0.003}_{- 0.007}$ & $-11.052^{+ 0.003}_{- 0.007}$ & +$0.250^{+ 0.012}_{- 0.011}$ & $13.3^{+ 1.4}_{- 1.2}$ & $0.42^{+ 0.06}_{- 0.05}$ & T,P \\
HS0741+3818 & 919963474904219776 & sdB & $29190^{+ 110}_{- 120}$ & +$5.31^{+ 0.02}_{- 0.02}$ & $-1.10^{+ 0.02}_{- 0.02}$ & +$0.081^{+ 0.006}_{- 0.030}$ & $-11.154^{+ 0.004}_{- 0.029}$ & +$0.247^{+ 0.020}_{- 0.019}$ & $40^{+ 7}_{- 6}$ & $0.46^{+ 0.09}_{- 0.08}$ & T,L \\
HS0743+6255 & 1088277608237901568 & eHe-sdO & $49010^{+ 290}_{- 290}$ & +$5.69^{+ 0.07}_{- 0.07}$ & +$0.77^{+ 0.08}_{- 0.10}$ & +$0.084^{+ 0.005}_{- 0.005}$ & $-11.669^{+ 0.003}_{- 0.005}$ & +$0.159^{+ 0.040}_{- 0.027}$ & $130^{+ 80}_{- 50}$ & $0.45^{+ 0.28}_{- 0.15}$ & T,N \\
HS0815+4243 & 915859960070594176 & sdOB & $34130^{+ 60}_{- 60}$ & +$5.98^{+ 0.01}_{- 0.01}$ & $-2.11^{+ 0.02}_{- 0.02}$ & +$0.055^{+ 0.007}_{- 0.014}$ & $-11.513^{+ 0.006}_{- 0.013}$ & +$0.135^{+ 0.021}_{- 0.017}$ & $22^{+ 8}_{- 6}$ & $0.63^{+ 0.22}_{- 0.15}$ & T,L,S,P \\
HS0836+6158 & 1042158494227683072 & eHe-sdOB & $38620^{+ 90}_{- 80}$ & +$5.81^{+ 0.05}_{- 0.04}$ & $>2.6$ & +$0.125^{+ 0.004}_{- 0.004}$ & $-11.116^{+ 0.004}_{- 0.004}$ & +$0.147^{+ 0.005}_{- 0.005}$ & $43^{+ 4}_{- 2.9}$ & $0.5^{+ 0.09}_{- 0.08}$ & T \\
HS1020+6926 & 1073281476521799296 & eHe-sdO & $47940^{+ 170}_{- 150}$ & +$5.84^{+ 0.03}_{- 0.03}$ & +$1.08^{+ 0.10}_{- 0.10}$ & +$0.037^{+ 0.003}_{- 0.004}$ & $-11.291^{+ 0.002}_{- 0.003}$ & +$0.184^{+ 0.011}_{- 0.010}$ & $160^{+ 19}_{- 17}$ & $0.86^{+ 0.16}_{- 0.13}$ & T \\
HS1030+6632 & 1059785555405762432 & eHe-sdO & $47020^{+ 260}_{- 240}$ & +$5.86^{+ 0.05}_{- 0.05}$ & +$0.86^{+ 0.12}_{- 0.12}$ & +$0.012^{+ 0.027}_{- 0.013}$ & $-11.490^{+ 0.018}_{- 0.015}$ & +$0.167^{+ 0.019}_{- 0.016}$ & $123^{+ 29}_{- 22}$ & $0.74^{+ 0.2}_{- 0.15}$ & T \\
HS1051+2933 & 731743061507469568 & iHe-sdOB & $36950^{+ 60}_{- 60}$ & +$5.95^{+ 0.01}_{- 0.01}$ & $-0.66^{+ 0.01}_{- 0.01}$ & +$0.025^{+ 0.004}_{- 0.009}$ & $-11.725^{+ 0.003}_{- 0.009}$ & +$0.096^{+ 0.030}_{- 0.019}$ & $16^{+ 11}_{- 6}$ & $0.31^{+ 0.22}_{- 0.11}$ & T,L,S \\
HS1203+6650 & 1681310444608684928 & eHe-sdO & $43810^{+ 50}_{- 50}$ & +$5.88^{+ 0.02}_{- 0.02}$ & $>2.5$ & +$0.030^{+ 0.006}_{- 0.006}$ & $-11.574^{+ 0.004}_{- 0.005}$ & +$0.187^{+ 0.029}_{- 0.023}$ & $120^{+ 40}_{- 26}$ & $1^{+ 0.4}_{- 0.24}$ & T,S,N \\
HS1236+4754 & 1543006141129419648 & sdB & $27710^{+ 40}_{- 40}$ & +$5.67^{+ 0.01}_{- 0.01}$ & $-2.57^{+ 0.03}_{- 0.03}$ & +$0.007^{+ 0.008}_{- 0.008}$ & $-11.322^{+ 0.007}_{- 0.009}$ & +$0.182^{+ 0.016}_{- 0.014}$ & $18^{+ 4}_{- 2.5}$ & $0.56^{+ 0.11}_{- 0.09}$ & B,L,S \\
HS1547+6312 & 1639999937328089088 & sdOB & $31700^{+ 400}_{- 400}$ & +$5.60^{+ 0.04}_{- 0.04}$ & $-3.98^{+ 0.37}_{- 0.54}$ & +$0.034^{+ 0.005}_{- 0.011}$ & $-11.234^{+ 0.009}_{- 0.012}$ & +$0.152^{+ 0.007}_{- 0.007}$ & $20.9^{+ 2.2}_{- 2}$ & $0.34^{+ 0.05}_{- 0.05}$ & T \\
HS1552+6333 & 1640097759502807936 & sdOB & $33900^{+ 400}_{- 500}$ & +$5.73^{+ 0.08}_{- 0.06}$ & $-2.24^{+ 0.08}_{- 0.08}$ & +$0.022^{+ 0.004}_{- 0.006}$ & $-11.541^{+ 0.006}_{- 0.007}$ & +$0.127^{+ 0.014}_{- 0.012}$ & $19^{+ 5}_{- 4}$ & $0.32^{+ 0.09}_{- 0.07}$ & T,N \\
HS1613+7253 & 1654392643319054464 & iHe-sdO & $46530^{+ 250}_{- 270}$ & +$5.85^{+ 0.04}_{- 0.07}$ & $-0.23^{+ 0.05}_{- 0.04}$ & +$0.014^{+ 0.006}_{- 0.007}$ & $-11.353^{+ 0.004}_{- 0.005}$ & +$0.132^{+ 0.007}_{- 0.006}$ & $74^{+ 8}_{- 7}$ & $0.45^{+ 0.07}_{- 0.07}$ & T \\
HS1638+6733 & 1648476656582002176 & eHe-sdO & $47300^{+ 500}_{- 600}$ & +$6.02^{+ 0.11}_{- 0.12}$ & +$1.77^{+ 0.35}_{- 0.70}$ & +$0.053^{+ 0.005}_{- 0.009}$ & $-11.583^{+ 0.004}_{- 0.008}$ & +$0.182^{+ 0.033}_{- 0.024}$ & $150^{+ 60}_{- 40}$ & $1.3^{+ 0.7}_{- 0.4}$ & T,N \\
HS1706+7424 & 1655490716133145216 & iHe-sdOB & $39950^{+ 90}_{- 90}$ & +$5.63^{+ 0.02}_{- 0.02}$ & $-0.41^{+ 0.02}_{- 0.02}$ & +$0.038^{+ 0.006}_{- 0.006}$ & $-11.263^{+ 0.005}_{- 0.005}$ & +$0.173^{+ 0.009}_{- 0.008}$ & $69^{+ 8}_{- 7}$ & $0.46^{+ 0.07}_{- 0.06}$ & T \\
HS1707+6121 & 1438423824907522688 & eHe-sdO & $50000^{+ 230}_{- 190}$ & +$5.78^{+ 0.05}_{- 0.04}$ & +$0.84^{+ 0.08}_{- 0.09}$ & +$0.019^{+ 0.008}_{- 0.008}$ & $-11.409^{+ 0.005}_{- 0.006}$ & +$0.219^{+ 0.023}_{- 0.019}$ & $270^{+ 60}_{- 50}$ & $1.07^{+ 0.27}_{- 0.21}$ & T \\
HS1710+1614 & 4546882216133354752 & sdOB & $35230^{+ 70}_{- 70}$ & +$5.84^{+ 0.01}_{- 0.01}$ & $-1.64^{+ 0.02}_{- 0.02}$ & +$0.089^{+ 0.010}_{- 0.010}$ & $-11.508^{+ 0.007}_{- 0.007}$ & +$0.130^{+ 0.015}_{- 0.012}$ & $24^{+ 6}_{- 5}$ & $0.43^{+ 0.11}_{- 0.08}$ & U \\
HS1717+6042 & 1437454188795350912 & sdB & $28850^{+ 210}_{- 150}$ & +$5.58^{+ 0.03}_{- 0.03}$ & $-2.49^{+ 0.04}_{- 0.03}$ & +$0.026^{+ 0.003}_{- 0.003}$ & $-11.095^{+ 0.003}_{- 0.004}$ & +$0.177^{+ 0.006}_{- 0.005}$ & $19.7^{+ 1.3}_{- 1.2}$ & $0.43^{+ 0.05}_{- 0.05}$ & T \\
HS1739+5244 & 1416603466898656384 & sdOB & $39110^{+ 220}_{- 180}$ & +$5.66^{+ 0.03}_{- 0.03}$ & $-3.90^{+ 0.24}_{- 0.29}$ & +$0.042^{+ 0.004}_{- 0.006}$ & $-11.355^{+ 0.005}_{- 0.007}$ & +$0.175^{+ 0.012}_{- 0.011}$ & $65^{+ 10}_{- 8}$ & $0.51^{+ 0.09}_{- 0.08}$ & T \\
HS1753+6618 & 1633291851249299968 & eHe-sdO & $46400^{+ 90}_{- 140}$ & +$5.61^{+ 0.03}_{- 0.02}$ & $>2.6$ & +$0.036^{+ 0.004}_{- 0.004}$ & $-10.931^{+ 0.003}_{- 0.003}$ & +$0.204^{+ 0.010}_{- 0.009}$ & $174^{+ 17}_{- 15}$ & $0.63^{+ 0.09}_{- 0.08}$ & T \\
HS1753+6911 & 1637779336156895360 & iHe-sdO & $43680^{+ 70}_{- 70}$ & +$5.81^{+ 0.02}_{- 0.02}$ & +$0.58^{+ 0.03}_{- 0.03}$ & +$0.047^{+ 0.020}_{- 0.021}$ & $-11.372^{+ 0.012}_{- 0.027}$ & +$0.184^{+ 0.015}_{- 0.014}$ & $111^{+ 19}_{- 17}$ & $0.79^{+ 0.17}_{- 0.14}$ & T \\
HS1756+7056 & 1638944818481870848 & sdB & $28700^{+ 400}_{- 600}$ & +$5.56^{+ 0.08}_{- 0.08}$ & $-3.42^{+ 0.14}_{- 0.17}$ & +$0.035^{+ 0.006}_{- 0.015}$ & $-11.366^{+ 0.007}_{- 0.016}$ & +$0.187^{+ 0.019}_{- 0.016}$ & $21^{+ 5}_{- 4}$ & $0.46^{+ 0.13}_{- 0.1}$ & T,N \\
HS1806+5024 & 2124662704347202944 & sdOB & $36790^{+ 250}_{- 400}$ & +$5.64^{+ 0.05}_{- 0.04}$ & $-1.47^{+ 0.03}_{- 0.03}$ & +$0.083^{+ 0.008}_{- 0.009}$ & $-11.623^{+ 0.007}_{- 0.009}$ & +$0.182^{+ 0.039}_{- 0.028}$ & $55^{+ 26}_{- 16}$ & $0.53^{+ 0.26}_{- 0.16}$ & T \\
HS1808+5647 & 2151694128996778880 & iHe-sdO & $51400^{+ 500}_{- 400}$ & +$5.68^{+ 0.07}_{- 0.06}$ & +$0.49^{+ 0.07}_{- 0.07}$ & +$0.041^{+ 0.008}_{- 0.012}$ & $-11.502^{+ 0.007}_{- 0.010}$ & +$0.261^{+ 0.043}_{- 0.033}$ & $430^{+ 160}_{- 100}$ & $1.2^{+ 0.5}_{- 0.4}$ & T \\
HS1813+7247 & 2266873808864525440 & sdB & $26760^{+ 250}_{- 500}$ & +$5.35^{+ 0.02}_{- 0.05}$ & $-2.67^{+ 0.08}_{- 0.09}$ & +$0.035^{+ 0.017}_{- 0.012}$ & $-11.260^{+ 0.010}_{- 0.010}$ & +$0.199^{+ 0.012}_{- 0.011}$ & $18.1^{+ 2.4}_{- 2.1}$ & $0.32^{+ 0.06}_{- 0.05}$ & T,P \\
HS1831+6432 & 2256432438426909952 & sdOB & $37570^{+ 200}_{- 160}$ & +$5.97^{+ 0.03}_{- 0.02}$ & $-1.63^{+ 0.02}_{- 0.02}$ & +$0.060^{+ 0.007}_{- 0.016}$ & $-11.371^{+ 0.006}_{- 0.016}$ & +$0.109^{+ 0.005}_{- 0.005}$ & $21.3^{+ 2.2}_{- 2.1}$ & $0.41^{+ 0.06}_{- 0.05}$ & T,S \\
HS1831+7647 & 2269474398679245312 & sdB & $22710^{+ 130}_{- 220}$ & +$5.40^{+ 0.02}_{- 0.02}$ & $-2.97^{+ 0.06}_{- 0.05}$ & +$0.087^{+ 0.004}_{- 0.004}$ & $-11.019^{+ 0.004}_{- 0.004}$ & +$0.226^{+ 0.006}_{- 0.006}$ & $12.1^{+ 0.8}_{- 0.8}$ & $0.46^{+ 0.05}_{- 0.05}$ & C,P \\
HS1832+6955 & 2259613463004902144 & iHe-sdO & $51230^{+ 240}_{- 230}$ & +$5.74^{+ 0.03}_{- 0.02}$ & +$0.56^{+ 0.05}_{- 0.05}$ & +$0.076^{+ 0.005}_{- 0.005}$ & $-11.207^{+ 0.003}_{- 0.003}$ & +$0.225^{+ 0.013}_{- 0.012}$ & $310^{+ 40}_{- 40}$ & $1.01^{+ 0.19}_{- 0.16}$ & T \\
HS1837+5913 & 2154810316748450432 & iHe-sdOB & $39530^{+ 60}_{- 90}$ & +$5.70^{+ 0.02}_{- 0.02}$ & $-0.21^{+ 0.02}_{- 0.02}$ & +$0.044^{+ 0.004}_{- 0.004}$ & $-11.242^{+ 0.004}_{- 0.004}$ & +$0.160^{+ 0.008}_{- 0.007}$ & $57^{+ 6}_{- 5}$ & $0.47^{+ 0.07}_{- 0.06}$ & T,L \\
HS1843+6343 & 2253326146279942144 & eHe-sdOB & $37910^{+ 120}_{- 90}$ & +$5.65^{+ 0.03}_{- 0.02}$ & $>2.6$ & +$0.062^{+ 0.005}_{- 0.013}$ & $-11.594^{+ 0.004}_{- 0.013}$ & +$0.171^{+ 0.027}_{- 0.021}$ & $54^{+ 19}_{- 13}$ & $0.48^{+ 0.17}_{- 0.12}$ & T,S,N \\
HS1843+6953 & 2259393595039224960 & sdO & $42100^{+ 500}_{- 500}$ & +$5.69^{+ 0.05}_{- 0.07}$ & $-3.34^{+ 0.37}_{- 0.60}$ & +$0.107^{+ 0.008}_{- 0.008}$ & $-11.354^{+ 0.008}_{- 0.007}$ & +$0.159^{+ 0.013}_{- 0.011}$ & $71^{+ 13}_{- 10}$ & $0.44^{+ 0.1}_{- 0.08}$ & C,N,R \\
HS1846+8149 & 2296351337648654848 & sdB & $25990^{+ 80}_{- 180}$ & +$5.52^{+ 0.01}_{- 0.02}$ & $-2.79^{+ 0.05}_{- 0.05}$ & +$0.061^{+ 0.004}_{- 0.003}$ & $-11.082^{+ 0.004}_{- 0.004}$ & +$0.193^{+ 0.006}_{- 0.006}$ & $15.2^{+ 1}_{- 0.9}$ & $0.45^{+ 0.05}_{- 0.04}$ & C \\
HS1859+6219 & 2252213921550077056 & sdB & $27140^{+ 280}_{- 250}$ & +$5.41^{+ 0.05}_{- 0.05}$ & $-2.52^{+ 0.05}_{- 0.08}$ & +$0.045^{+ 0.003}_{- 0.003}$ & $-11.230^{+ 0.004}_{- 0.005}$ & +$0.195^{+ 0.011}_{- 0.010}$ & $18.6^{+ 2.1}_{- 1.8}$ & $0.36^{+ 0.06}_{- 0.05}$ & T,N \\
HS1908+6534 & 2254558492658418048 & iHe-sdOB & $36200^{+ 800}_{- 800}$ & +$6.14^{+ 0.10}_{- 0.12}$ & +$0.21^{+ 0.06}_{- 0.06}$ & +$0.075^{+ 0.014}_{- 0.076}$ & $-11.713^{+ 0.010}_{- 0.078}$ & +$0.136^{+ 0.041}_{- 0.028}$ & $29^{+ 20}_{- 11}$ & $0.9^{+ 0.7}_{- 0.4}$ & T,N \\
HS1909+7004 & 2262587332720438400 & sdO & $40500^{+ 700}_{- 1000}$ & +$5.17^{+ 0.06}_{- 0.09}$ & $-2.88^{+ 0.15}_{- 0.15}$ & +$0.137^{+ 0.016}_{- 0.016}$ & $-11.375^{+ 0.016}_{- 0.016}$ & +$0.224^{+ 0.019}_{- 0.017}$ & $121^{+ 25}_{- 21}$ & $0.27^{+ 0.09}_{- 0.07}$ & T,R \\
HS1926+6947 & 2261817262264874752 & sdB & $25500^{+ 400}_{- 600}$ & +$5.27^{+ 0.06}_{- 0.08}$ & $-2.89^{+ 0.09}_{- 0.09}$ & +$0.185^{+ 0.004}_{- 0.004}$ & $-11.230^{+ 0.008}_{- 0.006}$ & +$0.208^{+ 0.012}_{- 0.011}$ & $16.3^{+ 2.1}_{- 1.9}$ & $0.29^{+ 0.07}_{- 0.06}$ & T,P \\
HS1939+6728 & 2248962562587900032 & eHe-sdO & $46290^{+ 230}_{- 600}$ & +$5.94^{+ 0.05}_{- 0.11}$ & +$2.06^{+ 0.33}_{- 1.24}$ & +$0.115^{+ 0.005}_{- 0.005}$ & $-11.284^{+ 0.004}_{- 0.004}$ & +$0.163^{+ 0.007}_{- 0.007}$ & $110^{+ 11}_{- 10}$ & $0.83^{+ 0.2}_{- 0.19}$ & T \\
HS2029+0301 & 4232792524791811968 & sdOB & $30420^{+ 110}_{- 100}$ & +$5.87^{+ 0.02}_{- 0.02}$ & $-2.18^{+ 0.03}_{- 0.03}$ & +$0.063^{+ 0.004}_{- 0.004}$ & $-11.302^{+ 0.003}_{- 0.004}$ & +$0.119^{+ 0.009}_{- 0.008}$ & $11^{+ 1.7}_{- 1.4}$ & $0.39^{+ 0.07}_{- 0.06}$ & T,L \\
HS2033+0507 & 4233159216213976832 & sdB & $27290^{+ 170}_{- 140}$ & +$5.43^{+ 0.03}_{- 0.02}$ & $-2.44^{+ 0.07}_{- 0.06}$ & +$0.110^{+ 0.004}_{- 0.004}$ & $-11.383^{+ 0.004}_{- 0.005}$ & +$0.330^{+ 0.098}_{- 0.062}$ & $50^{+ 40}_{- 19}$ & $1.1^{+ 0.8}_{- 0.4}$ & T \\
HS2033+0821 & 1749434329641481728 & sdOB & $32040^{+ 60}_{- 90}$ & +$6.00^{+ 0.01}_{- 0.02}$ & $-1.57^{+ 0.02}_{- 0.02}$ & +$0.084^{+ 0.003}_{- 0.003}$ & $-11.139^{+ 0.003}_{- 0.003}$ & +$0.124^{+ 0.004}_{- 0.004}$ & $14.6^{+ 1}_{- 0.9}$ & $0.56^{+ 0.06}_{- 0.06}$ & U \\
HS2035+0418 & 4233362621569490688 & sdB & $27510^{+ 200}_{- 220}$ & +$5.49^{+ 0.03}_{- 0.04}$ & $-2.52^{+ 0.04}_{- 0.05}$ & +$0.078^{+ 0.005}_{- 0.011}$ & $-11.116^{+ 0.006}_{- 0.011}$ & +$0.230^{+ 0.016}_{- 0.015}$ & $27^{+ 4}_{- 4}$ & $0.59^{+ 0.11}_{- 0.09}$ & T,R \\
HS2100+0650 & 1737015070927018368 & sdO & $60900^{+ 1400}_{- 2500}$ & +$6.04^{+ 0.06}_{- 0.06}$ & $-3.25^{+ 0.17}_{- 0.17}$ & +$0.071^{+ 0.004}_{- 0.008}$ & $-11.810^{+ 0.011}_{- 0.010}$ & +$0.075^{+ 0.017}_{- 0.012}$ & $70^{+ 40}_{- 21}$ & $0.23^{+ 0.12}_{- 0.07}$ & T \\
HS2100+1710 & 1764138098639251456 & sdOB & $35400^{+ 130}_{- 110}$ & +$5.81^{+ 0.02}_{- 0.02}$ & $-1.45^{+ 0.02}_{- 0.02}$ & +$0.101^{+ 0.016}_{- 0.016}$ & $-11.312^{+ 0.011}_{- 0.011}$ & +$0.154^{+ 0.010}_{- 0.009}$ & $33^{+ 5}_{- 4}$ & $0.56^{+ 0.09}_{- 0.08}$ & T,L \\
HS2108+1413 & 1759968097711823232 & eHe-sdO & $44020^{+ 90}_{- 100}$ & +$5.89^{+ 0.03}_{- 0.03}$ & +$1.64^{+ 0.15}_{- 0.19}$ & +$0.087^{+ 0.003}_{- 0.003}$ & $-11.166^{+ 0.003}_{- 0.003}$ & +$0.157^{+ 0.007}_{- 0.007}$ & $84^{+ 8}_{- 7}$ & $0.7^{+ 0.1}_{- 0.09}$ & T \\
HS2116+0045 & 2691204212499446272 & eHe-sdO & $45460^{+ 60}_{- 60}$ & +$5.63^{+ 0.02}_{- 0.02}$ & +$1.12^{+ 0.04}_{- 0.04}$ & +$0.073^{+ 0.004}_{- 0.004}$ & $-11.500^{+ 0.003}_{- 0.003}$ & +$0.186^{+ 0.038}_{- 0.027}$ & $130^{+ 60}_{- 40}$ & $0.54^{+ 0.26}_{- 0.16}$ & T,L,S \\
HS2121+1520 & 1783640205099886336 & eHe-sdO & $47510^{+ 170}_{- 160}$ & +$5.92^{+ 0.03}_{- 0.03}$ & +$1.04^{+ 0.09}_{- 0.11}$ & +$0.083^{+ 0.004}_{- 0.007}$ & $-11.687^{+ 0.003}_{- 0.007}$ & +$0.137^{+ 0.043}_{- 0.027}$ & $90^{+ 70}_{- 30}$ & $0.6^{+ 0.5}_{- 0.21}$ & T,L \\
HS2125+1105 & 1745787146491359104 & sdOB & $33390^{+ 80}_{- 80}$ & +$5.78^{+ 0.02}_{- 0.02}$ & $-1.71^{+ 0.02}_{- 0.02}$ & +$0.108^{+ 0.009}_{- 0.009}$ & $-11.540^{+ 0.007}_{- 0.008}$ & +$0.168^{+ 0.038}_{- 0.027}$ & $32^{+ 16}_{- 10}$ & $0.6^{+ 0.4}_{- 0.19}$ & U,P \\
HS2126+8320 & 2300497428494312960 & sdB & $27340^{+ 50}_{- 60}$ & +$5.46^{+ 0.01}_{- 0.01}$ & $-2.87^{+ 0.04}_{- 0.04}$ & +$0.099^{+ 0.008}_{- 0.008}$ & $-10.947^{+ 0.005}_{- 0.006}$ & +$0.204^{+ 0.005}_{- 0.005}$ & $20.9^{+ 1.1}_{- 1}$ & $0.44^{+ 0.05}_{- 0.04}$ & T,C \\
HS2143+8157 & 2297889558713251840 & sdB & $26980^{+ 300}_{- 200}$ & +$5.39^{+ 0.05}_{- 0.03}$ & $-2.60^{+ 0.07}_{- 0.08}$ & +$0.100^{+ 0.033}_{- 0.068}$ & $-11.256^{+ 0.016}_{- 0.053}$ & +$0.200^{+ 0.018}_{- 0.022}$ & $19^{+ 4}_{- 4}$ & $0.36^{+ 0.09}_{- 0.08}$ & T \\
HS2151+0214 & 2693917150066120704 & sdOB & $31680^{+ 230}_{- 220}$ & +$5.91^{+ 0.03}_{- 0.03}$ & $-2.31^{+ 0.06}_{- 0.06}$ & +$0.074^{+ 0.003}_{- 0.009}$ & $-11.487^{+ 0.004}_{- 0.009}$ & +$0.115^{+ 0.024}_{- 0.017}$ & $12^{+ 6}_{- 4}$ & $0.4^{+ 0.19}_{- 0.12}$ & T \\
HS2156+2517 & 1796522152051048064 & sdOB & $36770^{+ 270}_{- 400}$ & +$6.15^{+ 0.04}_{- 0.04}$ & $<-4.3$ & +$0.060^{+ 0.003}_{- 0.005}$ & $-11.323^{+ 0.006}_{- 0.006}$ & +$0.100^{+ 0.006}_{- 0.005}$ & $16.4^{+ 2.1}_{- 1.8}$ & $0.52^{+ 0.08}_{- 0.07}$ & C \\
HS2158+2137 & 1782619686510352128 & sdOB & $31710^{+ 100}_{- 80}$ & +$5.87^{+ 0.07}_{- 0.01}$ & $-2.24^{+ 0.03}_{- 0.04}$ & +$0.095^{+ 0.003}_{- 0.003}$ & $-11.066^{+ 0.003}_{- 0.003}$ & +$0.119^{+ 0.004}_{- 0.004}$ & $12.9^{+ 0.8}_{- 0.7}$ & $0.38^{+ 0.06}_{- 0.04}$ & C,L \\
HS2201+2610 & 1891930796082743680 & sdB & $28660^{+ 70}_{- 80}$ & +$5.57^{+ 0.02}_{- 0.02}$ & $-3.06^{+ 0.06}_{- 0.07}$ & +$0.054^{+ 0.003}_{- 0.003}$ & $-11.113^{+ 0.003}_{- 0.004}$ & +$0.226^{+ 0.012}_{- 0.011}$ & $31^{+ 4}_{- 2.8}$ & $0.7^{+ 0.09}_{- 0.08}$ & T,P \\
HS2206+2847 & 1893638956115949952 & sdOB & $31540^{+ 140}_{- 140}$ & +$5.83^{+ 0.02}_{- 0.02}$ & $-1.80^{+ 0.03}_{- 0.03}$ & +$0.094^{+ 0.004}_{- 0.039}$ & $-11.263^{+ 0.004}_{- 0.038}$ & +$0.154^{+ 0.010}_{- 0.012}$ & $21.1^{+ 2.9}_{- 4}$ & $0.58^{+ 0.1}_{- 0.09}$ & T \\
HS2208+2718 & 1892475088698125824 & sdB & $28880^{+ 160}_{- 150}$ & +$5.59^{+ 0.03}_{- 0.03}$ & $<-3.6$ & +$0.077^{+ 0.004}_{- 0.007}$ & $-11.241^{+ 0.004}_{- 0.007}$ & +$0.194^{+ 0.014}_{- 0.012}$ & $24^{+ 4}_{- 2.9}$ & $0.54^{+ 0.1}_{- 0.08}$ & T \\
HS2209+2840 & 1894911640824988288 & sdOB & $30640^{+ 100}_{- 150}$ & +$5.88^{+ 0.02}_{- 0.03}$ & $-1.76^{+ 0.03}_{- 0.03}$ & +$0.068^{+ 0.004}_{- 0.009}$ & $-11.392^{+ 0.004}_{- 0.009}$ & +$0.136^{+ 0.012}_{- 0.010}$ & $14.8^{+ 2.6}_{- 2.1}$ & $0.51^{+ 0.1}_{- 0.08}$ & T \\
HS2213+1336 & 2734200064947455232 & sdB & $21430^{+ 60}_{- 80}$ & +$4.95^{+ 0.01}_{- 0.01}$ & $-2.16^{+ 0.02}_{- 0.02}$ & +$0.109^{+ 0.007}_{- 0.010}$ & $-11.215^{+ 0.005}_{- 0.011}$ & +$0.378^{+ 0.058}_{- 0.045}$ & $27^{+ 10}_{- 7}$ & $0.46^{+ 0.16}_{- 0.11}$ & T,L,S \\
HS2218+0201 & 2703600484547535744 & sdB & $25020^{+ 100}_{- 110}$ & +$5.32^{+ 0.02}_{- 0.02}$ & $-2.85^{+ 0.06}_{- 0.06}$ & +$0.096^{+ 0.003}_{- 0.003}$ & $-10.945^{+ 0.003}_{- 0.003}$ & +$0.230^{+ 0.008}_{- 0.007}$ & $18.7^{+ 1.4}_{- 1.3}$ & $0.4^{+ 0.05}_{- 0.05}$ & T,P \\
HS2224+2618 & 1880674648872164224 & sdB & $21060^{+ 130}_{- 180}$ & +$4.80^{+ 0.02}_{- 0.03}$ & $-1.90^{+ 0.03}_{- 0.03}$ & +$0.087^{+ 0.003}_{- 0.003}$ & $-10.803^{+ 0.003}_{- 0.003}$ & +$0.477^{+ 0.022}_{- 0.020}$ & $40^{+ 4}_{- 4}$ & $0.52^{+ 0.07}_{- 0.06}$ & C \\
HS2225+2220 & 1875278658119889408 & sdOB & $31510^{+ 60}_{- 60}$ & +$5.85^{+ 0.01}_{- 0.01}$ & $-1.96^{+ 0.02}_{- 0.02}$ & +$0.058^{+ 0.004}_{- 0.003}$ & $-11.422^{+ 0.004}_{- 0.003}$ & +$0.131^{+ 0.014}_{- 0.012}$ & $15^{+ 4}_{- 2.6}$ & $0.44^{+ 0.11}_{- 0.08}$ & T,S \\
HS2225+2344 & 1875700252109494784 & eHe-sdO & $59900^{+ 600}_{- 500}$ & +$5.82^{+ 0.04}_{- 0.05}$ & +$0.77^{+ 0.08}_{- 0.07}$ & +$0.040^{+ 0.005}_{- 0.017}$ & $-11.372^{+ 0.004}_{- 0.016}$ & +$0.143^{+ 0.012}_{- 0.011}$ & $240^{+ 50}_{- 40}$ & $0.49^{+ 0.11}_{- 0.09}$ & T \\
HS2229+2628 & 1880738592345162752 & sdB & $19280^{+ 290}_{- 500}$ & +$4.41^{+ 0.07}_{- 0.07}$ & $-1.93^{+ 0.08}_{- 0.08}$ & +$0.074^{+ 0.004}_{- 0.010}$ & $-11.066^{+ 0.006}_{- 0.010}$ & +$0.677^{+ 0.107}_{- 0.082}$ & $56^{+ 20}_{- 14}$ & $0.43^{+ 0.17}_{- 0.12}$ & C \\
HS2233+1418 & 2732982626402453888 & sdOB & $35470^{+ 190}_{- 200}$ & +$5.17^{+ 0.02}_{- 0.03}$ & $-2.51^{+ 0.05}_{- 0.06}$ & +$0.068^{+ 0.003}_{- 0.003}$ & $-11.074^{+ 0.004}_{- 0.004}$ & +$0.211^{+ 0.023}_{- 0.019}$ & $63^{+ 15}_{- 11}$ & $0.24^{+ 0.06}_{- 0.05}$ & L \\
HS2233+2332 & 1876904938896587008 & sdB & $26610^{+ 50}_{- 50}$ & +$5.45^{+ 0.01}_{- 0.01}$ & $-2.93^{+ 0.03}_{- 0.03}$ & +$0.035^{+ 0.003}_{- 0.003}$ & $-11.056^{+ 0.003}_{- 0.004}$ & +$0.205^{+ 0.011}_{- 0.010}$ & $19.1^{+ 2.1}_{- 1.8}$ & $0.44^{+ 0.06}_{- 0.05}$ & T,C,P \\
HS2234+8457 & 2301069415059458560 & eHe-sdO & $45990^{+ 170}_{- 140}$ & +$5.93^{+ 0.03}_{- 0.04}$ & +$1.57^{+ 0.25}_{- 0.28}$ & +$0.337^{+ 0.017}_{- 0.017}$ & $-11.275^{+ 0.012}_{- 0.012}$ & +$0.193^{+ 0.011}_{- 0.010}$ & $150^{+ 17}_{- 15}$ & $1.15^{+ 0.21}_{- 0.18}$ & T \\
HS2240+0136 & 2654793575707334784 & sdOB & $30520^{+ 60}_{- 70}$ & +$5.95^{+ 0.01}_{- 0.01}$ & $-2.34^{+ 0.02}_{- 0.02}$ & +$0.075^{+ 0.005}_{- 0.005}$ & $-11.094^{+ 0.004}_{- 0.004}$ & +$0.117^{+ 0.004}_{- 0.004}$ & $10.7^{+ 0.7}_{- 0.6}$ & $0.44^{+ 0.05}_{- 0.04}$ & T,L \\
HS2240+1031 & 2717582699040219392 & sdOB & $35130^{+ 100}_{- 100}$ & +$5.80^{+ 0.02}_{- 0.01}$ & $-1.53^{+ 0.02}_{- 0.02}$ & +$0.068^{+ 0.004}_{- 0.005}$ & $-11.277^{+ 0.003}_{- 0.004}$ & +$0.141^{+ 0.009}_{- 0.008}$ & $27^{+ 4}_{- 2.9}$ & $0.46^{+ 0.07}_{- 0.06}$ & T,L \\
HS2242+3206 & 1890677009230168704 & sdB & $29160^{+ 40}_{- 40}$ & +$5.71^{+ 0.01}_{- 0.01}$ & $-2.93^{+ 0.02}_{- 0.02}$ & +$0.095^{+ 0.002}_{- 0.003}$ & $-10.974^{+ 0.003}_{- 0.004}$ & +$0.162^{+ 0.004}_{- 0.004}$ & $17.1^{+ 0.9}_{- 0.9}$ & $0.49^{+ 0.05}_{- 0.05}$ & T,L \\
HS2246+0158 & 2656170267344606464 & sdOB & $32800^{+ 700}_{- 700}$ & +$5.65^{+ 0.10}_{- 0.09}$ & $-1.76^{+ 0.07}_{- 0.07}$ & +$0.085^{+ 0.004}_{- 0.008}$ & $-11.373^{+ 0.007}_{- 0.010}$ & +$0.122^{+ 0.012}_{- 0.010}$ & $15^{+ 4}_{- 2.6}$ & $0.24^{+ 0.08}_{- 0.06}$ & T \\
HS2304+0118 & 2652331593079280000 & eHe-sdO & $49900^{+ 500}_{- 600}$ & +$5.66^{+ 0.12}_{- 0.11}$ & +$0.87^{+ 0.12}_{- 0.20}$ & +$0.054^{+ 0.008}_{- 0.008}$ & $-11.565^{+ 0.005}_{- 0.006}$ & +$0.116^{+ 0.016}_{- 0.013}$ & $76^{+ 23}_{- 16}$ & $0.23^{+ 0.09}_{- 0.07}$ & T,L \\
HS2307+3345 & 1911572540520290048 & eHe-sdO & $49500^{+ 400}_{- 250}$ & +$6.07^{+ 0.05}_{- 0.07}$ & +$0.69^{+ 0.09}_{- 0.08}$ & +$0.086^{+ 0.005}_{- 0.010}$ & $-11.443^{+ 0.003}_{- 0.009}$ & +$0.160^{+ 0.020}_{- 0.016}$ & $140^{+ 40}_{- 26}$ & $1.1^{+ 0.4}_{- 0.25}$ & T \\
HS2333+0014 & 2644373607090404480 & eHe-sdO & $61880^{+ 300}_{- 400}$ & +$5.65^{+ 0.02}_{- 0.02}$ & +$0.71^{+ 0.04}_{- 0.03}$ & +$0.051^{+ 0.004}_{- 0.005}$ & $-11.608^{+ 0.003}_{- 0.004}$ & +$0.185^{+ 0.055}_{- 0.035}$ & $500^{+ 400}_{- 160}$ & $0.6^{+ 0.4}_{- 0.2}$ & T,S \\
HS2357+2201 & 2847977322031768832 & sdB & $27336^{+ 20}_{- 30}$ & +$5.57^{+ 0.01}_{- 0.01}$ & $-2.57^{+ 0.02}_{- 0.02}$ & +$0.065^{+ 0.005}_{- 0.005}$ & $-11.013^{+ 0.004}_{- 0.004}$ & +$0.179^{+ 0.007}_{- 0.006}$ & $16.1^{+ 1.2}_{- 1.1}$ & $0.44^{+ 0.05}_{- 0.05}$ & U \\
HS2359+1942 & 2846319155418023424 & sdOB & $30860^{+ 90}_{- 70}$ & +$5.57^{+ 0.01}_{- 0.02}$ & $-3.80^{+ 0.08}_{- 0.08}$ & +$0.041^{+ 0.011}_{- 0.019}$ & $-11.355^{+ 0.007}_{- 0.017}$ & +$0.231^{+ 0.034}_{- 0.027}$ & $44^{+ 14}_{- 10}$ & $0.72^{+ 0.24}_{- 0.17}$ & U,1 \\
\\
\bottomrule
\end{longtable}

\tablefoot{  
Spectra used: DSAZ spectra \citep[see also][]{2003A&A...400..939E}: (T) 3.5m Twin, (C) 2.2m CAFOS, (B) 3.5m B\&C (used for comparison only), (C2) 2.2m, Cassegrain spectrograph (used for comparison only); ESO spectra used: (U) UVES$@$VLT; other spectra: (L) LAMOST, (S) SDSS, (AM) AM-BOK \citep{2025arXiv251102539L}; additional information: (P) Pulsator, (R) Reflection effect binary, (N) no IR (H-band) photometry, 
 (1) sdOB+WD  \citep[P=0.933d][]{2014A&A...562A..95G}: binaries with known orbit.
}

\end{landscape}
}

\begin{table*}
\caption{Results for stars without suitable parallaxes ($\delta \varpi/\varpi>0.25$).
For these objects, we only list their spectral classification, atmospheric parameters, $E(44-55)$, and $\Theta$. }\label{tab:bad_params}
\begin{center}
\renewcommand{\arraystretch}{1.2}
\setlength{\tabcolsep}{0.25em}
\begin{tabular}{lllllllllll}
\toprule\toprule
Name & \textit{Gaia} DR3 ID & Class & $T_\mathrm{eff}$ & $\log{g}$ & $\log{n(\mathrm{He}/\mathrm{H})}$ & $E(44-55)$ & $\log \Theta$ & comment\\
  &   &   & [K] & [cgs] &  & [mag] & [rad] &  \\
\midrule
HS0019+3944 & 380123129891300608 & eHe-sdO & $46610^{+ 170}_{- 240}$ & +$5.98^{+ 0.04}_{- 0.06}$ & +$0.83^{+ 0.11}_{- 0.05}$ & +$0.083^{+ 0.006}_{- 0.014}$ & $-11.791^{+ 0.004}_{- 0.014}$ & T \\
HS0024+3331 & 2863147593197468032 & iHe-sdO & $60300^{+ 500}_{- 500}$ & +$5.53^{+ 0.03}_{- 0.03}$ & $-0.18^{+ 0.02}_{- 0.03}$ & +$0.066^{+ 0.007}_{- 0.013}$ & $-11.793^{+ 0.006}_{- 0.013}$ & T,L \\
HS0110+3222 & 313700498585403776 & eHe-sdO & $45780^{+ 90}_{- 90}$ & +$6.04^{+ 0.03}_{- 0.04}$ & +$1.10^{+ 0.08}_{- 0.08}$ & +$0.081^{+ 0.005}_{- 0.007}$ & $-11.549^{+ 0.003}_{- 0.006}$ & T,2,N\\
HS0735+4026 & 923580185261224576 & eHe-sdO & $61600^{+ 400}_{- 500}$ & +$5.59^{+ 0.02}_{- 0.03}$ & +$0.66^{+ 0.04}_{- 0.04}$ & +$0.053^{+ 0.018}_{- 0.022}$ & $-11.720^{+ 0.018}_{- 0.025}$ & T,S \\
HS0941+4649 & 821716652060662144 & sdOB & $34200^{+ 160}_{- 280}$ & +$5.11^{+ 0.02}_{- 0.03}$ & $-2.54^{+ 0.05}_{- 0.08}$ & +$0.005^{+ 0.010}_{- 0.005}$ & $-11.661^{+ 0.009}_{- 0.012}$ & B,L,S,N \\
HS1000+4704 & 822338490313075712 & iHe-sdOB & $38860^{+ 120}_{- 140}$ & +$5.76^{+ 0.03}_{- 0.03}$ & +$0.04^{+ 0.02}_{- 0.02}$ & +$0.036^{+ 0.009}_{- 0.009}$ & $-11.857^{+ 0.008}_{- 0.008}$ & B,S\\
HS1320+2622 & 1447552691995517696 & sdB & $28600^{+ 100}_{- 100}$ & +$5.62^{+ 0.02}_{- 0.02}$ & $-2.74^{+ 0.08}_{- 0.08}$ & +$0.039^{+ 0.010}_{- 0.016}$ & $-11.617^{+ 0.008}_{- 0.021}$ & T,R \\
HS1732+7023 & 1638672002159650176 & sdO & $52000^{+ 1000}_{- 800}$ & +$5.52^{+ 0.06}_{- 0.06}$ & $-1.29^{+ 0.08}_{- 0.08}$ & +$0.052^{+ 0.004}_{- 0.016}$ & $-11.567^{+ 0.005}_{- 0.015}$ & T,N \\
HS1741+2133 & 4556063035156496768 & sdOB & $39300^{+ 400}_{- 500}$ & +$5.64^{+ 0.05}_{- 0.06}$ & $-2.43^{+ 0.08}_{- 0.09}$ & +$0.167^{+ 0.022}_{- 0.024}$ & $-11.052^{+ 0.014}_{- 0.018}$ & L,C2,1 \\
HS2149+0847 & 2725865595010422016 & sdOB & $36780^{+ 200}_{- 130}$ & +$5.80^{+ 0.03}_{- 0.02}$ & $-1.81^{+ 0.05}_{- 0.05}$ & +$0.046^{+ 0.005}_{- 0.020}$ & $-11.647^{+ 0.004}_{- 0.019}$ & AM,P \\
HS2205+0548 & 2720410333709020160 & eHe-sdO & $46320^{+ 120}_{- 120}$ & +$5.86^{+ 0.03}_{- 0.03}$ & +$1.00^{+ 0.08}_{- 0.08}$ & +$0.132^{+ 0.005}_{- 0.029}$ & $-11.655^{+ 0.003}_{- 0.021}$ & T,S \\
HS2231+0749 & 2710029363395012224 & eHe-sdO & $45800^{+ 500}_{- 500}$ & +$5.58^{+ 0.06}_{- 0.13}$ & $>2.4$ & +$0.166^{+ 0.009}_{- 0.009}$ & $-11.642^{+ 0.006}_{- 0.006}$ & T \\
\\
\bottomrule
\end{tabular}
\tablefoot{
Spectra used: DSAZ spectra \citep[see also][]{2003A&A...400..939E}: (T) 3.5m Twin, (C) 2.2m CAFOS, (B) 3.5m B\&C (used for comparison only), (C2) 2.2m, Cassegrain spectrograph (used for comparison only); other spectra: (L) LAMOST, (S) SDSS, (AM) AM-BOK \citep{2025arXiv251102539L}; additional information: (P) Pulsator, (R) Reflection effect binary, (N) no IR (H-band) photometry, (1) sdOB+WD \citep[P=0.2d,][]{2014ASPC..481..293K}, (2) Outlier
}
\end{center}
\end{table*}
\end{appendix}

\end{document}